\documentclass[12pt, draftclsnofoot, onecolumn]{IEEEtran}
\ifCLASSINFOpdf

\else

\fi

\usepackage{mathrsfs}%flower
\usepackage{diagbox}
\usepackage{amsmath}
\usepackage{ntheorem}
\allowdisplaybreaks

\usepackage{makeidx}  
\usepackage{algorithm}
\usepackage{algorithmic}
\usepackage{graphicx}
\usepackage{float}
\usepackage{subfigure}
\usepackage{epstopdf}
\usepackage{extarrows}
\usepackage{bm}
\usepackage{cite}
\usepackage{stfloats}

\usepackage{hyperref}

\usepackage{caption}
\usepackage{amssymb}
\newtheorem{theorem}{Theorem}

\usepackage{tabularx,booktabs}
\newcolumntype{C}{>{\centering\arraybackslash}X} 
\usepackage{lipsum}

\usepackage{makecell} 

\usepackage{color}

\newtheorem{corollary}{Corollary}
\usepackage{algorithm}
\usepackage{algorithmic}
\makeatletter
\newenvironment{breakablealgorithm}
{% \begin{breakablealgorithm}
		\begin{center}
			\refstepcounter{algorithm}% New algorithm
			\hrule height.8pt depth0pt \kern2pt% \@fs@pre for \@fs@ruled
			\renewcommand{\caption}[2][\relax]{% Make a new \caption
				{\raggedright\textbf{\ALG@name~\thealgorithm} ##2\par}%
				\ifx\relax##1\relax % #1 is \relax
				\addcontentsline{loa}{algorithm}{\protect\numberline{\thealgorithm}##2}%
				\else % #1 is not \relax
				\addcontentsline{loa}{algorithm}{\protect\numberline{\thealgorithm}##1}%
				\fi
				\kern2pt\hrule\kern2pt
			}
		}{% \end{breakablealgorithm}
		\kern2pt\hrule\relax% \@fs@post for \@fs@ruled
	\end{center}
}
\makeatletter

\begin{document}
\setlength{\abovedisplayskip}{4pt}  
\setlength{\belowdisplayskip}{4pt}
	
%\title{Two-Timescale Transmission Design for RIS-assisted Cell-Free Massive MIMO Systems With Hardware Impairments}
\title{RIS-assisted Cell-Free Massive MIMO Systems With Two-Timescale Design and Hardware Impairments}
%\title{RIS-assisted Cell-Free MIMO Systems With Two-Timescale CSI and Hardware Impairments: Power Scaling Analysis and Phase Shift Optimization}
%
%Performance Analysis and Optimization for 
%\textit{Fellow, IEEE}
\author{Jianxin Dai, Jin Ge, Kangda Zhi, Cunhua Pan, and Youguo Wang
	%and Cheng-Xiang Wang, \textit{Fellow, IEEE}
	\textit{}\thanks{
		This work of Jianxin Dai was supported in part by the open research fund of National Mobile Communications Research Laboratory, Southeast University. (No. 2023D03).
		
		Jianxin Dai is with the School of Science, Nanjing University of Posts and Telecommunications, Nanjing 210023, China, and also with the National Mobile Communications Research Laboratory, Southeast University, Nanjing 210096, China. (email: daijx@njupt.edu.cn).
		Jin Ge is with the College of Telecommunications and Information Engineering, Nanjing University of Posts and Telecommunications, Nanjing 210003, China. (email: 1221014211@njupt.edu.cn).
		Kangda Zhi is with the School of Electronic Engineering and Computer Science at Queen Mary University of London, London E1 4NS, U.K. (email: k.zhi@qmul.ac.uk).
		Cunhua Pan is with the National Mobile Communications Research Laboratory, Southeast University, Nanjing 210096, China. (email: cpan@seu.edu.cn).
		Youguo Wang is with the School of Science, Nanjing University of Posts and Telecommunications, Nanjing 210023, China. (email: wangyg@njupt.edu.cn).  
%		Cheng-Xiang Wang is with the National Mobile Communications Research	Laboratory, Southeast University, Nanjing 210096, China, and also with the Purple Mountain Laboratories, Pervasive Communication Research Center, Nanjing 211111, China. (e-mail: chxwang@seu.edu.cn).
%		Elza Erkip is with the Electrical and Computer Engineering Department, New York University Tandon School of Engineering, Brooklyn, NY 11201 USA. (e-mail: elza@nyu.edu).
%		(Corresponding author: Cunhua Pan).
	}}
\maketitle
	
\begin{abstract}
%	Due to the strong ability resist the inter-cell interference, cell-free (CF) network has been widely recognized as a promising technique to achieve high network capacity. To reduce the cost and power consumption caused by deploying base stations (BSs), we 
	Integrating the reconfigurable intelligent surface (RIS) into a cell-free massive multiple-input multiple-output (CF-mMIMO) system is an effective solution to achieve high system capacity with low cost and power consumption. However, existing works of RIS-assisted systems mostly assumed perfect hardware, while the impact of hardware impairments (HWIs) is generally ignored. In this paper, we consider the general Rician fading channel and uplink transmission of the RIS-assisted CF-mMIMO system under transceiver impairments and RIS phase noise. To reduce the feedback overhead and 
	power consumption, we propose a two-timescale transmission scheme to optimize the passive  beamformers at RISs with statistical channel state information (CSI), while transmit beamformers at access points (APs) are designed based on instantaneous CSI. Also, the maximum ratio combining (MRC) detection is applied to  the central processing unit (CPU). On this basis, we derive the closed-form approximate expression of the achievable rate, based on which the impact of HWIs and the power scaling laws are analyzed to draw useful theoretical insights. To maximize the users' sum rate or minimum rate, we first transform our rate expression into a tractable form, and then optimize the phase shifts of RISs based on an accelerated gradient ascent method. Finally, numerical results are presented to demonstrate the correctness of our derived expressions and validate the previous analysis, which provide some guidelines for the practical application of the imperfect RISs in the CF-mMIMO with transceiver HWIs.
\end{abstract}
\vspace{0cm}
\IEEEpeerreviewmaketitle
	
\section{Introduction}
Cell-free massive multiple-input multiple-output (CF-mMIMO) has been well recognized as a critical technique in future wireless communications \cite{9586055}, which connects multiple access points (APs) without cell boundaries to a central processing unit (CPU), and each AP coordinates with each other to provide uniform quality-of-service (QoS) to all users \cite{7421222,8599043,7827017}. Due to the cooperation among distributed APs, the interference between cells is effectively reduced, and thus the system capacity is improved. However, deploying lots of antennas at APs requires high hardware cost and power consumption, which limits the application of CF-mMIMO systems when the budget is limited.

Fortunately, reconfigurable intelligent surface (RIS) has emerged as an effective solution to the above issue by adopting low-cost passive elements instead of expensive radio frequency (RF) chains \cite{wu2020intelligent,8741198,di2020smart}. In particular, the RIS is an array composed of massive low-cost passive reflecting elements that does not require active RF chains and power amplifiers, which can achieve fine-grained passive beamforming by inducing phase shifts and amplitude changes with the aid of a controller. Therefore, it is promising to integrate RISs into CF-mMIMO systems, which achieves satisfactory performance at lower cost and power consumption \cite{pan2020multicell}.
 
Specifically, relying on instantaneous channel state information (CSI), Zhang \emph{et al.} \cite{9352948} proposed a hybrid beamforming scheme to maximize the energy efficiency, and Liu \emph{et al.} \cite{9448858} adopted an iterative precoding algorithm to maximize the energy efficiency of the worst user in the wideband scenario. Moreover, the authors of \cite{9771957} proposed a generalized superimposed channel estimation scheme assisted by an RIS to improve the spectral efficiency and wireless coverage of the uplink CF-mMIMO system. Besides, the authors of \cite{9806349} investigated the  uplink network throughput of RIS-assisted CF-mMIMO systems under the spatially correlated Rayleigh fading channel.

The above-mentioned contributions mainly considered to design the passive and active beamforming based on the rapidly-varying instantaneous CSI, which requires channel estimation in each coherence interval and results in high pilot overhead due to the large dimensionality of the RIS-related channels \cite{9847080}. In this regard, several works have been proposed to apply fully statistical CSI for RIS-assisted systems, which significantly decrease the pilot overhead and computational complexity \cite{9181610,9656609,9500188}. Nevertheless, using the fully statistical CSI for all beamformers may result in a serious rate degradation for the RIS-assisted system since the rapidly-varying environment information is not exploited at all. For the balance between the performance and overhead, a beamforming design named two-timescale design has been proposed in recent contributions, which applied the two-timescale scheme to cellular mMIMO systems aided by only one RIS \cite{9198125,9973349}.
 
In the two-timescale transmission, slowly-changing statistical CSI is used to design the RIS phase shifts, while the active beamformers at APs are designed according to instantaneous CSI. Compared with other two schemes, the two-timescale scheme reduces pilot overhead and power consumption while satisfying the requirement of QoS. For more complicated scenarios and insights, several works have applied the two-timescale design for RIS-assisted CF-mMIMO systems \cite{9665300,fullversion}. Specifically, the data rate of RIS-assisted CF-mMIMO systems over the spatially correlated channel has been investigated in \cite{9665300}, and the authors adopted a channel estimation approach to decrease the overhead from channel estimation. Moreover, the RIS-assisted CF-mMIMO system under the Rician channel and perfect hardware has been investigated in \cite{fullversion}, which conducted specific mathematical analysis and used conventional genetic algorithm to design the phase shifts of RISs. 

It is worth noting that all the aforementioned works of RIS-assisted systems assumed perfect hardware, while practical systems suffer from non-negligible transceiver hardware impairments (HWIs) and the phase noise caused by the limited precision of the RIS \cite{9295369}. 
As a result, several preliminary works have been proposed to reveal the impact of transceiver HWIs and RIS phase noise on various RIS-assisted systems \cite{9374557,9828539,9833357}. Specifically, the authors of \cite{9374557} considered the transceiver HWIs in an RIS-assisted MISO system and proposed a robust beamforming design to maximize the secrecy rate. In \cite{9828539}, the authors proposed two novel transmission schemes to improve the performance of RIS-assisted space-shift keying systems and investigated the impact of non-ideal transceivers. For RIS-assisted multi-user mMIMO systems, the authors of \cite{9833357} analyzed and optimized the achievable rate in the presence of the RIS phase noise and transceiver HWIs.

%9239335
%In general, transceiver HWIs can be divided into two main categories, i.e., the multiplicative and additive HWIs, where the impact of the additive transceiver HWIs are the focus of this work. In this direction, existing works with

Nevertheless, the transceiver HWIs and RIS phase noise under the cell-free architecture of RIS-assisted mMIMO systems have not been investigated \cite{fullversion}. Therefore, their impact and properties are still unknown. To balance the complexity and the rate performance, this paper proposes a two-timescale design for the uplink channel of RIS-assisted CF-mMIMO systems based on HWIs at both RISs and the transceiver. 
%We conduct mathematical analysis based on the derived rate expression to provide some valuable insights. Besides, we obtain a tractable rate expression by transforming formulas, and adopt a novel gradient ascent-based method to design the phase shifts of RISs.
%Based on that, we derive the closed-form approximate expression of the achievable rate and provide some useful insights for the RIS-assisted CF-mMIMO system with HWIs. Then, we propose a projected gradient ascent (PGA)-based method to optimize the phase shifts of RIS. Finally, we conduct some numerical results to analyze the system performance and verify our analysis. 
The main contributions are summarized as follows:
\begin{itemize}
	\item First, we model an uplink channel of RIS-assisted CF-mMIMO system with HWIs under the general Rician fading channel. Considering the complicated environment and the height of the RIS, both line-of-sight (LoS) and non-line-of-sight (NLoS) paths would exist in RIS-assisted channels, and we can gain some key insights into the practical application of RISs by adjusting Rician factors. Unlike the existing works \cite{9352948,9448858,9806349,9771957}, we propose a two-timescale-based design for RIS-assisted CF-mMIMO systems. Besides, in contrast to the assumption of perfect hardware, we consider the RIS phase noise and transceiver HWIs, which requires more complicated derivations and theoretical analysis of rate performance. 
	Based on these considerations, we can draw valuable insights into the benefits of deploying imperfect RIS in CF-mMIMO systems with transceiver HWIs by analyzing the rate expression and numerical results.
	
	\item Next, we derive the closed-form approximate expression of the achievable rate under the low-complexity maximum-ratio combining (MRC) detector, which characterizes the impacts of  key parameters on the rate. Aided by the derived results, we provide some analysis for the asymptotic behaviors of the rate, the impacts of HWIs, and the power scaling laws. 
	We find that the transmit power of users can be reduced at most by a factor $ \frac{1}{BR} $ while maintaining a satisfactory system performance, where $B$ and $R$ denote the number of AP antennas and RIS elements, respectively. Besides, the transceiver HWIs will restrict the gain brought by a large number of AP antennas or RIS elements. When there is no LoS path in the RIS-assisted channels, the RIS phase noise will not affect the rate performance.
	 These insights provide some useful guidelines for applying imperfect RISs to CF-mMIMO systems with transceiver HWIs.
	
	\item Then, due to the tightly coupled phase shifts in our complicated expressions, we transform our formulas to get a tractable rate expression, avoiding the intractable gradient for the whole phase shift vector of all RISs. Based on the tractable objective function, we optimize the RIS phase shifts using an accelerated gradient ascent-based method to solve users' sum rate and minimum rate maximization problems. The proposed gradient method avoids the suboptimality caused by the projection operation and effectively increases the convergence speed \cite{9973349}.
	For the phase shift optimization of each RIS, the proposed gradient ascent method avoids the rate loss caused by the projection operation.
	 since the objective function is periodic with the phase shifts vector and the unit modulus constraint holds for every phase shifts vector.
	
	\item Finally, we present extensive simulations based on our rate expression and the accelerated gradient ascent-based method. Our numerical results reveal that even though the RISs-APs links are in rich or less scattering environment, the balance between system capacity and user fairness can be maintained in RIS-assisted CF-mMIMO systems under HWIs. Then, we verify the gains of applying imperfect RISs to the CF-mMIMO system with transceiver HWIs. Meanwhile, we evaluate the impacts of HWIs and key parameters on the rate. Besides, we validate the power scaling laws for the number of RIS elements or AP antennas. The above analytical and numerical results provide valuable insights for deploying imperfect RIS in CF-mMIMO systems with transceiver HWIs.
\end{itemize}

The remainder of this paper is organized as follows. The channel model and uplink transmission of RIS-assisted CF-mMIMO systems with HWIs are described in Section \ref{section2}. Section \ref{section3} derives the closed-form approximate expression of the rate and discusses some special cases. The accelerated gradient ascent method for maximizing users' sum rate and minimum rate is presented in Section \ref{section4}. Section \ref{section5} gives numerical simulations, and Section \ref{section6} concludes this paper.

\emph{Notations}: Vectors and matrices are denoted by bold lowercase and uppercase letters, respectively. $\mathbf{A}^H$,  $\mathbf{A}^T$, and $\mathbf{A}^*$ respectively denote the conjugate transpose, transpose, and conjugate. $\left|\mathbf{a}\right|$  denotes the modulus of the complex number and  $\left\|\mathbf{a}\right\|$ denotes $l_2$-norm of the vector. 
The real and trace operators are denoted by Re$\left\{\cdot\right\}$ and $\mathrm{Tr}\left\{\cdot\right\}$, respectively. The expectation operators are denoted by $\mathbb{E}\left\{\cdot\right\}$. $\mathbf{I}_S$ and $\mathbf{0}$ respectively denote an $S \times S$ identity matrix and a zero matrix with appropriate dimension. $\mathbb{C}^{L\times S}$ denotes the space of $L\times S$ complex matrix. Besides, $ x\sim{\cal C}{\cal N}\left( {a,b} \right)$ is a complex Gaussian distributed random variable with mean $a$ and variance $b$. Operation $\left\lfloor n \right\rfloor $ and $\left\lceil n \right\rceil $ respectively denote the nearest integer smaller than and greater than $n$, and operation $\bmod$ means returning the remainder after division. $\widetilde{diag}\{\cdot\}$ denotes a diagonal matrix whose diagonal elements are the same as the original matrix. $\mathcal{O}$ denotes the standard big-O notation. 

\section{System Model}\label{section2} 
\begin{figure}[t]
	\setlength{\abovecaptionskip}{0pt}
	\setlength{\belowcaptionskip}{-10pt}
	\centering
	\includegraphics[width=3.8in]{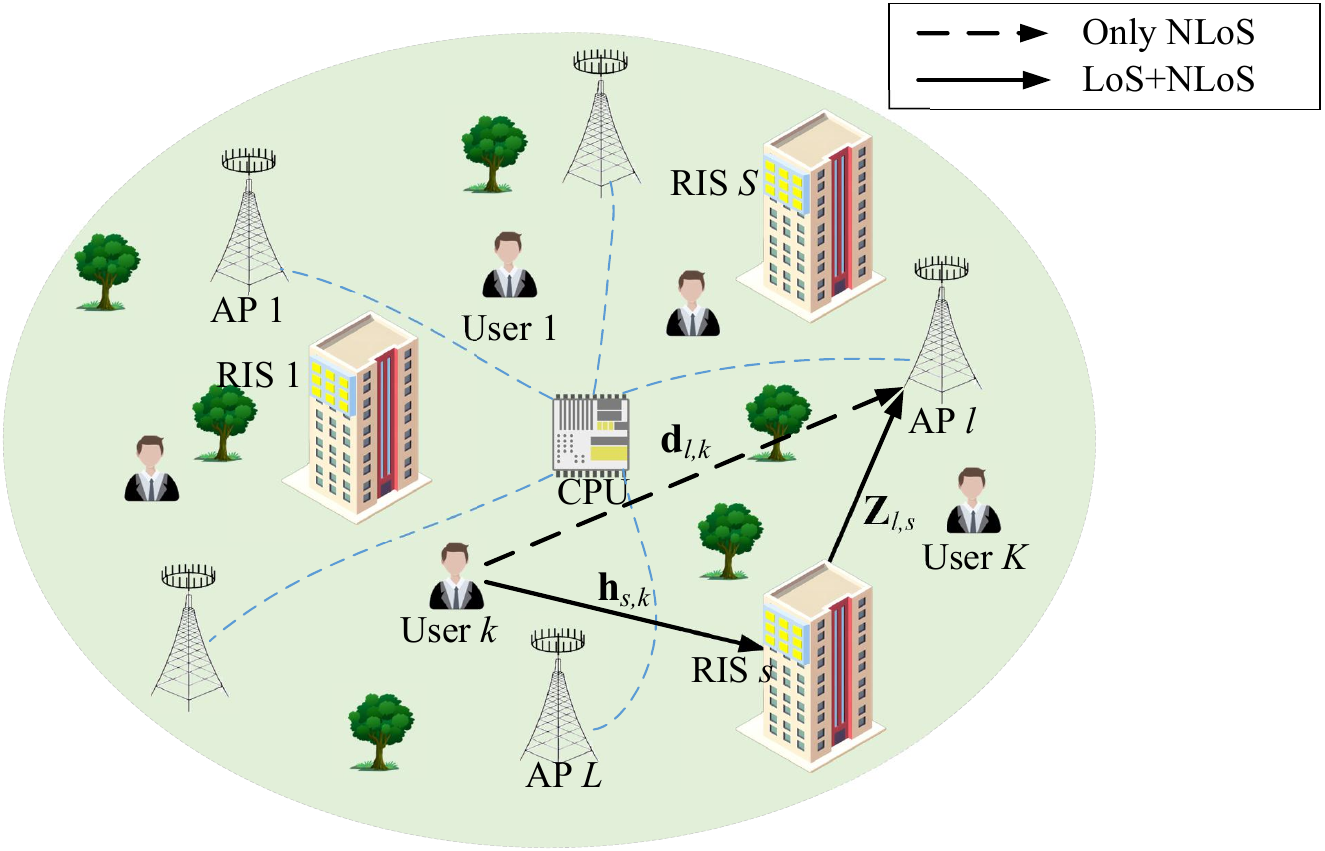}
	\DeclareGraphicsExtensions.
	\caption{The RIS-assisted CF-mMIMO system under uplink transmission.}
	\label{figure0}
\end{figure}

%\subsection{System Architecture}
As shown in Fig. \ref{figure0}, we consider the uplink transmission of a typical RIS-assisted CF-mMIMO system, where a CPU is deployed for system control and $L$ APs are connected to the CPU via optical cables, while the CPU uses wireless control to manage $S$ RISs, all RISs are deployed near the $K$ single-antenna users. Each RIS consists of $R$ reflecting elements, and each AP is equipped with $B$ antennas. For convenience, we denote the sets of RISs, reflecting elements, APs, antennas, and users as $\mathcal{S}=\left\{1, 2,\ldots,S\right\}$, $\mathcal{R}=\left\{1, 2,\ldots,R\right\}$,  $\mathcal{L}=\left\{1, 2,\ldots,L\right\}$, $\mathcal{B}=\left\{1, 2,\ldots,B\right\}$, and $\mathcal{K}=\left\{1, 2,\ldots,K\right\}$, respectively.
 The channels from user $k$ to RIS $s$, from RIS $s$ to AP $l$, and from user $k$ to AP $l$ are respectively denoted by $\mathbf {h}_{s,k} \in \mathbb{C}^{R\times 1}$,  $\mathbf {Z}_{l,s} \in \mathbb{C}^{B\times R}$, and $\mathbf {d}_{l,k} \in \mathbb{C}^{B\times 1}$, where $l \in \mathcal{L}$, $s \in \mathcal{S}$, $k \in \mathcal{K}$. Additionally, we define ${\bf {H}} = [{{\bf h}_1},{{\bf h}_2},...,{{\bf h}_K} ]$, ${{\bf h}_k^{T}} = \left[ {{{\bf h}_{1,k}^{T}},{{\bf h}_{2,k}^{T}},...,{{\bf h}_{S,k}^{T}}} \right]$, $\mathbf{Z}=\left[ \mathbf{Z}_1,\mathbf{Z}_2,\ldots,\mathbf{Z}_S\right]$, $ \mathbf{Z}_s^{T} = [{{\bf Z}_{1,s}^{T}},{{\bf Z}_{2,s}^{T}},...,{{\bf Z}_{L,s}^{T}} ]$, $ {\bf {D}} = [{{\bf d}_1},{{\bf d}_2},...,{{\bf d}_K} ] $, $ \mathbf{d}_k^{T} = [{{\bf d}_{1,k}^{T}},{{\bf d}_{2,k}^{T}},...,{{\bf d}_{L,k}^{T}} ]$.

\subsection{Channel Model}
Since the environment blocking objects may block the LoS path between APs and users, following \cite{pan2020multicell,9355404,han2019large}, we use the Rayleigh fading model to model user $k$-AP $l$ channel as follows
\begin{align}\label{rate_user_k}
	{{\bf d}_{l,k}} = \sqrt{\gamma_{l,k}} \widetilde{\mathbf{d}}_{l,k}, \forall l \in \mathcal{L}, k \in \mathcal{K},
\end{align} 
where $\gamma_{l,k}$ denotes large-scale path-loss. The entries of $\widetilde{\mathbf{d}}_{l,k}$ are independent and identically distributed (i.i.d.) complex Gaussian random variables, i.e., $ \widetilde{\mathbf{d}}_{l,k} \sim \mathcal{CN}\left({\bf 0},\mathbf{I}_{B}\right)$.

RISs are integrated into CF-mMIMO systems to provide extra reflection links for users, in which only the signals reflected by the RIS for the first time are considered, and the signals reflected by the RIS for two or more times are ignored\cite{9387559}. The phase shift matrix of ideal RISs can be written as $ \mathbf\Phi  = \mathrm{diag}\Big\{\mathrm{diag}^{T}\Big\{{{\mathbf\Phi}_1}\Big\},\mathrm{diag}^{T}\Big\{{\mathbf\Phi}_2\Big\},...,\mathrm{diag}^{T}\Big\{{\mathbf\Phi}_S\Big\} \Big\}\in {{\mathbb{C}}^{SR \times SR}}$, ${\bf \Phi }_s = {\mathrm{diag}} \left\{ {{e^{j{\theta _{s,1}}}},{e^{j{\theta _{s,2}}}},...,{e^{j{\theta _{s,R}}}}} \right\}\in {{\mathbb{C}}^{R \times R}}$, where ${\theta _{s,r}}\in [0,2\pi)$ represents the phase shift of the $r$-th element of the $s$-th RIS. Considering that the infinite precision configuration of RIS elements is infeasible, the phase shifts of the RIS can only be a finite number of discrete values, which causes the phase noise at the RIS \cite{9534477}. The phase noise matrix of RISs can be expressed as $ \widetilde{\mathbf\Phi } = {\mathrm{diag}}\left\{\mathrm{diag}^{T}\left\{{\widetilde{\mathbf\Phi}_1}\right\},\mathrm{diag}^{T}\left\{\widetilde{\mathbf\Phi}_2\right\},...,\mathrm{diag}^{T}\left\{\widetilde{\mathbf\Phi}_S\right\} \right\}\in {{\mathbb{C}}^{SR \times SR}}$, $\widetilde{\bf \Phi }_s = {\rm{diag}}\big\{ e^{j{\widetilde{\theta}_{s,1}}},{e^{j{\widetilde{\theta}_{s,2}}}},...,{e^{j{\widetilde{\theta}_{s,R}}}} \big\}\in {\mathbb{C}}^{\mathit{R \times R}}$, where $\widetilde{\theta}_{s,r}$ represents the phase noise uniformly distributed in $[-\kappa_{r}\pi,\kappa_{r}\pi]$, $\kappa_{r}=1/2^{b}$ measures the severity of the residual HWI at RISs, and $b$ is the number of quantization bits \cite{9295369}.  Therefore, the phase shift matrix of imperfect RISs is given by $ \widehat{\mathbf\Phi } = \mathrm{diag}\Big\{\mathrm{diag}^{T}\Big\{{\widehat{\mathbf\Phi}_1}\Big\},\mathrm{diag}^{T}\Big\{\widehat{\mathbf\Phi}_2\Big\},...,\mathrm{diag}^{T}\Big\{\widehat{\mathbf\Phi}_S\Big\} \Big\}\in {\mathbb{C}}^{SR \times SR}$, $\widehat{\bf \Phi }_s = {\rm{diag}}\big\{ {e^{j{\widehat{\theta}_{s,1}}}},{e^{j{\widehat{\theta}_{s,2}}}},...,{e^{j{\widehat{\theta}_{s,R}}}} \big\}\in {{\mathbb{C}}^{R \times R}}$, where $ \widehat{\mathbf\Phi }={\mathbf\Phi }\widetilde{\mathbf\Phi }$, $\widehat{\theta}_{s,r}={\theta}_{s,r}+\widetilde{\theta}_{s,r}$.

The RIS is often installed on the facades of high-rise buildings and placed near users, which implies that LoS components are likely to be present in RIS-assisted channels. Therefore, we adopt the Rician fading model to model the user $k$-RIS $s$ and RIS $s$-AP $l$ channels as follows
\begin{align}
	&{{\bf h}_{s,k}} = \sqrt {{\alpha _{s,k}}} \left( {\sqrt {\frac{{{\varepsilon _{s,k}}}}{{{\varepsilon _{s,k}} + 1}}} {{\bf  \overline h}_{s,k}} + \sqrt {\frac{1}{{{\varepsilon _{s,k}} + 1}}} {{\bf \widetilde h}_{s,k}}} \right), \label{rician2}\\
	&{{\bf Z}_{l,s}} = \sqrt {\beta_{l,s}} \left( {\sqrt {\frac{{{\delta _{l,s}}}}{{{\delta _{l,s}} + 1}}} {{\bf \overline Z}_{l,s}} + \sqrt {\frac{1}{{\delta _{l,s} + 1}}} {{\bf \widetilde Z}_{l,s}}} \right),  \label{rician3}
\end{align}
$\forall l \in \mathcal{L}$, $s \in \mathcal{S}$, $k \in \mathcal{K}$, where ${\alpha _{s,k}}$ and $\beta _{l,s}$ represent the path-loss factors, ${\varepsilon _{s,k}}$ and ${\delta _{l,s}}$ are the Rician factors. ${{\bf \overline h}_{s,k}} \in {{\mathbb{C}}^{R \times 1}}$ and ${{\bf \overline Z}_{l,s}} \in {{\mathbb{C}}^{B \times R}}$ are deterministic LoS components, ${{\bf \widetilde h}_{s,k}} \in {{\mathbb{C}}^{R \times 1}}$ and ${{\bf \widetilde Z}_{l,s}} \in {{\mathbb{C}}^{B \times R}}$ are the corresponding NLoS components, whose elements are i.i.d. complex Gaussian random variables and follow $ {\cal C}{\cal N}\left( {0,1} \right)$ \cite{5281762,6094142}. 
Furthermore, by adopting the uniform square planar array (USPA) model \cite{9079457}, the LoS paths of RISs and APs can be respectively modeled as follows
\begin{align}
	{{\bf \overline h}_{s,k}} &= {{\bf a}_{R}}\left( {\varphi _{s,k}^a,\varphi _{s,k}^e} \right) \triangleq \mathbf{a}_{R}(s,k),\\
	{{\bf \overline Z}_{l,s}}& = {{\bf a}_{B}}\left( {\phi _{l,s}^a,\phi _{l,s}^e} \right){\bf a}_{R}^H\left( {\varphi _{l,s}^a,\varphi _{l,s}^e} \right)\triangleq \mathbf{a}_{B}(l,s) \mathbf{a}_{R}^{H}(l,s),
\end{align}
where ${{\bf a}_X}\left( {\vartheta _{}^a,\vartheta _{}^e} \right) \in {{\mathbb{C}}^{X \times 1}}$ is the array response vector, whose $x$-th entry is 
\begin{align}\label{upa}
	\left[ {{\bf a}_X}\left( {\vartheta _{}^a,\vartheta _{}^e} \right) \right]_{ x} = e^{\left\{j2\pi \frac{d}{\lambda }
		\left( {   \lfloor \left({ x} - 1 \right)/\sqrt{X}\rfloor \sin \vartheta _{}^e\sin \vartheta _{}^a
			+ \left(\left({x}-1\right)\bmod \sqrt{X}\right)  \cos \vartheta _{}^e} \right) \right\}},
\end{align}
where $d$ and $\lambda$ denote the element spacing and carrier wavelength, respectively. $\varphi _{s,k}^a$  and $\varphi _{s,k}^e$ respectively represent the azimuth and elevation angles of arrival (AoA) of the incident signal at RIS $s$ from user $k$. $\phi _{l,s}^a$ and $\phi_{l,s}^e$ respectively denote the AoA at AP $l$ from RIS $s$. $\varphi _{l,s}^a$ and $\varphi _{l,s}^e$ represent the azimuth and elevation angles of departure (AoD) reflected by RIS $s$ towards AP $l$, respectively.  

Based on the above definitions, the user $k$-RIS-AP channel can be expressed as $\widehat{\mathbf{g}}_{k} =[{\widehat{\mathbf{g}}_{1,k}^{T}},{\widehat{\mathbf{g}}_{2,k}^{T}},\\...,{\widehat{\mathbf{g}}_{L,k}^{T}} ]^{T}= \mathbf{Z \widehat{\mathbf\Phi }{h}}_{k} \in \mathbb{C}^{LB \times 1}$, where $\mathbf{Z} \in \mathbb{C}^{LB \times SR}$ denotes the channel between RISs and APs, ${\bf {h}}_{k} \in \mathbb{C}^{SR \times 1}$ represents user $k$-RIS channel, and $\widehat{\mathbf{g}}_{l,k}\in {{\mathbb{C}}^{B \times 1}}$ represents user $k$-RIS-AP $l$ channel. Furthermore, the channels of all users through the RISs are collected in the matrix $\widehat{\mathbf{G}} = [{\widehat{\mathbf{g}}_1},{\widehat{\mathbf{g}}_2},...,{\widehat{\mathbf{g}}_K} ] = \mathbf{Z\widehat{\mathbf\Phi }{H}} \in \mathbb{C}^{LB \times K}$, where ${\bf {H}} \in \mathbb{C}^{SR \times K}$ represents the channel between users and RISs. For ease of exposition, let $\widehat{\mathbf{Q}} = \widehat{\mathbf{G}}+\mathbf{D}=[\widehat{\mathbf{q}}_1,\widehat{\mathbf{q}}_2,\ldots,\widehat{\mathbf{q}}_K]  \in \mathbb{C}^{LB\times K}$ denotes the aggregated channel matrix from users to APs, where $\widehat{\mathbf{q}}_{k}=\left[ {{\widehat{\mathbf{q}}_{1,k}^{T}},{\widehat{\mathbf{q}}_{2,k}^{T}},...,{\widehat{\mathbf{q}}_{L,k}^{T}}} \right]^{T}\in {{\mathbb{C}}^{LB \times 1}}$, and $\widehat{\mathbf{q}}_{l,k}\in {{\mathbb{C}}^{B \times 1}}$ denotes the aggregated channel between user $k$ and AP $l$.

Then, the aggregated channel $\widehat{\mathbf{q}}_{k}$ from user $k$ to APs can be decomposed as
\begin{align}\label{exact_channel}
	\widehat{\mathbf{q}}_{k}=\widehat{\mathbf{g}}_{k}+\mathbf{d}_{k}=\left[ {{\widehat{\mathbf{q}}_{1,k}^{T}},{\widehat{\mathbf{q}}_{2,k}^{T}},...,{\widehat{\mathbf{q}}_{L,k}^{T}}} \right]^{T}, {\widehat{\mathbf{q}}_{l,k}}=\widehat{\mathbf{g}}_{l,k}+\mathbf{d}_{l,k}= \sum\limits_{s=1}^{S}\widehat{\mathbf{g}}_{l,s,k}+\sqrt{\gamma_{l,k}} \widetilde{\mathbf{d}}_{l,k} ,
\end{align}
\begin{align}\label{g_mnk}
	\begin{array}{l}
		\widehat{\mathbf{g}}_{l,s,k}=\!\mathbf{Z}_{l,s} \widehat{\mathbf\Phi}_{s} \mathbf{h}_{s,k}=\! \underbrace{\sqrt{c_{l,s,k}\delta_{l,s} \varepsilon_{s,k}} \overline{\mathbf{Z}}_{l,s} \widehat{\mathbf\Phi}_{s} \overline{\mathbf{h}}_{s,k}}_{\widehat{\mathbf{g}}_{l,s,k}^{1}} 
		+\underbrace{\sqrt{c_{l,s,k}\delta_{l,s}} \overline{\mathbf{Z}}_{l,s} \widehat{\mathbf\Phi}_{s} \widetilde{\mathbf{h}}_{s,k}}_{\widehat{\mathbf{g}}_{l,s,k}^{2}}
		\\+
		\underbrace{\sqrt{c_{l,s,k}\varepsilon_{s,k}} \widetilde{\mathbf{Z}}_{l,s} \widehat{\mathbf\Phi}_{s} \overline{\mathbf{h}}_{s,k}}_{\widehat{\mathbf{g}}_{l,s,k}^{3}}+\underbrace{\sqrt{c_{l,s,k}}\widetilde{\mathbf{Z}}_{l,s} \widehat{\mathbf\Phi }_{s} \widetilde{\mathbf{h}}_{s,k}}_{\widehat{\mathbf{g}}_{l,s,k}^{4}},
	\end{array}
\end{align}
where $c_{l,s,k}\triangleq\frac{\beta_{l,s}\alpha_{s,k}}{(\delta_{l,s}+1)\left(\varepsilon_{s,k}+1\right)}$ and $\widehat{\mathbf{g}}_{l,s,k}\in {{\mathbb{C}}^{B \times 1}}$ represents the cascaded user $k$-RIS $s$-AP $l$ channel. Besides, it is worth noting that $\widehat{\mathbf{g}} _{k} $ and ${\mathbf{d}}_{k}$ are mutually independent.

\subsection{Uplink Transmission}
Unlike the unrealistic assumption of perfect hardware, we consider that the additive distortions at the transceiver are Gaussian distributed with average powers proportional to the average powers of the corresponding transceivers \cite{9534477}. Specifically, the distortion noise at users is denoted by ${\bm{\eta}}_t=\left[{\mathbb{\eta}}_{1,t},{\mathbb{\eta}}_{2,t},...,{\mathbb{\eta}}_{K,t}\right]^{T}\in {{\mathbb{C}}^{K\times 1}}$, whose $k$-th element is i.i.d. random variables following $ \mathcal{CN} (0, \kappa_u^{2} p_k)$, $\forall k$, where $\kappa_u$ is the error vector magnitude (EVM) and measures the severity of transmitter HWI, and $p_k$ is the transmit power of user $k$. The distortion noise at AP $l$ is denoted by  ${\bm{\eta}}_{l,r}\in {{\mathbb{C}}^{B\times 1}}$ and follows $ \mathcal{CN} \big(0,\kappa_b^{2}\sum_{i=1}^{K}\widetilde{diag}\{\mathbb{E}\{\widehat{\mathbf{y}}_{l,i}\widehat{\mathbf{y}}_{l,i}^H\}\}\big)$, where $\kappa_b$ is the EVM and measures the severity of receiver HWI,  $\widehat{\mathbf{y}}_{l,i}=\widehat{\mathbf{q}}_{l,i}(\sqrt{p_i}x_i + {\mathbb{\eta}}_{i,t})$ denotes the received signal at AP $l$ from user $i$, and ${\mathbb{\eta}}_{i,t}$ represents the distortion noise at user $i$. In the CPU-based centralized operation, the received signals at APs are fully processed at the CPU \cite{9586055}. Therefore, the distortion noise at all APs is collected in the matrix   ${\bm{\eta}}_r=\left[{\bm{\eta}}_{1,r},{\bm{\eta}}_{2,r},...,{\bm{\eta}}_{L,r}\right]^{T}\in {{\mathbb{C}}^{LB\times 1}}$, following $ \mathcal{CN} \big(0,\kappa_b^{2}\sum_{i=1}^{K}\widetilde{diag}\{\mathbb{E}\{\widehat{\mathbf{y}}_{i}\widehat{\mathbf{y}}_{i}^H\}\}\big)$, where $\widehat{\mathbf{y}}_{i}=\widehat{\mathbf{q}}_{i}(\sqrt{p_i}x_i + {\mathbb{\eta}}_{i,t})$.
Then, the collective received signal is
\begin{align}
		{\mathbf y} = {\widehat{\mathbf{Q}} \left(\mathbf{ P} {\mathbf{x}} +{\bm{\eta}}_{\it t} \right)+{\bm{\eta}}_{\it r}+ {\mathbf{n}}} =  \sum\nolimits_{k=1}^{K} \widehat{\mathbf{q}}_k \left(\sqrt{p_{k}} x_k + {\mathbb{\eta}}_{\it k,t} \right) +{\bm{\eta}}_{\it r} +\mathbf{n},
\end{align}
where $\mathbf{P}={\rm{diag}}\left\{\sqrt{p_1},\sqrt{p_2},...,\sqrt{p_K}\right\}$, and $p_k$ is the transmit power of user $k$. ${\bf x} = {\left[ {{x_1},{x_2},...,{x_K}} \right]^T}$ $ \in {{\mathbb{C}}^{K \times 1}}$ represents the transmit symbols with $\mathbb{E}\left\{\left|x_{k}\right|^{2}\right\}=1$. ${\bf n} \sim {\cal C}{\cal N}\left( {0,{\sigma ^2}{\bf I}_{LB}} \right)$ denotes the additional white Gaussian noise vector.

As considered in \cite{zhang2014ArRank}, the CPU applies the MRC technique to achieve a low-complexity implementation in practical systems. We assume that the CSI of overall channels is perfectly estimated at APs and fully acquired at the CPU. Thus, under fully centralized processing, the CPU performs MRC by multiplying the received signal $ \bf{y} $ with ${\bf {Q}}^{H}={\mathbf{H}}^{H}{\mathbf \Phi}^H {\mathbf Z}^H+\mathbf{D}^H$ as follows
\begin{align}
	{{\bf r} = {{{\bf {Q}}}^H}{\bf y} }=  { {{{\bf {Q}}}^H}{{\widehat{\bf {Q}}}\left({\mathbf{ P}} {\bf x} +{\bm{\eta}}_{\it t} \right)+{{\bf {Q}}}^H}{\bm{\eta}}_{\it r} + {{{\bf {Q}}}^H}\bf n},
\end{align}
and the detected signal corresponding to user $k$ can be expressed as
\begin{align}
	r_{k}=\underbrace{\sqrt{p_{k}} x_{k} {\mathbf{q}}_{k}^{H} \widehat{\mathbf{q}}_{k}}_{\text{Desired signal}} + \underbrace{\sum\nolimits_{i=1, i \neq k}^{K} \sqrt{p_{i}} x_{i}{\mathbf{q}}_{k}^{H} \widehat{\mathbf{q}}_{i}}_{\text{Multi-user interference}}+\underbrace{\sum\nolimits_{i=1}^{K}{\mathbf{q}}_{k}^{H}\widehat{\mathbf{q}}_{i}{\mathbb{\eta}}_{\it i,t}+{\mathbf{q}}_{k}^{H}{\bm{\eta}}_{\it r}}_{\text{HWIs}}+\underbrace{{\mathbf{q}}_{k}^{H} \mathbf{n}}_{\text{Noise}}, \;\; k \in \mathcal{K},
\end{align}
where ${\mathbf{q}}_{k}= \mathbf{Z {\mathbf\Phi }{h}}_{k} + \mathbf{d}_{k} \in \mathbb{C}^{LB\times 1}$ is similar to $\widehat{\mathbf{q}}_{k}$.

Finally, we can formulate the achievable rate of user $k$ as $R_{k}=\mathbb{E}\left\{\mathrm{log_{2}}\left(1+\mathrm{SINR}_{k}\right)\right\}$, where the signal-to-interference-plus-noise ratio (SINR) is given by 
\begin{align}\label{rek}
	\mathrm{SINR}_{k} = \frac{{p_k}  \left|{{\bf q}}_k^{H} \widehat{\mathbf{q}}_k\right|^{2}  }{ \sum\limits_{i=1 \atop i\neq k}^{K}{p_i}\left|{\bf q}_k^{H} \widehat{\mathbf{q}}_i\right|^{2}+\sum\limits_{i=1}^{K}\left|{\mathbf{q}}_{k}^{H}\widehat{\mathbf{q}}_{i}{\mathbb{\eta}}_{\it i,t}\right|^{2}+\left|{\mathbf{q}}_{k}^{H}{\bm{\eta}}_{\it r}\right|^{2}+\sigma^{2}\left\|{\bf q}_k\right\|^{2}}.
\end{align}

\section{Uplink Achievable Rate Analysis}\label{section3}
We derive the closed-form approximate expression of the rate and provide valuable guidelines for RIS-assisted CF-mMIMO systems with HWIs. 

\begin{theorem}\label{lemma1}
Based on (\ref{rek}), the closed-form approximate expression of the achievable rate for user $k$ is given by
\begin{align}\label{rate}
	R_{k} \approx \log _{2}\left(1+\frac{p_k {E}_{k}^{(\mathrm{signal})}({\bf\Phi})}{ \sum\limits_{i=1 \atop i \neq k}^{K} p_i I_{k i}({\bf\Phi})+E_{k}^{(\mathrm{hwi})}({\bf\Phi})+\sigma^{2} E_{k}^{(\mathrm{\mathrm{noise}})}({\bf\Phi})}\right),
\end{align}
where ${E}_{k}^{(\mathrm{signal})}({\bf\Phi})$ = $\mathbb{E}\big\{\left|{{\bf q}}_k^{H} \widehat{\mathbf{q}}_k\right|^{2}\big\}$, $E_{k}^{(\mathrm{noise})}({\bf\Phi})$ = $\mathbb{E}\big\{\left\|{\bf q}_k\right\|^{2}\big\}$, $E_{k}^{(\mathrm{hwi})}({\bf\Phi})=\mathbb{E}\bigg\{\sum\limits_{i=1}^{K}\left|{\mathbf{q}}_{k}^{H}\widehat{\mathbf{q}}_{i}{\mathbb{\eta}}_{\it i,t}\right|^{2}+\left|{\mathbf{q}}_{k}^{H}{\bm{\eta}}_{\it r}\right|^{2}\bigg\}$, and $I_{k i}({\bf\Phi}) = \mathbb{E}\big\{\left|{\bf{q}}_k^H\widehat{\mathbf{q}}_i\right|^2\big\} $ represent the terms of the desired signal, noise, HWI, and multi-user interference, respectively. The closed-form expressions are given by (\ref{noise}), (\ref{signal}), (\ref{interference}), and (\ref{HWI_begin}), where some parameters and functions are defined as follows
\begin{align}\label{f_k_Phi_1}
	f_{l,s,k}({\bf\Phi}) \triangleq \mathbf{a}_{R}^{H}(l,s) {\bf\Phi}_{s} \overline{\mathbf{h}}_{s,k}=\sum\nolimits_{r=1}^{R} e^{j\left(\zeta_{r}^{l,s,k}+\theta_{s,r}\right)}
\end{align}
and
\begin{align}\label{f_k_Phi_2}
	\zeta_{r}^{l,s,k}=&2 \pi \frac{d}{\lambda}  \left(   \lfloor(r-1) / \sqrt{R}\rfloor  \left(\sin \varphi_{s,k}^{e} \sin \varphi_{s,k}^{a}-\sin \varphi_{l,s}^{e} \sin \varphi_{l,s}^{a}\right)\right.\nonumber\\
	&\left.+\big((r-1) \bmod \sqrt{R}\big)\left( \cos \varphi_{s,k}^{e}- \cos \varphi_{l,s}^{e}\right)\right).
\end{align}
\end{theorem}
\itshape \textbf{Proof:}  \upshape It follows some decoupling operations and tools from probability theory. The final expressions and the detailed proof are given in Appendix \ref{appA}.  \hfill $\blacksquare$
%Using the same method as \cite[Theorem 1]{fullversion}. To save space, we only present the final results

As shown in Theorem \ref{lemma1}, by averaging over the fast-changing variables, the expression $R_k$ in (\ref{rate}) only relies on the slow-varying statistical CSI, i.e., Rician factors, large-scale path-loss factors, the AoA, and AoD. Therefore, the rate expression enables us to optimize the RIS phase shifts with low computational complexity and overhead.
Also, unlike a time-consuming Monte Carlo (MC) simulation that requires $10^{5}$ repeated calculations to obtain the expectation. The derived expression enables us to evaluate the rate performance quickly.  
Furthermore, the derived expression of rate analytically characterizes the impacts of some key parameters, i.e., $B$, $R$, $p_{k}$, $\kappa_{r}$, $\kappa_{u}$, and $\kappa_{b}$. Although our analytical expressions may seem cumbersome and verbose, they can provide clear insights from these system parameters and can be analyzed for some special cases after some simplifications. 

For instance, we find that $E_{k}^{({\mathrm{noise}})}({\bf\Phi})$ scales as $\mathcal{O}(BR)$, $I_{k i}({\bf\Phi})$ scales as $\mathcal{O}(B^2R)$, and ${E}_{k}^{({\mathrm{signal}})}({\bf\Phi})$ scales as $\mathcal{O}(B^2R^2)$. This implies that the desired signal has the same order of magnitude with respect to $ B $ as the interference term. Even if HWIs are not considered, the benefits from increasing the number of BS antennas will be limited by multi-user interference. However, it is expected that considerable performance improvement can be achieved by continuously increasing the number of RIS elements and optimizing phase shifts. 
Also, we find that the HWI term ${E}_{k}^{(\mathrm{hwi})}({\bf\Phi})$ affects the rate as the denominator of the SINR. In particular, the HWI at the transmitter introduces a part of the HWI term, which is proportional to the signal and the interference terms, where the transmitter's HWI coefficient $\kappa_{u}$ determines the severity of this part of HWI. Similarly, the receiver's HWI coefficient $\kappa_{b}$ affects the rate performance as a measure factor in another part of the HWI term.
To provide clear insights, we utilize some special cases below to show the benefits of RISs, the power scaling laws, and the impacts of HWIs. To begin with, we first present the rate expression without RISs as a baseline.

\begin{corollary}\label{corollary1}
	The rate of RIS-free mMIMO systems with HWIs can be obtained by setting $ S=0$, which is $R_{k}^{(\mathrm{w})} \triangleq \log _{2}\Big(1+\mathrm{SINR}_{k}^{(\mathrm{w})}\Big)$ with
	\begin{align}\label{SINK_w}
		&\mathrm{SINR}_{k}^{(\mathrm{w})} 
		\nonumber\\&\approx\frac{p_k \bigg(\bigg(\sum\limits_{l=1}^{L}\gamma_{l,k}B\bigg)^{2}+\sum\limits_{l=1}^{L}\gamma_{l,k}^{2}B\bigg)}      {p_k\kappa_{u}^{2}\bigg(\sum\limits_{l=1}^{L}\gamma_{l,k}B\bigg)^{2}+ {\sum\limits_{l=1}^L \gamma_{l,k} B\left(\sigma^{2}-p_k\gamma_{l,k}\right) }    +(1+\kappa_b^{2})(1+\kappa_u^{2})\sum\limits_{i=1}^{K}  {\sum\limits_{l=1}^L p_i \gamma_{l,k}\gamma_{l,i}B}}.
	\end{align}
\end{corollary}

For the RIS-free mMIMO systems in the presence of transceiver HWIs \cite{8891922}, i.e., the number of RISs $S$ is zero, the SINR in (\ref{SINK_w}) will converge to the constant $1/\kappa_{u}^{2}$ when $B\rightarrow\infty$. However, under the assumption of the ideal transceiver, i.e., $\kappa_{u}=\kappa_{b}=0$, the rate will increase without bound as $B$ increases. This result implies that the transmitter HWI limits the rate improvement brought by massive AP antennas. Besides, it is expected that the integration of the RIS into cellular mMIMO and CF-mMIMO systems can bring significant gains.
However, in the presence of RIS phase noise, we can find that the signal term ${E}_{k}^{(\mathrm{signal})}({\bf\Phi})$ in (\ref{signal}), the interference term $I_{k i}({\bf\Phi})$ in (\ref{interference}), and the HWI term $E_{k}^{(\mathrm{hwi})}({\bf\Phi})$ in (\ref{HWI_begin}) scale as $\mathcal{O}(B^2)$, then when $B\rightarrow\infty$, the rate will converge to a constant containing HWI factors, i.e., $\kappa_{r},\kappa_{u}^{2},\kappa_{b}^{2}$. Thus, a natural question is whether HWIs severely limit the gains of RISs or not. To answer this question and get more insights, we resort to asymptotic results in some special cases.

\begin{corollary}\label{corollary2}
	When the channels between RISs and APs are only NLoS (i.e., $\delta_{l,s}=0$, $ \forall l,s$), the rate is given by $R_{k}^{(\mathrm{NL})} \triangleq \log _{2}\Big(1+\mathrm{SINR}_{k}^{(\mathrm{NL})}\Big)$, where
	\begin{align}\label{SINK_NL}
		\mathrm{SINR}_{k}^{(\mathrm{NL})}\approx 
		\frac{p_k {E}_{k}^{(\mathrm{signal},\mathrm{NL})}}{ \sum\limits_{i=1, i \neq k}^{K} p_i I_{k i}^{(\mathrm{NL})}+E_{k}^{(\mathrm{hwi},\mathrm{NL})}+\sigma^{2} E_{k}^{(\mathrm{noise},\mathrm{NL})}},
	\end{align}
	with
	\begin{align}\label{noise_NL}
			E_{k}^{(\mathrm{noise},\mathrm{NL})} =  
			{\sum\limits_{l=1}^L \sum\limits_{s=1}^S   {\beta_{l,s}\alpha_{s,k}} BR} +    {\sum\limits_{l=1}^L \gamma_{l,k}B },
	\end{align}
	\begin{align}\label{signal power_NL}
	&E_{k}^{(\mathrm{signal},\mathrm{NL})}\nonumber\\
	&=
	{\sum\limits_{l_{1}=1}^{L} \sum\limits_{l_{2}=1}^{L} \sum\limits_{s_{1}=1}^{S} \sum\limits_{s_{2}=1}^{S} }{{\beta_{l_{1},s_{1}}\beta_{l_{2},s_{2}}\alpha_{s_{1},k}\alpha_{s_{2},k}}}B^{2}R^{2}\mathrm{sinc}^{2}\left(\kappa_r\pi\right)
	+ %式子9
	{\sum\limits_{l=1}^{L} \sum\limits_{s_{1}=1}^{S} \sum\limits_{s_{2}=1}^{S}} {\beta_{l,s_{1}}\beta_{l,s_{2}}\alpha_{s_{1},k}\alpha_{s_{2},k}} B R^{2}\nonumber
	\\&+ 
	{\sum\limits_{l_{1}=1}^{L} \sum\limits_{l_{2}=1}^{L} \sum\limits_{s=1}^{S}}    B^{2}R\Big(2\beta_{l_{1},s}\alpha_{s,k}\gamma_{l_{2},k}\mathrm{sinc}\left(\kappa_r\pi\right)+c_{l_{1},s,k}^{(\mathrm{NL})}c_{l_{2},s,k}^{(\mathrm{NL})}\left(2\varepsilon_{s,k}+1\right)+\beta_{l_{1},s}\beta_{l_{2},s}\alpha_{s,k}^{2}\left(1-\mathrm{sinc}^{2}\left(\kappa_r\pi\right)\right)\Big)\nonumber
	\\&+{\sum\limits_{l=1}^{L} \sum\limits_{s=1}^{S}}  BR \bigg(2\beta_{l,s}\alpha_{s,k}\gamma_{l,k}+\left(c_{l,s,k}^{(\mathrm{NL})}\right)^{2}\left(2\varepsilon_{s,k}+1\right)\bigg)
	 + %式子13
	\left(\sum\limits_{l=1}^{L}\gamma_{l,k}B\right)^{2}+ \sum\limits_{l=1}^{L} \gamma_{l,k}^{2}B,
	\end{align}
	\begin{align}\label{interference_NL}
	&I_{ki}^{(\mathrm{NL})}={\sum\limits_{l_{1}=1}^{L} \sum\limits_{l_{2}=1}^{L} \sum\limits_{s_{1}=1}^{S} \sum\limits_{s_{2}=1}^{S} }\sqrt{c_{l_{1},s_{1},k}^{(\mathrm{NL})}c_{l_{1},s_{2},i}^{(\mathrm{NL})}c_{l_{2},s_{2},i}^{(\mathrm{NL})}c_{l_{2},s_{1},k}^{(\mathrm{NL})} \varepsilon_{s_{1},k}\varepsilon_{s_{1},i}\varepsilon_{s_{2},i}\varepsilon_{s_{2},k}} B^{2}\mathrm{sinc}^{2}\left(\kappa_r\pi\right)\overline{\mathbf{h}}_{s_{1},k}^{H}\overline{\mathbf{h}}_{s_{1},i} \overline{\mathbf{h}}_{s_{2},i}^{H} \overline{\mathbf{h}}_{s_{2},k}\nonumber\\
	&+
	{\sum\limits_{l_{1}=1}^{L} \sum\limits_{l_{2}=1}^{L} \sum\limits_{s=1}^{S}} \sqrt{c_{l_{1},s,k}^{(\mathrm{NL})}c_{l_{1},s,i}^{(\mathrm{NL})} c_{l_{2},s,i}^{(\mathrm{NL})}c_{l_{2},s,k}^{(\mathrm{NL})}}B^{2} R \left(\varepsilon_{s,k}\left(1+\varepsilon_{s,i}\left(1-\mathrm{sinc}^{2}\left(\kappa_r\pi\right)\right)\right)+\varepsilon_{s,i}+1\right) \nonumber\\
	&+  %式子10
	{\sum\limits_{l=1}^{L} \sum\limits_{s_{1}=1}^{S} \sum\limits_{s_{2}=1}^{S}}  \beta_{l,s_{1}}\beta_{l,s_{2}}\alpha_{s_{1},k} \alpha_{s_{2},i} B R^{2}+ 
	{\sum\limits_{l=1}^L \sum\limits_{s=1}^S} 
	{\beta_{l,s}B R \left(\alpha_{s,k} \gamma_{l,i}+\alpha_{s,i}\gamma_{l,k} \right)}+ 
	{\sum\limits_{l=1}^L \gamma_{l,k}\gamma_{l,i}B},
	\end{align}
\begin{align}\label{hwi_NL}
	&E_{k}^{(\mathrm{hwi},\mathrm{NL})} =\kappa_u^{2}  \Bigg(p_{k}{E}_{k}^{(\mathrm{signal},\mathrm{NL})}+\sum\limits_{i=1,i\neq k}^{K}p_{i}I_{k i}^{(\mathrm{NL})}\Bigg)\nonumber\\
	&+\kappa_b^{2}\left(1+\kappa_u^{2}\right)\sum\limits_{l=1}^{L}\sum\limits_{i=1}^{K}p_{i}
	\Bigg({\sum\limits_{s=1}^S \beta_{l,s}\alpha_{s,k} BR} +  \gamma_{l,k}B\Bigg)\Bigg(\gamma_{l,i}+\sum\limits_{s=1}^{S}\beta_{l,s}\alpha_{s,i} R \Bigg),
\end{align}
	and
\begin{align}\label{C_mnk_NL}
	c_{l,s,k}^{(\mathrm{NL})} \triangleq \frac{\beta_{l,s}\alpha_{s,k}}{\varepsilon _{s,k}+1} .
\end{align}
\end{corollary}

\itshape \textbf{Proof:}  \upshape  Setting $\delta_{l,s}=0$ and using  $c_{l,s,k}^{(\mathrm{NL})}\left(\varepsilon_{s,k}+1\right)=\beta_{l,s}\alpha_{s,k}$, $\forall l,s,k$, the proof follows after some simplifications.  \hfill $\blacksquare$

Corollary \ref{corollary2} corresponds to a special case where the RISs-APs channels are characterized by rich scattering, such that the Rician fading channel degrades to the Rayleigh fading channel. 
In this case, the RISs-APs links may be blocked by a lot of environmental blocking objects, which may limit the gains of RISs and reduce the users' sum rate.
Besides, we can see that the achievable rate does not rely on the phase shifts of RISs, indicating that the phase shifts can be set arbitrarily and can no longer be optimized to bring gain.

Then, we re-examine the order of the magnitude of the rate when RISs are deployed. As $B,R\rightarrow\infty$, the SINR in (\ref{SINK_NL}) will converge to the constant $1/\kappa_{u}^{2}$, which indicates that the rate will increase without limit under the assumption of the ideal transceiver. Therefore, even if the LoS links do not exist, deploying RISs with massive reflecting elements can significantly improve the performance of CF-mMIMO systems, but transceiver HWIs will restrict the gain. Note that the signal and interference terms contain the HWI coefficient $\kappa_{r}$ but are not proportional to it, and as can be seen from their expressions, the RIS phase noise does not have a noticeable effect on the rate in this case.
Based on the above analysis, we prove that the rates of all users will increase with $B,R$ and approach saturation when $B,R\rightarrow\infty$. 
This implies that user fairness can be implicitly guaranteed in the special case where the RISs-APs channels are rich-scattering. Based on Corollary \ref{corollary2}, we next give two power scaling laws for $B$ and $R$, respectively to gain more insights on the properties of HWIs.

\begin{corollary}\label{corollary3}
In the case of $\delta_{l,s}=0$, $ \forall l,s$, we assume that the transmit power is scaled as $p_{k}=p/B$, $\forall k$, as $B\rightarrow\infty$, the rate converges to ${R}_{k}^{\mathrm{(NL)}}\rightarrow R_{k}^{(\mathrm{NL})(B)} \triangleq \log _{2}\Big(1+\mathrm{SINR}_{k}^{(\mathrm{NL})(B)}\Big)$, where
\begin{align}
	\mathrm{SINR}_{k}^{(\mathrm{NL})(B)}\approx\frac{p {E}_{k}^{(\mathrm{signal},\mathrm{NL})(B)}}{ \sum\limits_{i=1, i \neq k}^{K} p I_{k i}^{(\mathrm{NL})(B)}+E_{k}^{(\mathrm{hwi},\mathrm{NL})(B)}+\sigma^{2} E_{k}^{(\mathrm{noise},\mathrm{NL})(B)}},
\end{align}
with
	\begin{align}
		E_{k}^{(\mathrm{noise},\mathrm{NL})(B)} =  
		{\sum\limits_{l=1}^L \sum\limits_{s=1}^S   \beta_{l,s}\alpha_{s,k} R} +    {\sum\limits_{l=1}^L \gamma_{l,k} },
\end{align}
\begin{align}
		&E_{k}^{(\mathrm{signal},\mathrm{NL})(B)} \nonumber\\
		&=
		{\sum\limits_{l_{1}=1}^{L} \sum\limits_{l_{2}=1}^{L} \sum\limits_{s_{1}=1}^{S} \sum\limits_{s_{2}=1}^S }
		{\beta_{l_{1},s_{1}}\beta_{l_{2},s_{2}}\alpha_{s_{1},k}\alpha_{s_{2},k}}R^{2}\mathrm{sinc}^{2}\left(\kappa_r\pi\right) 
		+ 
		{\sum\limits_{l_{1}=1}^{L} \sum\limits_{l_{2}=1}^{L} \sum\limits_{s=1}^{S}}    R\Big(2\beta_{l_{1},s}\alpha_{s,k}\gamma_{l_{2},k}\mathrm{sinc}\left(\kappa_r\pi\right)\nonumber\\
		&+c_{l_{1},s,k}^{(\mathrm{NL})}c_{l_{2},s,k}^{(\mathrm{NL})}\left(2\varepsilon_{s,k}+1\right)+\beta_{l_{1},s}\beta_{l_{2},s}\alpha_{s,k}^{2}\left(1-\mathrm{sinc}^{2}\left(\kappa_r\pi\right)\right)\Big)
		+ %式子13
		\left(\sum\limits_{l=1}^{L}\gamma_{l,k}\right)^{2},
\end{align}
\begin{align}
		&I_{ki}^{(\mathrm{NL})(B)} =
		{\sum\limits_{l_{1}=1}^{L} \sum\limits_{l_{2}=1}^{L} \sum\limits_{s_{1}=1}^{S} \sum\limits_{s_{2}=1}^{S} }
		\sqrt{c_{l_{1},s_{1},k}^{(\mathrm{NL})}c_{l_{1},s_{2},i}^{(\mathrm{NL})}c_{l_{2},s_{2},i}^{(\mathrm{NL})}c_{l_{2},s_{1},k}^{(\mathrm{NL})}\varepsilon_{s_{1},k}\varepsilon_{s_{1},i}\varepsilon_{s_{2},i}\varepsilon_{s_{2},k}}\mathrm{sinc}^{2}\left(\kappa_r\pi\right) \overline{\mathbf{h}}_{s_{1},k}^{H}\overline{\mathbf{h}}_{s_{1},i} \overline{\mathbf{h}}_{s_{2},i}^{H} \overline{\mathbf{h}}_{s_{2},k}\nonumber\\
		&+
		{\sum\limits_{l_{1}=1}^{L} \sum\limits_{l_{2}=1}^{L} \sum\limits_{s=1}^{S}} \sqrt{c_{l_{1},s,k}^{(\mathrm{NL})}c_{l_{1},s,i}^{(\mathrm{NL})} c_{l_{2},s,i}^{(\mathrm{NL})}c_{l_{2},s,k}^{(\mathrm{NL})}} R\left(\varepsilon_{s,k}\left(1+\varepsilon_{s,i}\left(1-\mathrm{sinc}^{2}\left(\kappa_r\pi\right)\right)\right)+\varepsilon_{s,i}+1\right) ,
\end{align}
and
\begin{align}
	E_{k}^{(\mathrm{hwi},\mathrm{NL})(B)} 
		=\kappa_u^{2} p \Bigg({E}_{k}^{(\mathrm{signal},\mathrm{NL})(B)}+\sum\limits_{i=1,i\neq k}^{K} I_{k i}^{(\mathrm{NL})(B)}\Bigg).
\end{align}
\end{corollary}

\itshape \textbf{Proof:}  \upshape After substituting $p_{k}=p/B$, $\forall k$ into the expression (\ref{SINK_NL}), as $B\rightarrow\infty$, we can complete the proof by retaining the significant terms whose asymptotic behavior is $\mathcal{O}(B)$.  \hfill $\blacksquare$

As shown in Corollary \ref{corollary3}, if the transmit power is reduced inversely proportional to $B$, the SINR will converge to a non-zero value containing the factors $\kappa_{u}^{2}$ and $\kappa_{r}$ as $B\rightarrow\infty$, which proves the existence of the power scaling law with respect to the RIS.
Similarly, in RIS-free mMIMO systems with HWIs, when the transmit power is also scaled as $p_k=p/B$, $\forall k$, the SINR converges to $\mathrm{SINR}_{k}^{(\mathrm{w})(B)} \approx  {p \Big(\sum\nolimits_{l=1}^{L}\gamma_{l,k}\Big)^{2}}\Big/{\Big(p\kappa_{u}^{2}\Big(\sum\nolimits_{l=1}^{L}\gamma_{l,k}\Big)^{2}+\sigma^{2}{\sum\nolimits_{l=1}^L\gamma_{l,k}}\Big)}$, which is a non-zero and finite value even if $\kappa_{u}=0$. 
However, with a large number of RIS elements, i.e., $R\to\infty$, the rate ${R}_{k}^{\mathrm{(NL)}{(B)}}$ will converge to the constant $\log _{2}\big(1+1/\kappa_{u}^{2}\big)$, which shows that when $\kappa_{u}=0$, the rate will increase without bound for $R\rightarrow\infty$, which demonstrates the promising feature of applying RISs to CF-mMIMO systems even under HWI.

\begin{corollary}\label{corollary4}
	In the case of $\delta_{l,s}=0$, $ \forall l,s$, if the transmit power is scaled down with the number of RIS elements according to $p_{k}=p/R$, $\forall k$, as $R\rightarrow\infty$, the rate  will converge to ${R}_{k}^{\mathrm{(NL)}}\rightarrow R_{k}^{(\mathrm{NL})(R )} \triangleq \log _{2}\Big(1+\mathrm{SINR}_{k}^{(\mathrm{NL})(R)}\Big)$, where
\begin{align}\label{key}
	\mathrm{SINR}_{k}^{(\mathrm{NL})(R)}\approx\frac{p {E}_{k}^{(\mathrm{signal},\mathrm{NL})(R)}}{ \sum\limits_{i=1, i \neq k}^{K} p I_{k i}^{(\mathrm{NL})(R)}+E_{k}^{(\mathrm{hwi},\mathrm{NL})(R)}+\sigma^{2} E_{k}^{(\mathrm{noise},\mathrm{NL})(R)}},
\end{align}
with
\begin{align}
		E_{k}^{(\mathrm{noise},\mathrm{NL})(R)} =  
		{\sum\limits_{l=1}^L \sum\limits_{s=1}^S   {\beta_{l,s}\alpha_{s,k}} B},
\end{align}
\begin{align}
		I_{ki}^{(\mathrm{NL})(R)}=
		{\sum\limits_{l=1}^{L} \sum\limits_{s_{1}=1}^{S} \sum\limits_{s_{2}=1}^{S}}  \beta_{l,s_{1}}\beta_{l,s_{2}}\alpha_{s_{1},k} \alpha_{s_{2},i} B,
\end{align}
\begin{align}
		&E_{k}^{(\mathrm{signal},\mathrm{NL})(R)}\nonumber\\
		&=
		{\sum\limits_{l_{1}=1}^{L} \sum\limits_{l_{2}=1}^{L} \sum\limits_{s_{1}=1}^{S} \sum\limits_{s_{2}=1}^{S} }
		{{\beta_{l_{1},s_{1}}\beta_{l_{2},s_{2}}\alpha_{s_{1},k}\alpha_{s_{2},k}}}B^{2}\mathrm{sinc}^{2}\left(\kappa_r\pi\right)
		+ %式子9
		{\sum\limits_{l=1}^{L} \sum\limits_{s_{1}=1}^{S} \sum\limits_{s_{2}=1}^{S}} {\beta_{l,s_{1}}\beta_{l,s_{2}}\alpha_{s_{1},k}\alpha_{s_{2},k}} B,
\end{align}
and 
\begin{align}
		&E_{k}^{(\mathrm{hwi},\mathrm{NL})(R)} =\kappa_u^{2} p \Bigg({E}_{k}^{(\mathrm{signal},\mathrm{NL})(R)}+\sum\limits_{i=1,i\neq k}^{K}I_{k i}^{(\mathrm{NL})(R)}\Bigg)\nonumber\\
		&+\kappa_b^{2}\left(1+\kappa_u^{2}\right)p\sum\limits_{i=1}^{K}\sum\limits_{l=1}^{L}
		{\sum\limits_{s_{1}=1}^S\sum\limits_{s_{2}=1}^{S}\beta_{l,s_{1}}\beta_{l,s_{2}}\alpha_{s_{1},k}\alpha_{s_{2},i}B}.
\end{align}
\end{corollary}

\itshape \textbf{Proof:}  \upshape  After substituting $p_{k}=p/R$, $\forall k$ into the expression in (\ref{SINK_NL}), as $R\rightarrow\infty$, we can complete the proof by retaining the significant terms whose asymptotic behavior is $\mathcal{O}(R)$.   \hfill $\blacksquare$

Corollary \ref{corollary4} reveals another power scaling law with respect to $R$. Compared with the scaling law with $B$, this new scaling law brought by RISs is more promising, since the hardware cost and power consumption of RIS elements are much less than that of AP antennas. We can find that the asymptotic SINR in (\ref{key}) is a non-zero constant containing the factors $\kappa_{u}^{2}$, $\kappa_{b}^{2}$ and $\kappa_{r}$, and the rate $R_{k}^{(\mathrm{NL})(R )}$ does not rely on the Rician factors of users-RISs channels, meaning that the rates for LoS-only ($\varepsilon_{s,k}\to\infty, \forall s,k$) and NLoS-only ($\varepsilon_{s,k}=0$) users-RISs channels are the same.

Additionally, based on Corollary \ref{corollary2} and Corollary \ref{corollary4}, we further consider the case that the transmit power is scaled down by $p_{k}=\frac{p}{BR}$, $\forall k$. When $B,R\rightarrow\infty$, the rate will converge to ${R}_{k}^{\mathrm{(NL)}}\rightarrow R_{k}^{(\mathrm{NL})(BR )} \triangleq \log _{2}\Big(1+\mathrm{SINR}_{k}^{(\mathrm{NL})(BR)}\Big)$, where
\begin{align}\label{BR}
	\mathrm{SINR}_{k}^{(\mathrm{NL})(BR)}\approx 
	\frac{p {E}_{k}^{(\mathrm{signal},\mathrm{NL})(BR)}}{ E_{k}^{(\mathrm{hwi},\mathrm{NL})(BR)}+\sigma^{2} E_{k}^{(\mathrm{noise},\mathrm{NL})(BR)}},
\end{align}
with
\begin{align}
	E_{k}^{(\mathrm{noise},\mathrm{NL})(BR)} =  
	{\sum\limits_{l=1}^L \sum\limits_{s=1}^S   {\beta_{l,s}\alpha_{s,k}}},
\end{align}
\begin{align}
	E_{k}^{(\mathrm{signal},\mathrm{NL})(BR)}=
	{\sum\limits_{l_{1}=1}^{L} \sum\limits_{l_{2}=1}^{L} \sum\limits_{s_{1}=1}^{S} \sum\limits_{s_{2}=1}^{S} }
	\beta_{l_{1},s_{1}}\beta_{l_{2},s_{2}}\alpha_{s_{1},k}\alpha_{s_{2},k}\mathrm{sinc}^{2}\left(\kappa_r\pi\right),
\end{align}
and
\begin{align}
		E_{k}^{(\mathrm{hwi},\mathrm{NL})(BR)} 
		=\kappa_u^{2} p{E}_{k}^{(\mathrm{signal},\mathrm{NL})(BR)}.
\end{align}

According to the above results, we can conclude that in the environment with NLoS-only RISs-APs channels, $ p_{k}$ can be reduced to $\frac{p}{BR}$ at most, while keeping a non-zero value of the rate when $B,R\to\infty$. 
Note that this power scaling law also holds for the RIS-assisted CF-mMIMO system without HWIs.

\begin{corollary}\label{corollary6}
When the RIS-aided channels only exist LoS paths (i.e., $\varepsilon_{s,k}$ and  $\delta_{l,s}\to\infty$, $ \forall l,s,k$), and $p_{k}$ scales as $p_{k}$, $ \forall k$, as $B\to\infty$, the rate converges to ${R}_{k}\rightarrow R_{k}^{(\mathrm{OL})(B)} \triangleq \log _{2}\Big(1+\mathrm{SINR}_{k}^{(\mathrm{OL})(B)}\Big)$, where 
	\begin{align}\label{SINK_OL_B}
		\mathrm{SINR}_{k}^{(\mathrm{OL})(B)}\approx \frac{p {E}_{k}^{(\mathrm{signal},\mathrm{OL})(B)}({\bf\Phi})}{ \sum\limits_{i=1, i \neq k}^{K} p I_{k i}^{(\mathrm{OL})(B)}({\bf\Phi})+E_{k}^{(\mathrm{hwi},\mathrm{OL})(B)}({\bf\Phi})+\sigma^{2} E_{k}^{(\mathrm{noise},\mathrm{OL})(B)}({\bf\Phi})}, 
	\end{align}
	with
	\begin{align}\label{noise_OL_B}
		E_{k}^{(\mathrm{noise},\mathrm{OL})(B)}({\bf\Phi}) =  
		{\sum\limits_{l=1}^L \sum\limits_{s=1}^S \beta_{l,s}\alpha_{s,k}}\left|{f_{l,s,k}({\bf\Phi})}\right|^{2} +    {\sum\limits_{l=1}^L \gamma_{l,k} },
	\end{align}
	\begin{align}\label{signal power_OL_B}
	&E_{k}^{(\mathrm{signal},\mathrm{OL})(B)}({\bf\Phi}) ={\sum\limits_{l_{1}=1}^L\sum\limits_{l_{2}=1}^L\sum\limits_{s_{1}=1 }^S \sum\limits_{s_{2}=1 }^S } \beta_{l_{1},s_{1}}\beta_{l_{2},s_{2}}\alpha_{s_{1},k}\alpha_{s_{2},k}\mathrm{sinc}^{2}\left(\kappa_r\pi\right)\left|{f_{l_{1},s_{1},k}({\bf\Phi})}\right|^{2}\left|{f_{l_{2},s_{2},k}({\bf\Phi})}\right|^{2}\nonumber\\
	&+{\sum\limits_{l_{1}=1}^{L} \sum\limits_{l_{2}=1}^{L} \sum\limits_{s=1}^{S}}   \Big(2\beta_{l_{1},s}\alpha_{s,k}\gamma_{l_{2},k}\mathrm{sinc}\left(\kappa_r\pi\right)\left|{f_{l_{1},s,k}({\bf\Phi})}\right|^{2}+\beta_{l_{1},s}\left(1-\mathrm{sinc}^{2}\left(\kappa_r\pi\right)\right)\beta_{l_{2},s}\alpha_{s,k}^{2}\nonumber\\
	&{f_{l_{1},s,k}^{H}({\bf\Phi})}{f_{l_{2},s,k}({\bf\Phi})}{\mathbf{a}_{R}^{H}(l_{1},s)} {\mathbf{a}_{R}(l_{2},s)}\Big),
	\end{align}
	\begin{align}\label{interference_OL_B}
	&I_{ki}^{(\mathrm{OL})(B)}({\bf\Phi}) \nonumber\\
	&={\sum\limits_{l_{1}=1}^L\sum\limits_{l_{2}=1}^L\sum\limits_{s_{1}=1 }^S \sum\limits_{s_{2}=1 }^S } \sqrt{\alpha_{s_{1},k}\alpha_{s_{1},i}\alpha_{s_{2},i}\alpha_{s_{2},k}}\beta_{l_{1},s_{1}}\beta_{l_{2},s_{2}}\mathrm{sinc}^{2}\left(\kappa_r\pi\right) f_{l_{1},s_{1},k}^{H}({\bf\Phi})f_{l_{1},s_{1},i}({\bf\Phi})f_{l_{2},s_{2},i}^{H}({\bf\Phi})f_{l_{2},s_{2},k}({\bf\Phi})\nonumber\\
	&+{\sum\limits_{l_{1}=1}^{L} \sum\limits_{l_{2}=1}^{L} \sum\limits_{s=1}^{S}}   
	\beta_{l_{1},s}\beta_{l_{2},s}\alpha_{s,k}\alpha_{s,i}\left(1-\mathrm{sinc}^{2}\left(\kappa_r\pi\right)\right){f_{l_{1},s,k}^{H}({\bf\Phi})}{f_{l_{2},s,k}({\bf\Phi})}{\mathbf{a}_{R}^{H}(l_{1},s)} {\mathbf{a}_{R}(l_{2},s)},
	\end{align}
\begin{align}\label{hwi_OL_B}
	E_{k}^{(\mathrm{hwi},\mathrm{OL})(B)} ({\bf\Phi})=\kappa_u^{2} p \Bigg({E}_{k}^{(\mathrm{signal},\mathrm{OL})}({\bf\Phi})+\sum\limits_{i=1,i\neq k}^{K}I_{k i}^{(\mathrm{OL})}({\bf\Phi})\Bigg).
\end{align}
\end{corollary}

\itshape \textbf{Proof:}  \upshape  As $\varepsilon_{s,k}$ and $\delta_{l,s}\to\infty$, we set $c_{l,s,k}=0$ and $c_{l,s,k}\delta_{l,s}\varepsilon_{s,k}=\alpha_{s,k}\beta_{l,s}$, $\forall l,s,k$. When $p_{k}=p/B$, $B\rightarrow\infty$, we complete the proof by retaining the significant terms whose asymptotic behavior is $\mathcal{O}(B)$. Note that ${\mathbf{a}_{B}^{H}(l_{1},s_{1})} {\mathbf{a}_{B}(l_{2},s_{2})}=B$ when $l_{1}=l_{2}$, $s_{1}=s_{2}$.   \hfill $\blacksquare$

In Corollary \ref{corollary6}, we consider the ideal case where the RIS-aided channel is in a scenario with almost no scattering. We can find that when $p_{k}=p/B$ and as $B\rightarrow\infty$, the SINR in (\ref{SINK_OL_B}) will converge to a constant containing only the factors $\kappa_{r}$ and $\kappa_{u}^{2}$, meaning that the rate will be affected by the receiver HWI and RIS phase noise even in the case of LoS channels. Also, this result proves that $p_{k}$ can scale as $p/B$ while keeping a non-zero value of the rate when $B\rightarrow\infty$. Nevertheless, we can compensate for this rate degradation by properly optimizing the phase shifts of RISs \cite[Corollary 1]{9743440}. For instance, we can align the phase shift towards target user $k$, and the rate for user $k$ can be greatly improved compared to the reduced interference from other users, indicating the potential for optimizing $\bf\Phi$ in less scattering environments.

Besides, we can observe that the above expressions in (\ref{signal power_OL_B}) $\sim$ (\ref{hwi_OL_B}) contain the term with respect to the variable of phase shift matrix ${\bf\Phi}_{s}$, i.e., ${f_{l,s,k}({\bf\Phi})}$. Due to these terms involve RIS phase shifts, it is difficult to receval the power scaling law with respect to the number of RIS elements $R$. Therefore, we consider the special case where the phase shifts of RISs are optimized to serve multi-user communication. In this case, we assume that the phase shifts are not aligned to arbitrary user to maintain sufficient system capacity, which corresponds to the condition that $0< \left|{f_{l,s,k}({\bf\Phi})}\right|<  R$, $\forall l,s,k$ \cite[Lemma 3]{9973349}. Then, we have $\frac{\left|{f_{l,s,k}({\bf\Phi})}\right|}{R}\to 0$ as $R\rightarrow\infty$, $\forall l,s,k$. In this case, we obtain that the noise term in (\ref{noise_OL_B}) scales as $\mathcal{O}(R)$, the desired signal of user $k$ in (\ref{signal power_OL_B}) scales as $\mathcal{O}(R^2)$, the multi-user interference suffered by user $k$ in (\ref{interference_OL_B}) scales as $\mathcal{O}(R^2)$, and the HWI in (\ref{hwi_OL_B}) scales as $\mathcal{O}(R^2)$. Note that this result holds for any user, which means that user fairness can be guaranteed in this case. Then, when the cascaded channels are pure LoS, the transmit power can be scaled down by $p_{k}=\frac{p}{BR}$, $\forall k$, while maintaining a non-zero rate for $B,R\rightarrow\infty$.
 
\section{Phase Shifts Design}\label{section4}
In this section, we aim to maximize the rate performance based on the rate expression $R_{k}(\mathbf{\Phi})$ in (\ref{rate}) by designing optimization problem and optimizing the RIS phase shifts. Since the rate (\ref{rate}) only relies on long-term statistical CSI, the phase shifts of RISs need to be updated for a long time, and the channel estimation requires a lower overhead compared to instantaneous CSI-based schemes. On the one hand, to improve the system capacity of the RIS-aided CF-mMIMO system, an optimization problem that maximizes users' sum rate can be formulated as
\begin{subequations}\label{p1}
	\begin{equation}\label{objective1}
		\;\;\;\;\max\limits_{\mathbf{\Phi}}  \;\; \sum\limits_{k=1}^{K} {R_{k}(\mathbf{\Phi})},\qquad\qquad\quad
	\end{equation}
	\begin{equation}\text {s.t. } \;\;{\left|\left[\mathbf{\Phi}_{s}\right]_{r,r}\right|=1, \forall s,r.} \label{constraint1}
	\end{equation}
\end{subequations}
 On the other hand, to provide uniform QoS for users, we should guarantee user fairness in phase shift optimization. Thus, we consider the users' minimum rate maximization problem and formulate it as
\begin{subequations}\label{p2}
	\begin{equation}\label{objective2}
		\max\limits_\mathbf{\Phi}  \;\; \min\limits_{k} \;R_{k}(\mathbf{\Phi}),\qquad\;\;\;\end{equation}
	\begin{equation}\text {s.t. } \;\;(\text{\ref{constraint1}}) \; .\nonumber\qquad\quad\end{equation}
\end{subequations}

%Considering the object function and the unit modulus constraint are non-convex, we can conclude that the optimization problems (\ref{p1}) and (\ref{p2}) are non-convex. Also, 
The two objective functions contain the complicated rate expression $R_{k}(\mathbf{\Phi})$, where the phase shift matrices of different RISs are tightly coupled. Therefore, it is difficult to solve the two problems by exploiting conventional optimization methods, such as semi-definite programming (SDP) and the majorization-minimization (MM). In this regard, the genetic algorithm (GA)-based methods have been proposed to tackle the similar optimization problems in RIS-assisted mMIMO systems and even CF-mMIMO systems \cite{9743440,fullversion}. However, the GA-based method could converge slowly. In contrast, without considering the coupled phase shifts, the projected gradient ascent-based method can converge faster \cite{kammoun2020asymptotic}. In this paper, we utilize an accelerated gradient ascent-based method, which avoids the suboptimality caused by the projection operation and effectively increases the convergence speed.

According to the objective function containing the rate expression (\ref{rate}), we can find that the phase shift matrix $\mathbf{\Phi}_{s}$ of each RIS $s$ is tightly coupled to each other, which makes it difficult to obtain the gradient with respect to the whole phase shift of all RISs. To solve this problem, we transform the summation of matrix products containing matrix $\mathbf{\Phi}_{s}$ into the matrix product only containing matrix $\mathbf{\Phi}$. The detailed transformation steps and results can be found in Appendix \ref{appB}.

After the above transformation, we can obtain a tractable rate expression $r_{k}(\mathbf{\Phi})$ only containing the whole phase shift matrix $\mathbf{\Phi}$, where the result of $r_{k}(\mathbf{\Phi})$ is given in (\ref{r_k}). For ease of expression and processing, we introduce the vectors $\boldsymbol{\theta}_{s}=[\theta_{s,1},\theta_{s,2},\ldots,\theta_{s,R}]^T$, $\boldsymbol{\theta}=[\boldsymbol{\theta}_{1}^{T},\boldsymbol{\theta}_{2}^{T},\ldots,\boldsymbol{\theta}_{S}^{T}]^T$, $\boldsymbol{v}_{s} = [e^{j \theta_{s,1}},e^{j \theta_{s,2}},\ldots,e^{j \theta_{s,R}}]^T$, and $\boldsymbol{v} = [\boldsymbol{v}_{1}^{T},\boldsymbol{v}_{2}^{T},\ldots,\boldsymbol{v}_{S}^{T}]^T$, so that  $\boldsymbol{v} = e^ {j \boldsymbol{\theta}}$ and $\mathbf{\Phi}=\mathrm{diag}  \left(  \boldsymbol{v}  \right)$. Then, we can obtain tractable objective functions with  $r_{k}(\boldsymbol{\theta})$ by substituting $\mathbf{\Phi}=\mathrm{diag}  \left( e^ {j \boldsymbol{\theta}}  \right)$ into $r_{k}(\mathbf{\Phi})$, and the optimization problem and the gradient with respect to the phase shift $\boldsymbol{\theta}$ can be given next. Since the objective function in (\ref{p2}) includes the min function, which is not differentiable, we utilize the method in \cite{9973349} to approximate the objective function in (\ref{p2}) as
\begin{align}\label{r_k_min}
	&\min _{k} r_{ k} (\boldsymbol{\theta})\approx  -\frac{1}{\mu} \ln \left\{\sum\limits_{k=1}^{K} \exp \left\{-\mu r_{ k}(\mathbf{\boldsymbol{\theta}})\right\}\right\}  \triangleq  \underline{r}_{ k}(\mathbf{\boldsymbol{\theta}}),
\end{align}
where $\mu$ is the constant that controls the accuracy of the approximation. Thus, the two optimization problems in (\ref{p1}) and (\ref{p2}) can be recast as 
\begin{subequations}\label{Problem}
	\begin{align}
		&\max _{   \boldsymbol{\theta}  }\; \;  \sum\limits_{k=1}^{K}r_{ k} (\boldsymbol{\theta}) \text{ or } \underline{r}_{ k}(\mathbf{\boldsymbol{\theta}}) , \\\label{constraint_f}
		&\text { s.t. } \quad  0 \leq \theta_{s,r} < 2 \pi, \forall s, r.
	\end{align}
\end{subequations}

We then calculate the gradients of $r_{ k} (\boldsymbol{\theta})$ and $\underline{r}_{ k}(\mathbf{\boldsymbol{\theta}})$ with respect to $\mathbf{\boldsymbol{\theta}}$, and the detailed calculation steps and results of $\frac{\partial r_{k}(\boldsymbol{\theta})}{\partial \boldsymbol{\theta}}$ and $\frac{\partial \underline{r}_{k}(\boldsymbol{\theta})}{\partial \boldsymbol{\theta}}$ are given in Appendix \ref{appC}. Based on the derived gradients of $r_{k}(\boldsymbol{\theta})$ and $\underline{r}_{ k}(\mathbf{\boldsymbol{\theta}})$ in (\ref{gradient_r_k_min}) and (\ref{gradient_r_k}), we adopt the accelerated gradient ascent-based method to obtain the optimal phase shift $\boldsymbol{\theta}^*$ of all RISs. For completeness, the procedure of the proposed method is presented in Algorithm \ref{algorithm1}, which shows the steps of optimizing $\underline{r}_{ k}(\mathbf{\boldsymbol{\theta}})$, and the optimization process for $r_{ k} (\boldsymbol{\theta})$ is similar and thus omitted here. 
\begin{breakablealgorithm}
	\caption{Accelerated Gradient Ascent-based Method}
	\begin{algorithmic}[1]\label{algorithm1}
		\STATE Initialize $\boldsymbol{\theta}_0$ randomly, $i=0$, $a_0=1$, $\boldsymbol{x}_{-1} =   \boldsymbol{\theta}_0  $;
		\WHILE{1}
		\STATE Calculate the gradient vector $  \underline{\boldsymbol{r}}_{k}^{\prime}(   \boldsymbol{\theta}_i   )= \left.    \frac{\partial \underline{r}_{k}(\boldsymbol{\theta})}{\partial \boldsymbol{\theta}}     \right| _  {\boldsymbol{\theta} = \boldsymbol{\theta}_i}   $;
		\STATE Obtain the step size $t_i$ based on the backtracking line search;
		\STATE $\boldsymbol{x}_{i} =  \boldsymbol{\theta}_i +t_i  \underline{\boldsymbol{r}}_{k}^{\prime}(   \boldsymbol{\theta}_i   )  $;
		\STATE $a_{i+1}=(1+\sqrt{4 a_i^2+1}) / 2$;
		\STATE $\boldsymbol{\theta}_{i+1} =\boldsymbol{x}_{i}+\left(a_i-1\right)\left(     \boldsymbol{x}_{i}  - \boldsymbol{x}_{i-1}       \right) / a_{i+1}$;
		\IF {$\underline{r}_k( \boldsymbol{\theta}_{i+1}  ) - \underline{r}_k( \boldsymbol{\theta}_{i}  ) < 10^{-5}$}
		\STATE $\boldsymbol{\theta}^*= \boldsymbol{\theta}_{i+1}$, break;
		\ENDIF
		\STATE $i=i+1$;
		\ENDWHILE
	\end{algorithmic}
\end{breakablealgorithm}
%The proposed method will stop if the change of the function value is less than $\xi$.

\section{Numerical Results}\label{section5}
\begin{figure}
	\setlength{\abovecaptionskip}{0pt}
	\setlength{\belowcaptionskip}{-10pt}
	\centering
	\includegraphics[width=3.8in]{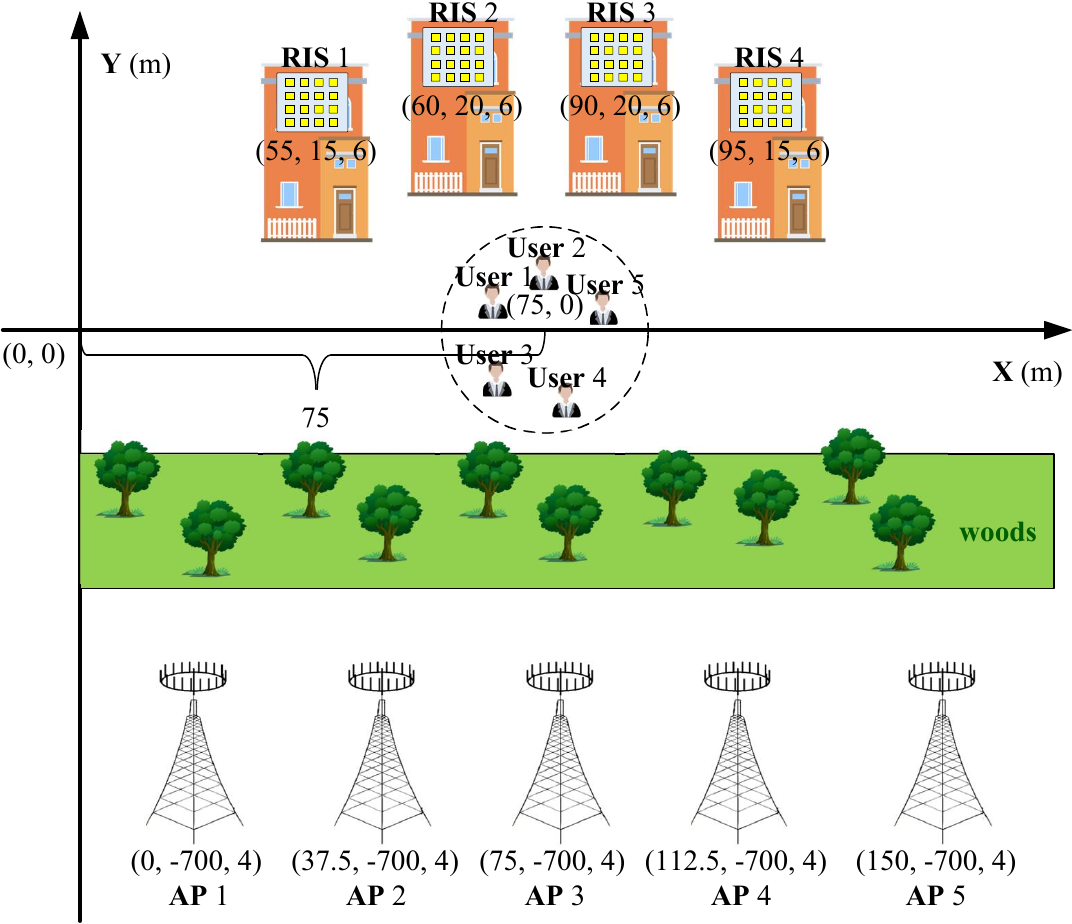}
	\DeclareGraphicsExtensions.
	\caption{The simulation scenario of the RIS-assisted CF-mMIMO system.}
	\label{sim_fig}
\end{figure}

In this section, several numerical simulations are provided to evaluate the rate performance of the RIS-assisted CF-mMIMO system with HWIs and validate our analytical conclusions. 
Unless otherwise specified, we adopt an RIS-assisted CF-mMIMO system with the topology given in Fig. \ref{sim_fig}. 
%In this setup, there are $L = 5$ APs serving $K = 5$ users simultaneously in the cell-free network, while the network throughput is limited due to the obstacles such as green belts. To address this issue, we deploy $N = 4$ RISs on the facade of four buildings separately, which are high enough to construct extra reflection links. 
In this setup, $K = 5$ users are randomly located in a circle with a radius of $4$ m, and $S = 4$ RISs on the facade of high buildings construct extra reflection links to assist $L = 5$ APs to communicate with users.
The AoA and AoD of APs, RISs, and users are generated randomly from $[0,2\pi]$\cite{pan2020intelligent}, and these angles will be fixed after the initial generation. The large-scale path-loss factors are set as $ \alpha_{s,k}=10^{-3} \big(d^{\mathrm{UR}}_{s,k}\big)^{-\alpha_{\mathrm{UR}}} $, $ \beta_{l,s}=10^{-3} \big(d^{\mathrm{RB}}_{l,s}\big)^{-\beta_{\mathrm{RB}}}  $, and $\gamma_{l,k}=10^{-3}\big(d_{l,k}^{\mathrm{UB}}\big)^{-\gamma_{\mathrm{UB}}}$ \cite{pan2020multicell}, where $d^{\mathrm{UR}}_{s,k}$, $d^{\mathrm{RB}}_{l,s}$ and $d_{l,k}^{\mathrm{UB}}$ respectively represent the distances of user $k$-RIS $s$, RIS $s$-AP $l$, and user $k$-AP $l$, the path-loss exponents are $\alpha_{\mathrm{UR}}=2$, $\beta_{\mathrm{RB}}=2.5$, and $\gamma_{\mathrm{UB}}=4$ \cite{wu2019intelligent}. 
Moreover, the severity of the residual HWIs at the transmitter and the receiver are set equal, i.e., $\kappa_{u}^{2}=\kappa_{b}^{2}=0.3^{2}$. Also, we set $b=2$, then $\kappa_{r}=0.25$. The other simulation parameters are given in Table \ref{tab1}. The MC simulation is achieved by averaging $ 10^{5} $ random channel realizations.
\begin{table}[t]
	\captionsetup{font={small}}
	\caption{Simulation Parameters\label{tab1}}
	\centering 	
	\begin{tabular}{|c|c|c|c|c|c|c}
		\hline
		{\color{black}AP} antennas & {\color{black}$B=9$} & RIS elements &	$R=36$\\	
		\hline
		\color{black}Antenna spacing & \color{black}$d=\lambda/2$ & 		
		Noise power& {$\sigma^2=-104$ dBm}\\
		\hline
		Transmit power & \multicolumn{3}{c| } {\color{black}$p_{k}=P=30$ dBm, $\forall k$}\\  
		\hline
		Rician factors & \multicolumn{3}{c|}{$\delta_{l,s}=\delta=1$, $\varepsilon_{s,k}=\varepsilon=10,\forall l,s,k$}\\
		\hline
		Approximation factor &\multicolumn{3}{c| } {$\mu=100$}\\
		\hline
	\end{tabular}
	\vspace{-10pt}
\end{table}

\begin{figure}
	\setlength{\abovecaptionskip}{0pt}
	\setlength{\belowcaptionskip}{-20pt}
	\centering
	\includegraphics[width=3.8in]{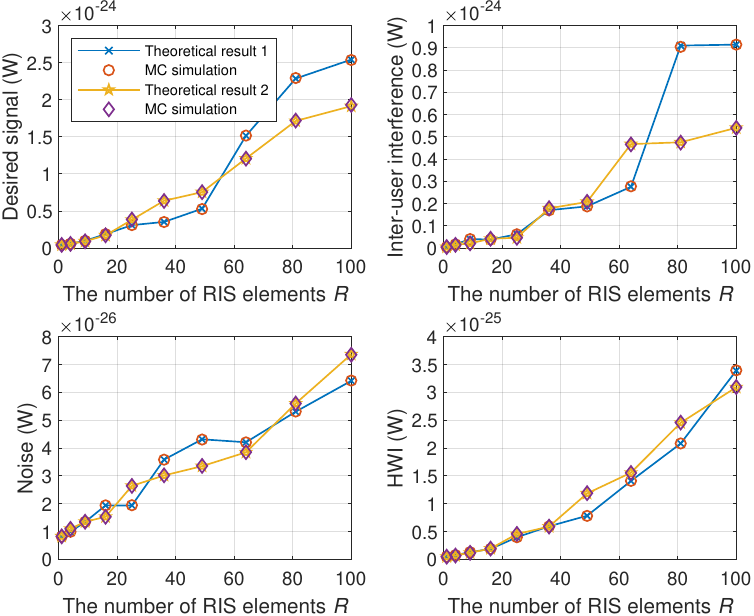}
	\DeclareGraphicsExtensions.
	\caption{Desired signal power, inter-user interference power, noise power and HWI power for user $k$ under random channel realizations.}
	\label{figure1}
\end{figure}
To begin with, we validate the accuracy of the derivations of our closed-form expressions in Theorem \ref{lemma1}. Fig. \ref{figure1} shows the theoretical and numerical results of the desired signal $ p_k {E}_{k}^{(\mathrm{signal})}({\bf\Phi})$, the HWI $E_{k}^{(\mathrm{hwi})}({\bf\Phi})$, the inter-user interference $\sum\nolimits_{i=1, i \neq k}^{K} p_i I_{k i}({\bf\Phi})$, and noise $\sigma^{2}E_{k}^{(\mathrm{noise})}({\bf\Phi})$ for user $k$ under two independent random realizations. To adequately demonstrate the accuracy of our derivations, we adopt random channel realizations and phase shifts for different number of RIS elements. The theoretical results are obtained by substituting the random phase shifts into our derived expressions. Similarly, based on the same channel and phase shifts, we obtain the numerical results with the MC method. As shown in Fig. \ref{figure1}, the theoretical results calculated by the derived expressions of the four terms in (\ref{rate}) exactly match the MC simulations, which verifies the correctness of our derived expressions in Theorem \ref{lemma1}. 

In the following, we utilize the gradient ascent-based method to optimize the phase shifts of RISs based on the closed-form approximate rate expression in Theorem \ref{lemma1}. To this end, the sum rate and minimum rate maximization problems in (\ref{Problem}) are solved respectively to get the optimized phase shifts $\boldsymbol{\theta}^*_{\mathrm{sum}}$ and $\boldsymbol{\theta}^*_{\mathrm{min}}$. Then, we calculate the rate performance by substituting the optimized phase shifts into objective functions, i.e., obtaining the sum user rate $\sum_{k=1}^{K} r_k\left(\boldsymbol{\theta}^*_{\mathrm{sum}}\right)$ and the minimum user rate $\min\limits_{k} r_k\left(\boldsymbol{\theta}^*_{\mathrm{min}}\right)$.

\subsection{The Balance between System Capacity and User Fairness}
%\begin{figure*}[ht]
%	\setlength{\abovecaptionskip}{0pt}
%	\setlength{\belowcaptionskip}{-10pt}
%	\centering
%	\begin{minipage}[b]{0.4\linewidth}
%		\centering
%		\includegraphics[width=2in]{fig_ric.pdf}
%		\DeclareGraphicsExtensions.
%		\caption{The rate performance versus the Rician factor of RISs-APs channels. }
%		\label{figure_Ri}
%	\end{minipage}%
%	\begin{minipage}[b]{0.62\linewidth}
%		\centering
%		\subfigure[Sum user rate versus transmit power $P$.] {\includegraphics[width=2in]{fig2-a.pdf}
%			\label{figure2_a}}
%		\subfigure[Minimum user rate versus transmit power $P$.] {\includegraphics[width=2in]{fig2-b.pdf}
%			\label{figure2_b}}
%		\DeclareGraphicsExtensions.
%		\caption{Achievable rate versus transmit power $P$.}
%		\label{figure2}
%	\end{minipage}
%	\vspace{-10pt}
%\end{figure*}
For comparison, we adopt six phase shift designs based on RIS-assisted CF-mMIMO systems. We denote the minimum rate and sum rate obtained from the optimized phase shift $\boldsymbol{\theta}^*_{\mathrm{sum}}$ as ``min rate by max-sum'' and ``sum rate by max-sum'', respectively. The minimum rate and sum rate obtained from the optimized phase shift $\boldsymbol{\theta}^*_{\mathrm{min}}$ are denoted as ``min rate by max-min'' and ``sum rate by max-min'', respectively. In addition, we propose a non-optimized RIS design in which the rate is obtained by averaging over $ 10^{5} $ results calculated by random phase shifts. 

\begin{figure}
	\setlength{\abovecaptionskip}{0pt}
	\setlength{\belowcaptionskip}{-20pt}
	\centering
	\includegraphics[width=3.8in]{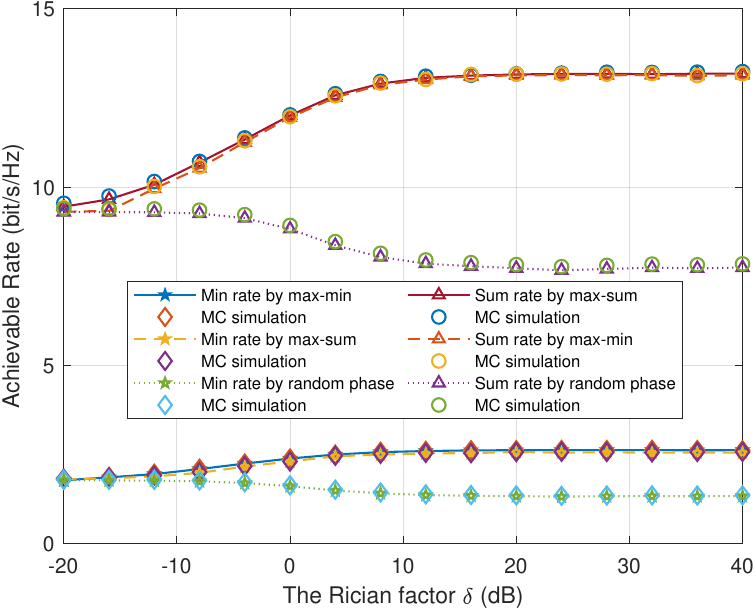}
	\DeclareGraphicsExtensions.
	\caption{The rate performance versus the Rician factor of RISs-APs channels. }
	\label{figure_Ri}
\end{figure}
Fig. \ref{figure_Ri} evaluates the rate versus the Rician factor $\delta$ for six kinds of RIS design ($\delta_{l,s}=\delta, \forall l,s$). These results show that there is a small gap between the analytical result and the MC simulation since the rate expression in (\ref{rate}) is an approximation of the achievable rate. In all results, the approximate rate matches well with the MC simulation, which validates the correctness of our derived expressions. It can be seen that as $\delta$ grows, there is a significant rate loss for the random phase shift-based design. However, with the increase of $\delta$, the sum rate and minimum rate keep increasing for solving the sum rate and minimum rate maximization problems in (\ref{Problem}), indicating that it is important to optimize the phase shifts of RISs in scenarios with less scatters.
Meanwhile, the rate improvement obtained by solving the two maximization problems are almost equal, which means that high system capacity and user fairness can be maintained at the same time in a scenario with less scatters.
This result differs from the RIS-assisted mMIMO systems with the cellular architecture \cite{9743440}. The reason is that the direct link can provide sufficient spatial multiplexing gains and an additional communication link. Besides, the single RIS $s$-AP $l$ channel is highly correlated for different users, but the channels between different APs and RISs are independent. Specifically, when $\delta\rightarrow\infty$, the rank of the cascaded channel $\mathbf{G}$ will approach $LS$, which shows that when $K<LS$, the system can support the communication of $K$ users.
%, which demonstrates the superiority of the proposed architecture.

\subsection{The Interplay between RISs and CF-mMIMO in the Presence of HWIs}
\begin{figure}[!t]
	\setlength{\abovecaptionskip}{0pt}
	\setlength{\belowcaptionskip}{-20pt}
	\centering
	\subfigure[Sum user rate versus transmit power $P$.] {\includegraphics[width=3in]{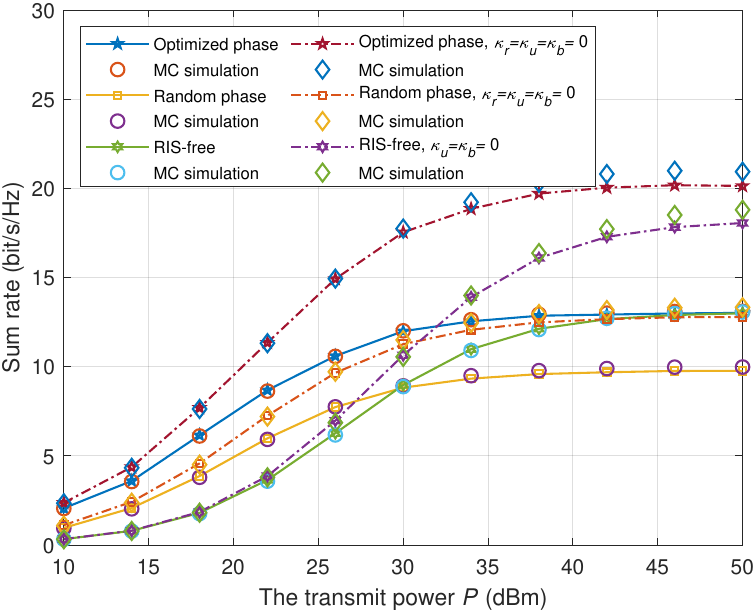}
		\label{figure2_a}}
	\subfigure[Minimum user rate versus transmit power $P$.] {\includegraphics[width=3in]{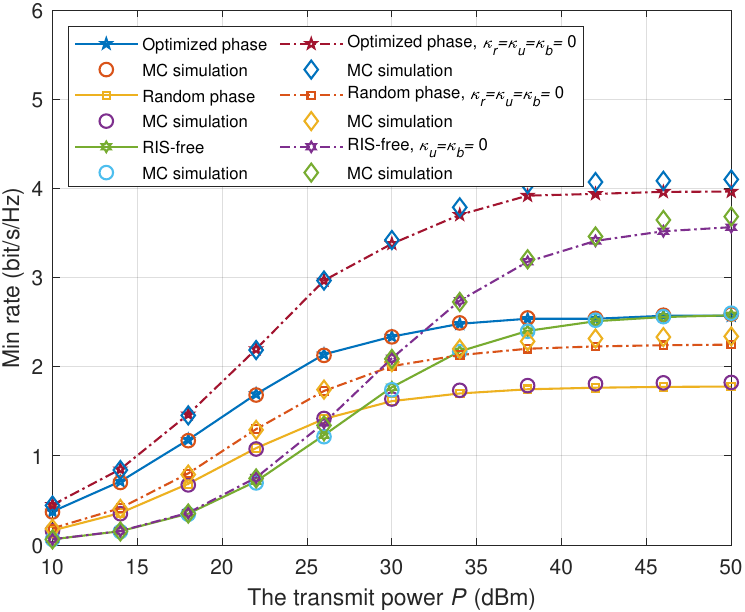}
		\label{figure2_b}}
	\DeclareGraphicsExtensions.
	\caption{Achievable rate versus transmit power $P$.}
	\label{figure2}
\end{figure}
In this subsection, we explore the gains of applying imperfect RISs to the CF-mMIMO system with HWIs. We consider the RIS-assisted system with $R=25$ and $B=9$ as a benchmark system. We adopt the RIS design based on the optimized phase shift, and refer to it as ``optimized phase'', in which the sum rate is calculated by $\boldsymbol{\theta}^*_{\mathrm{sum}}$, and the minimum rate is calculated by $\boldsymbol{\theta}^*_{\mathrm{min}}$. Also, we consider the RIS design based on the random phase shift, and refer to it as ``random phase''.
For comparison, we consider an RIS-free mMIMO system \cite{8891922}, meaning that there is no RIS in the system to provide additional communication links to users, i.e., $S = 0$.

Fig. \ref{figure2} illustrates the rate performance versus transmit power $P$. Here, we also present the results with $\kappa_{u}=\kappa_{b}=0$ as the case of ideal transceiver hardware, and $\kappa_{r}=0$ corresponds to the RIS with infinite quantization precision. We can observe that in the small $P$ region, optimized phase-based RIS effectively improves the rate performance of CF-mMIMO systems, which unveils the gain of RIS at low transmit power. 
Nevertheless, as $P$ increases, the rate of the system with ideal hardware greatly outperforms that of the system with HWIs, which shows the significant performance deterioration caused by HWIs. This result agrees with our derived rate expression in Theorem \ref{lemma1}, where the HWIs restrict the rate via the denominator of SINR in (\ref{rate}). Besides, RIS-free systems will gradually outperform the random phase-based RIS systems as $P$ increases, which shows the gain of optimizing phase shifts of RISs. With a very large $P$, the RIS-free system even approaches the optimized phase-based RIS systems.
This is because the rate will be limited by multi-user interference in the large $P$ region, which aggravates the negative impacts of additional interference caused by RISs.

\begin{figure}
	\setlength{\abovecaptionskip}{0pt}
	\setlength{\belowcaptionskip}{-20pt}
	\centering
	\includegraphics[width=3.8in]{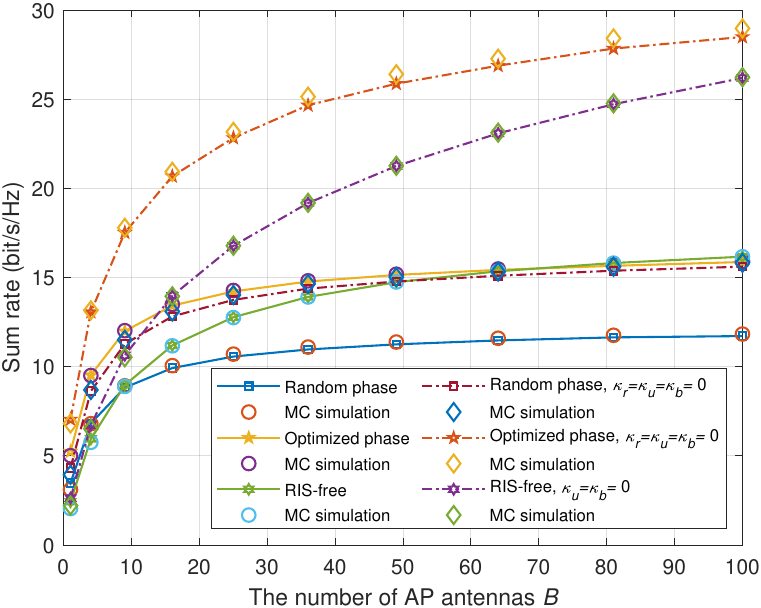}
	\DeclareGraphicsExtensions.
	\caption{Sum rate versus the number of AP antennas $B$.}
	\label{figure4}
\end{figure}
To achieve a larger system capacity, in the following, we only consider the users' sum rate obtained by RIS-free systems or RIS-assisted systems.
In Fig. \ref{figure4}, we evaluate the sum rate as a function of the number of AP antennas $B$. We can see that in RIS-assisted systems or the RIS-free system with HWIs, the sum rate increases significantly as $B$ increases and approaches saturation as $B\rightarrow\infty$ due to the multi-user interference and HWIs. Nevertheless, this feature is no longer available for RIS-free systems without HWIs. It is observed that the rate keeps growing in the RIS-free system when $\kappa_{u}=\kappa_{b}=0$, which is consistent with our analysis in Corollary \ref{corollary1}. The HWIs reduce the rate performance of all systems, and this performance degradation is more significant in the region with a high value of $B$. 
To reduce hardware cost and power consumption, RISs can be deployed instead of APs to improve rate performance. Fig. \ref{figure4} has shown the sum rate of the RIS-assisted system with $R=25$ elements and the RIS-free system. We can see that even in the presence of HWIs, the rate of the RIS-assisted system using $16$ AP antennas can be obtained by the RIS-free system using $36$ AP antennas. This result shows that we can achieve the same rate performance with fewer AP antennas with the help of RIS. Since the RIS with passive elements does not require expensive hardware, it is promising to integrate RISs into CF-mMIMO systems to maintain the system capacity at lower cost. 

\begin{figure}
	\setlength{\abovecaptionskip}{0pt}
	\setlength{\belowcaptionskip}{-20pt}
	\centering
	\includegraphics[width=3.8in]{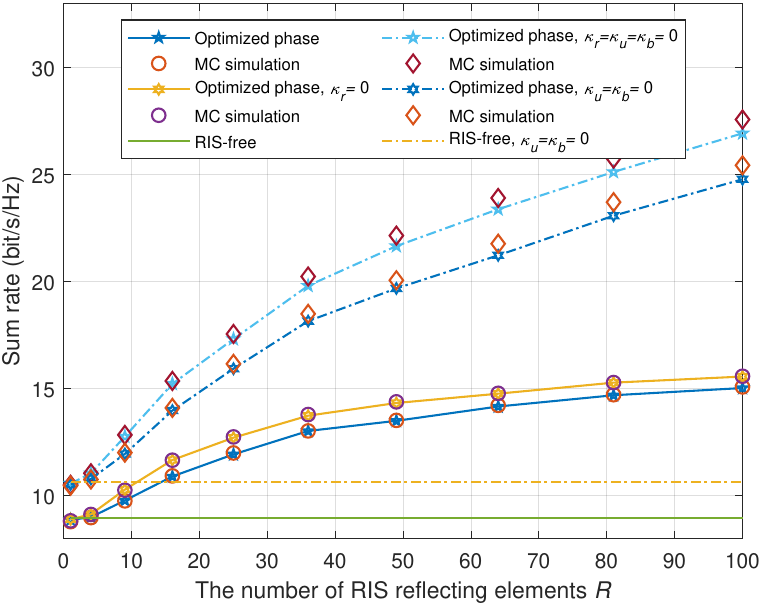}
	\DeclareGraphicsExtensions.
	\caption{Sum user rate versus the number of RIS reflecting elements $R$.}
	\label{figure3}
\end{figure}
Fig. \ref{figure3} shows the sum user rate versus the number of RIS reflecting elements $R$. As $R$ increases, regardless of the presence of HWIs, the optimized phase shifts-based RIS can significantly improve the rate performance of CF-mMIMO systems.  
We can see that the RIS-assisted CF-mMIMO systems without three kinds of HWIs have better performance than those with HWIs, clearly showing the impact of different HWIs on system performance. Specifically, in the region of large $R$, the transceiver HWIs will bring a more significant rate degradation to the optimized phase shifts-based system than that caused by the RIS phase noise. The result indicates that the benefit of RIS is still substantial even if the imperfect RIS is applied in cell-free systems with low quantization precision. 

\subsection{The Impact of HWIs}

To further investigate the effect of HWIs on the achievable rate, we evaluate the rate performance by varying the values of the HWI coefficients (i.e., $\kappa_{u}, \kappa_{b}, \kappa_{r}$). 
%We only consider the RIS-assisted system based on optimized phase shifts as the benchmark system in this subsection.

\begin{figure}
	\setlength{\abovecaptionskip}{0pt}
	\setlength{\belowcaptionskip}{-20pt}
	\centering
	\includegraphics[width=3.8in]{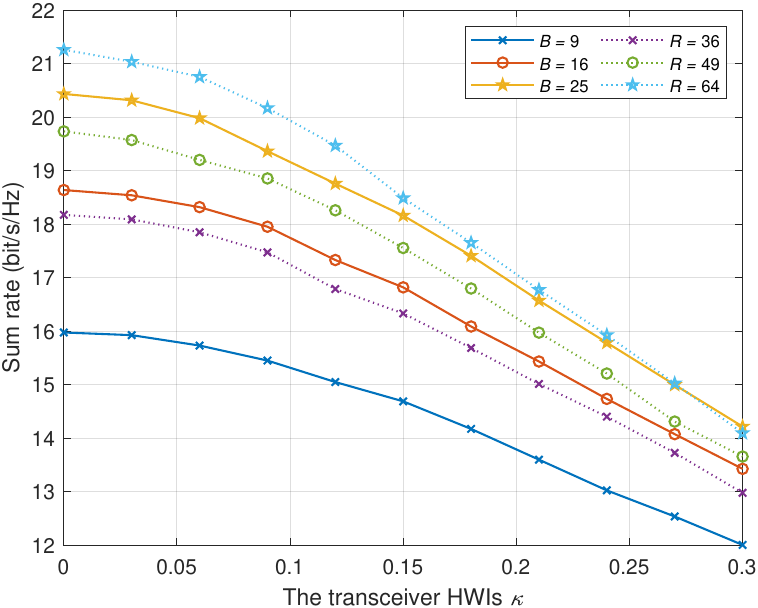}
	\DeclareGraphicsExtensions.
	\caption{Sum user rate versus the transceiver HWIs coefficient $\kappa$ under different AP antennas $B$ and RIS elements $R$}
	\label{figure5}
\end{figure}
Fig. \ref{figure5} shows the performance of the users' sum rate versus the transceiver HWIs coefficient $\kappa$, where $\kappa_{u}=\kappa_{b}= \kappa$. Here, the effect of RIS phase noise is not considered since we only explore the effect of the transceiver HWIs. In addition, we change the number of AP antennas or RIS elements under the variation of the coefficient $\kappa$. We can see that the rate performance of the systems will decrease as the transceiver HWIs become severer. Besides, we can observe that the number of AP antennas $B$ has a larger impact on the rate performance than the number of RIS elements $R$ due to the larger number of APs and the presence of HWI at RISs. However, the rate gap between systems with different numbers of AP antennas or RIS elements will become smaller as $\kappa$ increases, especially regarding the number of RIS elements.

\begin{figure}
	\setlength{\abovecaptionskip}{0pt}
	\setlength{\belowcaptionskip}{-20pt}
	\centering
	\includegraphics[width=3.8in]{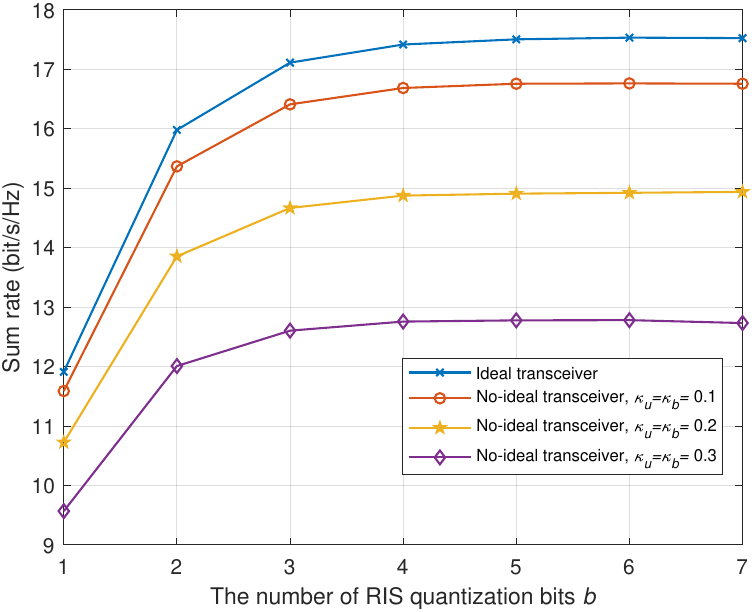}
	\DeclareGraphicsExtensions.
	\caption{The impact of RIS quantization bits $b$}
	\label{figure6}
\end{figure}
The above results show the detrimental impacts of increasing the transceiver HWI level, especially compared to the case of ideal hardware. Considering the quantization errors caused by discrete phase shifts at RISs, we present Fig. \ref{figure6} to investigate the effect of RIS phase noise under different quantization bits. Note that a larger $\kappa_{u}$ or $\kappa_{b}$ indicates more severe transceiver distortion. It can be seen that RIS phase noise has a more significant impact on the rate when the transceiver has lower impairments. For different  transceiver HWI levels, only the serious phase noise, i.e., $b=1$ and $b=2$, will bring an obvious performance degradation to the system. Then, even with ideal transceiver hardware, the RISs with a $3$-bit quantizer are sufficient to achieve near-optimal rate performance. This observation verifies that without using high-quality hardware, low-precision RISs can still bring good enough performance to the CF-mMIMO system with transceiver HWIs.

\subsection{The Power Scaling Laws}
Finally, we examine the promising properties of RIS-assisted CF-mMIMO systems when the transmit power is scaled down according to certain laws in the presence of HWIs. 
%We consider the RIS-assisted system with optimized phase shifts as the benchmark system, which adopts the Rician channel and imperfect hardware.

\begin{figure}
	\setlength{\abovecaptionskip}{0pt}
	\setlength{\belowcaptionskip}{-20pt}
	\centering
	\includegraphics[width=3.8in]{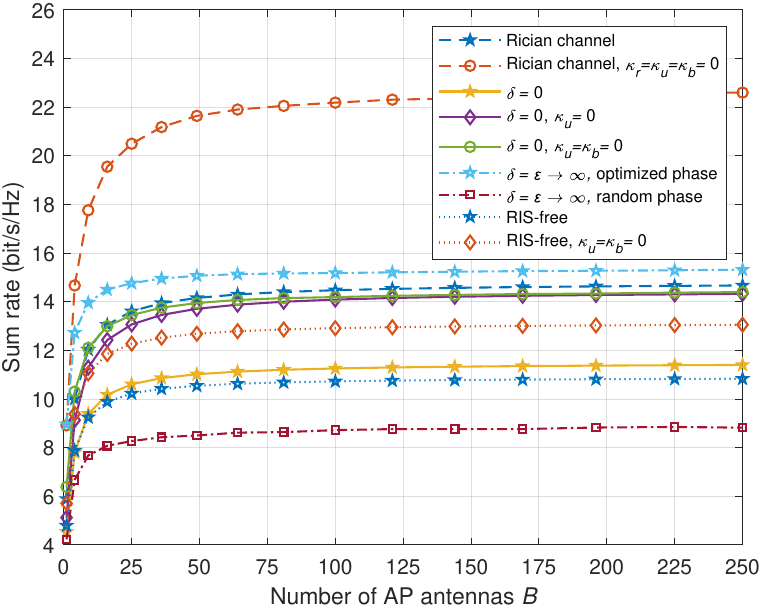}
	\DeclareGraphicsExtensions.
	\caption{Sum user rate versus the number of AP antennas $B$, with scaled transmit power $p_{k}=p/B, \forall k$, where $p=40$ dBm.}
	\label{figure_p_B}
\end{figure}
In Fig. \ref{figure_p_B}, we investigate the power scaling law as a function of $B$ in the RIS-free system or the RIS-assisted system. We consider the RIS-assisted link under the general Rician fading channel ($\varepsilon_{s,k}$ and $\delta_{l,s}$ are the default values in Table \ref{tab1}) and other fading channels of $\delta_{l,s}=\delta=0$, $\varepsilon_{s,k}=\varepsilon=0$, or $\delta=\varepsilon\to\infty$, $\forall l,s$. In agreement with our corollaries, it is observed that in all systems, the sum rate can maintain a non-zero value when the transmit power is scaled proportionally to $p/B$ as $B\to\infty$. Compared with RIS-free mMIMO systems, integrating RISs into the system effectively improves the performance limit when $B\to\infty$. Meanwhile, it can be found that the presence of the HWIs does not break the power scaling laws. Besides, in the NLoS-only RISs-APs channels ($\delta_{l,s}=\delta=0, \forall l,s$), the receiver HWI has almost no effect on system performance when $B$ is large. In the environment with LoS-only RIS-assisted channels ($\delta=\varepsilon\to\infty$), the imperfect RIS based on optimized phase shifts provides a considerable rate improvement for the CF-mMIMO system with transceiver HWIs. This observation emphasizes the importance of optimizing phase shifts even if RISs are imperfect.

\begin{figure}
	\setlength{\abovecaptionskip}{0pt}
	\setlength{\belowcaptionskip}{-20pt}
	\centering
	\includegraphics[width=3.8in]{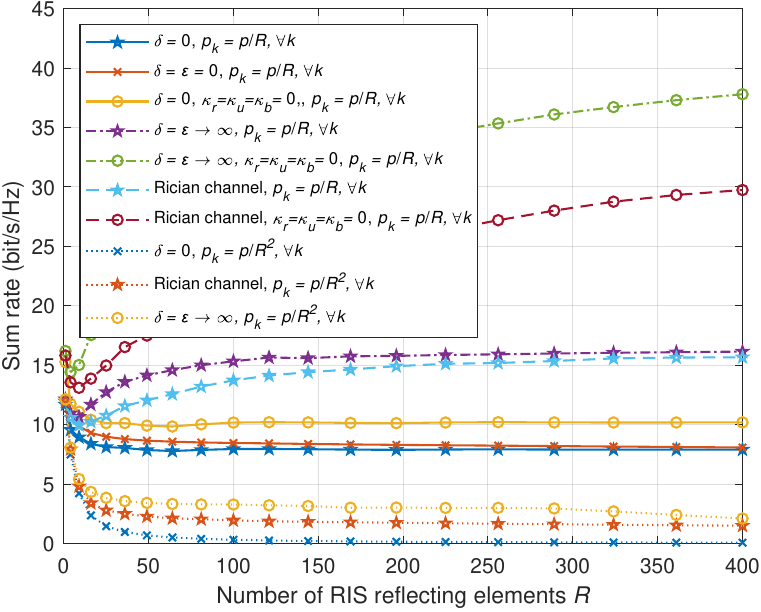}
	\DeclareGraphicsExtensions.
	\caption{Sum user rate versus the number of RIS reflecting elements $R$. The transmit power is scaled as $p_{k}=p/R$ or $p_{k}=p/R^{2}, \forall k$, where $p=40$ dBm.}
	\label{figure_p_R}
\end{figure}
Fig. \ref{figure_p_R} illustrates the power scaling law as a function of $R$ in RIS-assisted CF-mMIMO systems. As proved in Corollary \ref{corollary4}, if we scale the transmit power as $p_{k}=p/R, \forall k$, the rate can maintain a non-zero value when $R\to\infty$ for pure NLoS RISs-APs channels ($\delta_{l,s}=\delta=0, \forall l,s$). It's worth noting that when $\delta=0$ and $R\to\infty$, the rate will converge to the same value for the NLoS-only users-RISs channel ($\varepsilon_{s,k}=\varepsilon=0, \forall s,k$) and the Rician users-RISs channel ($\varepsilon$ is the default value $10$). Meanwhile, in pure LoS RIS-assisted channels ($\delta=\varepsilon\to\infty$) and Rician cascaded channels, when $R$ is small, the sum rate will decrease as $R$ increases, but with a large enough $R$, the rate will increase and approach a non-zero value as $R\to\infty$. This result corresponds to the general case of serving multi-user and maintaining system capacity, which agrees with our analysis in Corollary \ref{corollary6}. However, if the transmit power is scaled proportionally to $p/R^{2}$, the rate will reduce to a small value as $R$ increases, and this value will tend to zero when $R\to\infty$. Similarly, in the Rician cascaded and NLoS RISs-APs channels, scaling the power proportionally to $p/R^{2}$ reduces the rate to zero.  
	
\section{Conclusion}\label{section6}
In this paper, we investigated the rate performance of an uplink RIS-assisted CF-mMIMO system with transceiver HWIs and RIS phase noise under the two-timescale design. For the aggregated channel, we adopted the Rician fading model and the MRC detection. To begin with, we derived the closed-form analytical expression of the approximate rate in the general case. To draw valuable insights, we discussed some special cases, revealed the power scaling laws, and investigated the impacts of HWIs on the rate. 
These insights have provided clear guidelines for applying imperfect RISs to the CF-mMIMO system with transceiver HWIs.
To reduce the computational time and avoid falling into the local optimum, we transformed our formulas to get a tractable objective function, and optimized  the RIS phase shifts using the accelerated gradient ascent-based method to maximize users' sum rate and minimum rate, respectively.
Finally, numerical results verified the tightness of our derived expressions and revealed the maintainable balance between system capacity and user fairness with cell-free architectures. 
We have investigated the benefits of the imperfect RISs in the CF-mMIMO system with transceiver HWIs. 
The phase noise of RIS has a small impact on the rate, and the RIS elements are expected to provide the same rate increase as AP antennas at a lower cost.
Besides, we presented numerical results to verify the power scaling laws and the impacts of HWIs on rate performance, which also can provide some guidelines for the benefits of RIS-assisted CF-mMIMO with HWIs.

\begin{appendices}
	\section{}\label{appA}
	As known in Section \ref{section2}, by applying \cite[Lemma 1]{zhang2014ArRank}, the achievable rate expression of user $k$ can be approximated as
	\begin{align}\label{Rk}
		&R_{k} \approx \nonumber\\
		&\mathrm{log_{2}}\left(1+\frac{{p_k} \mathbb{E}\left\{ \left|{{\bf q}_k^{H}\widehat{\bf q}_k}\right|^{2} \right\} }{ \sum\limits_{i=1, i\neq k}^{K}{p_i}\mathbb{E}\left\{\left|{\bf q}_k^{H} \widehat{\bf q}_i\right|^{2}\right\}+\sum\limits_{i=1}^{K}\mathbb{E}\left\{\left|{\mathbf{q}}_{k}^{H}\widehat{\mathbf{q}}_{i}{\mathbb{\eta}}_{\it i,t}\right|^{2}\right\}+\mathbb{E}\left\{\left|{\mathbf{q}}_{k}^{H}{\bm{\eta}}_{\it r}\right|^{2}\right\}+\sigma^{2}\mathbb{E}\left\{\left\|{\bf q}_k\right\|^{2}\right\}}\right).
	\end{align}
	\subsection{Derivations of $\mathbb{E}\left\{\left\|{\bf q}_k\right\|^{2}\right\}$}\label{subsection1}
	To derive the closed-form expression in (\ref{Rk}), we first derive the the noise term $E_{k}^{(\mathrm{noise})}({\bf\Phi}) =\mathbb{E}\left\{\left\|{\bf q}_k\right\|^{2}\right\}$. 
	Based on the prior work in \cite{fullversion}, we can obtain the expression of the term $ \mathbb{E}\left\{\left\|\mathbf{q}_{k}\right\|^{2}\right\}$ given in \cite[Theorem 1]{fullversion}. Therefore, the noise term $\mathbb{E}\left\{\left\|{\bf q}_k\right\|^{2}\right\}$ can be given by
	\begin{align}\label{noise}
		&E_{k}^{(\mathrm{noise})}({\bf\Phi})=\mathbb{E}\left\{\left\|{\bf q}_k\right\|^{2}\right\}=\mathbb{E}\left\{\left\|\mathbf{g}_{k}+\mathbf{d}_{k}\right\|^{2}\right\}\nonumber\\
		&={\sum\limits_{l=1}^{L}\sum\limits_{s_{1}=1 }^S \sum\limits_{s_{2}=1 }^S} {\sqrt{c_{l,s_{1},k}c_{l,s_{2},k}\delta _{l,s_{1}}\delta _{l,s_{2}}\varepsilon _{s_{1},k}\varepsilon _{s_{2},k}} {f_{l,s_{1},k}^{H}({\bf\Phi})}{f_{l,s_{2},k}({\bf\Phi})}}{\mathbf{a}_{B}^{H}(l,s_{1})} {\mathbf{a}_{B}(l,s_{2})}\nonumber \\ 
		& + { \sum\limits_{l=1}^{L}\sum\limits_{s=1}^S c_{l,s,k} B R (\delta _{l,s}+\varepsilon _{s,k}+1)}+\sum\limits_{l=1}^{L}\gamma_{l,k} B.
	\end{align} 
	Note that $ \mathbb{E}\left\{\left\|\mathbf{g}_{k}\right\|^{2}\right\}=\sum\limits_{l=1}^L\mathbb{E}\left\{\left\|\mathbf{g}_{l,k}\right\|^{2}\right\}$, where
	\begin{align}\label{ggk}
		\begin{array}{l}
			\mathbb{E}\left\{\left\|{\bf g}_{l,k}\right\|^{2}\right\} =  {\sum\limits_{s_{1}=1 }^S \sum\limits_{s_{2}=1 }^S} {\sqrt{c_{l,s_{1},k}c_{l,s_{2},k}\delta _{l,s_{1}}\delta _{l,s_{2}}\varepsilon _{s_{1},k}\varepsilon _{s_{2},k}} {f_{l,s_{1},k}^{H}({\bf\Phi})}{f_{l,s_{2},k}({\bf\Phi})}} \\{\mathbf{a}_{B}^{H}(l,s_{1})} {\mathbf{a}_{B}(l,s_{2})} + 
			{ \sum\limits_{s=1}^S c_{l,s,k} B R (\delta _{l,s}+\varepsilon _{s,k}+1)}.
		\end{array}
	\end{align}
	
	Considering the independence of the phase noise variable $\widetilde{\bf\Phi}$, we can also substitute $\widehat{\bf\Phi}$ into ${\bf\Phi}$ in the result of (\ref{ggk}) to obtain the expression of the term $\mathbb{E}\left\{\left\|\widehat{\mathbf{g}}_{k}\right\|^{2}\right\} $.  Then, the term $\mathbb{E}\left\{\left\|\widehat{\mathbf{g}}_{k}\right\|^{2}\right\} $ can be given by 
	\begin{align}\label{g^g^k}
		\begin{array}{l}
			\mathbb{E}\left\{\left\|\widehat{\mathbf{g}}_{k}\right\|^{2}\right\}=\sum\limits_{l=1}^L\mathbb{E}\left\{\left\|\widehat{\mathbf{g}}_{l,k}\right\|^{2}\right\}\\=  {\sum\limits_{l=1}^L \sum\limits_{s_{1}=1 }^S \sum\limits_{s_{2}=1 }^S} {\sqrt{c_{l,s_{1},k}c_{l,s_{2},k}\delta _{l,s_{1}}\delta _{l,s_{2}}\varepsilon _{s_{1},k}\varepsilon _{s_{2},k}} \mathbb{E}\left\{{f_{l,s_{1},k}^{H}(\widehat{\bf\Phi})}{f_{l,s_{2},k}(\widehat{\bf\Phi})}\right\}} \\{\mathbf{a}_{B}^{H}(l,s_{1})} {\mathbf{a}_{B}(l,s_{2})} + 
			{\sum\limits_{l=1}^L \sum\limits_{s=1}^S c_{l,s,k} B R (\delta _{l,s}+\varepsilon _{s,k}+1)}.
		\end{array}
	\end{align}
	
	\subsection{Derivations of $\mathbb{E}\left\{\left|{{\bf q}_k^{H}\widehat{\bf q}_k}\right|^{2} \right\}$ and $\mathbb{E}\left\{\left|{{\bf q}_k^{H}\widehat{\bf q}_i}\right|^{2} \right\}$}\label{subsection2}
	In this subsection, we derive the terms ${E}_{k}^{(\mathrm{signal})}=\mathbb{E}\left\{\left|{{\bf q}_k^{H}\widehat{\bf q}_k}\right|^{2} \right\}$ and $I_{k i}({\bf\Phi})=\mathbb{E}\left\{\left|{{\bf q}_k^{H}\widehat{\bf q}_i}\right|^{2} \right\}$. To begin with, we can derive the preliminary expression by expanding and simplifying the mathematical expectation terms. Specially, note that $ {\mathbf{d}}_{k}$, ${\mathbf{d}}_{i}$ and ${\mathbf{g}}_{k}$ are independent of each other, $\forall i \neq k$, and $ {\mathbf{d}}_{k}$ is composed of i.i.d. random variables with zero mean.  Since $\widehat{\mathbf{g}}_{k}$ is independent of $ {\mathbf{d}}_{k}$, we can derive the signal term $\mathbb{E}\left\{\left|{{\bf q}_k^{H}\widehat{\bf q}_k}\right|^{2} \right\}$ as
	\begin{align}\label{E4}
		\begin{array}{l}
			\mathbb{E}\left\{\left|{{\bf q}_k^{H}\widehat{\bf q}_k}\right|^{2} \right\}=\mathbb{E}\left\{\left|({\bf g}_k+{\bf d}_k)^{H} (\widehat{\bf g}_k+{\bf d}_k)\right|^{2}\right\} \\
			=\mathbb{E}\left\{\left|\mathbf{g}^{H}_{k}\widehat{\mathbf{g}}_{k}+\mathbf{d}_{k}^{H} \widehat{\mathbf{g}}_{k}+\mathbf{g}_{k}^{H}\mathbf{d}_{k} +\left\|\mathbf{d}_{k}\right\|^{2}\right|^{2}\right\} \\
			=\mathbb{E}\left\{\left|\mathbf{g}^{H}_{k}\widehat{\mathbf{g}}_{k}\right|^{2}\right\}+\mathbb{E}\left\{\left|\mathbf{d}_{k}^{H} \widehat{\mathbf{g}}_{k}\right|^{2}\right\}+ \mathbb{E}\left\{\left|\mathbf{g}_{k}^{H}\mathbf{d}_{k}\right|^{2}\right\}
			+\mathbb{E}\left\{\left\|\mathbf{d}_{k}\right\|^{4}\right\}+2 \mathbb{E}\left\{\operatorname{Re}\left\{\mathbf{g}^{H}_{k}\widehat{\mathbf{g}}_{k}\right\}\left\|\mathbf{d}_{k}\right\|^{2}\right\}.
		\end{array}
	\end{align}
	Note that ${\left[\mathbf{d}_{k}\right]_{l,b}=\left[\mathbf{d}_{l,k}\right]_{b}}$, where ${{\bf d}_{l,k}} = \sqrt{\gamma_{l,k}} \tilde{\mathbf{d}}_{l,k} $ and $ \tilde{\mathbf{d}}_{l,k} \sim \mathcal{CN}\left({\bf 0},\mathbf{I}_{B}\right)$. Therefore, we have
	\begin{align}\label{E4_1}
		\begin{array}{l}
			\mathbb{E}\left\{\left\|\mathbf{d}_{k}\right\|^{4}\right\}\\=\mathbb{E}\left\{\left(\sum\limits_{l=1}^{L}\sum\limits_{b=1}^{B}\left|\left[\mathbf{d}_{k}\right]_{l,b}\right|^{2}\right)^{2}\right\} \\
			=\mathbb{E} \! \left\{\sum\limits_{l=1}^{L}\sum\limits_{b=1}^{B}\left|\left[\mathbf{d}_{k}\right]_{l,b}\right|^{4}\right\}  \! + \!  \mathbb{E} \!  \left\{	\mathop{\sum\limits_{l_1=1}^{L} \sum\limits_{b_1=1}^{B}{\sum\limits_{l_2=1}^{L}\sum\limits_{b_2=1}^{B}} }\limits_{(l_{2},b_{2})\neq(l_{1},b_{1})}
			\!  \!  \left|\left[\mathbf{d}_{k}\right]_{l_1,b_1}\right|^{2}\left|\left[\mathbf{d}_{k}\right]_{l_2,b_2}\right|^{2}\right\} \\
			=2\sum\limits_{l=1}^{L}  \gamma_{l,k}^{2}B+\left(\sum\limits_{l=1}^{L} \gamma_{l,k}B\right)^{2}-\sum\limits_{l=1}^{L}  \gamma_{l,k}^{2}B	=\sum\limits_{l=1}^{L}  \gamma_{l,k}^{2}B+\left(\sum\limits_{l=1}^{L} \gamma_{l,k}B\right)^{2}.
		\end{array}
	\end{align}
	We assume that the phase shift of each element is adjusted independently in each updated interval. Based on the symmetry of the odd function $\mathrm{sin}\left(\widetilde{\theta}_{s,r}\right)$ with $\widetilde{\theta}_{s,r}$ and the probability density function of $\widetilde{\theta}_{s,r}$, we have $\mathbb{E}\left\{e^{\pm j\widetilde{\theta}_{s,r}}\right\}=\mathbb{E}\left\{\mathrm{cos}\left(\widetilde{\theta}_{s,r}\right)\right\}=\mathrm{sinc}\left(\kappa_r\pi\right)$, $\forall s, r$ \cite{9295369}. By exploiting the  identity $\mathbb{E}\big\{\widetilde{\bf\Phi}\big\}=\mathbb{E}\big\{\widetilde{\bf\Phi}^{H}\big\}=\mathrm{sinc}\left(\kappa_r\pi\right)\mathbf{I}_{SR}$ and the independence of $\widetilde{\bf\Phi}$, we have
	\begin{align}\label{E4_2}
		\begin{array}{l}
			\mathbb{E}\left\{\operatorname{Re}\left\{\mathbf{g}^{H}_{k}\widehat{\mathbf{g}}_{k}\right\}\left\|\mathbf{d}_{k}\right\|^{2}\right\}=\mathbb{E}\left\{\operatorname{Re}\left\{\mathbf{g}^{H}_{k}\widehat{\mathbf{g}}_{k}\right\}\right\}\mathbb{E}\left\{\left\|\mathbf{d}_{k}\right\|^{2}\right\}
			\\=\mathbb{E}\left\{\operatorname{Re}\left\{\mathbf{h}_{k}^{H}{\mathbf\Phi }^{H}\mathbf{Z}^{H}\mathbf{Z}  {\mathbf\Phi }\mathbb{E}\left\{\widetilde{\mathbf\Phi }\right\}\mathbf{h}_{k}\right\}\right\}\mathbb{E}\left\{\left\|\mathbf{d}_{k}\right\|^{2}\right\}
			\\=   \mathbb{E}\left\{\left\|\mathbf{g}_{k}\right\|^{2}\right\}\mathbb{E}\left\{\left\|\mathbf{d}_{k}\right\|^{2}\right\}\mathrm{sinc}\left(\kappa_r\pi\right)  =  \sum\limits_{l=1}^{L}\gamma_{l,k} B \mathbb{E}\left\{\left\|\mathbf{g}_{k}\right\|^{2}\right\}\mathrm{sinc}\left(\kappa_r\pi\right).
		\end{array}
	\end{align}
	Also, the terms $ \mathbb{E}\left\{\left|\mathbf{d}_{k}^{H} \widehat{\mathbf{g}}_{k}\right|^{2}\right\}$ and $ \mathbb{E}\left\{\left|\mathbf{g}_{k}^{H}\mathbf{d}_{k}\right|^{2}\right\} $ can be derived as
	\begin{align}
		&\mathbb{E}\left\{\left|\mathbf{d}_{k}^{H} \widehat{\mathbf{g}}_{k}\right|^{2}\right\}  =\mathbb{E}\left\{\widehat{\mathbf{g}}_{k}^{H} \mathbb{E}\left\{\mathbf{d}_{k} \mathbf{d}_{k}^{H}\right\} \widehat{\mathbf{g}}_{k}\right\} 
		=\sum\limits_{l=1}^{L}\gamma_{l,k} \mathbb{E}\left\{\left\|\widehat{\mathbf{g}}_{l,k}\right\|^{2}\right\}, \label{E4_3}
		\\&\mathbb{E}\left\{\left|\mathbf{g}_{k}^{H}\mathbf{d}_{k}\right|^{2}\right\} =\mathbb{E}\left\{\mathbf{g}_{k}^{H} \mathbb{E}\left\{\mathbf{d}_{k} \mathbf{d}_{k}^{H}\right\} \mathbf{g}_{k}\right\} 
		=\sum\limits_{l=1}^{L}\gamma_{l,k} \mathbb{E}\left\{\left\|\mathbf{g}_{l,k}\right\|^{2}\right\}. \label{E4_4}
	\end{align}
	Substituting (\ref{E4_1}) $\sim$ (\ref{E4_4}) into (\ref{E4}), we get the expression of the signal term as follows
	\begin{align}\label{signal_term}
		\begin{array}{l}
			\mathbb{E}\left\{\left|{{\bf q}_k^{H}\widehat{\bf q}_k}\right|^{2} \right\}\\
			=\mathbb{E}\left\{\left|\mathbf{g}^{H}_{k}\widehat{\mathbf{g}}_{k}\right|^{2}\right\}+\sum\limits_{l=1}^{L}\gamma_{l,k}\left( \mathbb{E}\left\{\left\|\mathbf{g}_{l,k}\right\|^{2}\right\}+\mathbb{E}\left\{\left\|\widehat{\mathbf{g}}_{l,k}\right\|^{2}\right\}+2B \mathbb{E}\left\{\left\|\mathbf{g}_{k}\right\|^{2}\right\}\mathrm{sinc}\left(\kappa_r\pi\right)\right)
			\\+\sum\limits_{l=1}^{L}  \gamma_{l,k}^{2}B+\left(\sum\limits_{l=1}^{L} \gamma_{l,k}B\right)^{2}.
		\end{array}
	\end{align}
	
	Next, the interference term $\mathbb{E}\left\{\left|{\bf q}_k^{H} \widehat{\bf q}_i\right|^{2}\right\} $ can be expanded as
	\begin{align}\label{interference_expend}
		\begin{array}{l}
			\mathbb{E}\left\{\left|{\bf q}_k^{H} \widehat{\bf q}_i\right|^{2}\right\}=\mathbb{E}\left\{\left|\left(\mathbf{g}_{k}^{H}+\mathbf{d}_{k}^{H}\right)\left(\widehat{\mathbf{g}}_{i}+\mathbf{d}_{i}\right)\right|^{2}\right\} \\
			=\mathbb{E}\left\{\left|\mathbf{g}_{k}^{H} \widehat{\mathbf{g}}_{i}+\mathbf{g}_{k}^{H} \mathbf{d}_{i}+\mathbf{d}_{k}^{H} \widehat{\mathbf{g}}_{i}+\mathbf{d}_{k}^{H} \mathbf{d}_{i}\right|^{2}\right\} \\
			=\mathbb{E}\! \left\{\left|\mathbf{g}_{k}^{H} \widehat{\mathbf{g}}_{i}\right|^{2}\! \right\}   \!+\!   \mathbb{E}\! \left\{\left|\mathbf{g}_{k}^{H} \mathbf{d}_{i}\right|^{2}\! \right\}
			\!+\!   \mathbb{E}\! \left\{\left|\mathbf{d}_{k}^{H} \widehat{\mathbf{g}}_{i}\right|^{2}\! \right\}    \!+\!   \mathbb{E}\! \left\{\left|\mathbf{d}_{k}^{H} \mathbf{d}_{i}\right|^{2}\! \right\},
		\end{array}
	\end{align}
	where
	\begin{align}
		&\mathbb{E}\left\{\left|\mathbf{g}_{k}^{H} \mathbf{d}_{i}\right|^{2}\right\}
		=\mathbb{E}\left\{\mathbf{g}_{k}^{H} \mathbb{E}\left\{\mathbf{d}_{i} \mathbf{d}_{i}^{H}\right\} \mathbf{g}_{k}\right\} 
		=\sum\limits_{l=1}^{L}\gamma_{l,i} \mathbb{E}\left\{\left\|\mathbf{g}_{l,k}\right\|^{2}\right\},\label{interference_term_re1}\\
		&\mathbb{E}\left\{\left|\mathbf{d}_{k}^{H} \widehat{\mathbf{g}}_{i}\right|^{2}\right\} 
		=\mathbb{E}\left\{\widehat{\mathbf{g}}_{i}^{H} \mathbb{E}\left\{\mathbf{d}_{k} \mathbf{d}_{k}^{H}\right\}\widehat{\mathbf{g}}_{i}\right\} 
		=\sum\limits_{l=1}^{L}\gamma_{l,k} \mathbb{E}\left\{\left\|\widehat{\mathbf{g}}_{l,i}\right\|^{2}\right\},\label{interference_term_re2}\\
		&\mathbb{E}\left\{\left|\mathbf{d}_{k}^{H} \mathbf{d}_{i}\right|^{2}\right\} 
		=\mathbb{E}\left\{\mathbf{d}_{k}^{H} \mathbb{E}\left\{\mathbf{d}_{i} \mathbf{d}_{i}^{H}\right\} \mathbf{d}_{k}\right\} 
		=\sum\limits_{l=1}^{L} \gamma_{l,i} \gamma_{l,k} B.\label{interference_term_re3}
	\end{align}
	Substituting (\ref{interference_term_re1}) $\sim$ (\ref{interference_term_re3}) into (\ref{interference_expend}), we can derive the interference term as follows
	\begin{align}\label{interference_term}
		\mathbb{E}\left\{\left|{\bf q}_k^{H} \widehat{\bf q}_i\right|^{2}\right\}=\mathbb{E} \left\{\left|\mathbf{g}_{k}^{H} \widehat{\mathbf{g}}_{i}\right|^{2} \right\}   +\sum\limits_{l=1}^{L}\left(\gamma_{l,i}\left(\gamma_{l,k}B+\mathbb{E}\left\{\left\|\mathbf{g}_{l,k}\right\|^{2}\right\}\right)+\gamma_{l,k}\mathbb{E}\left\{\left\|\widehat{\mathbf{g}}_{l,i}\right\|^{2}\right\}\right).
	\end{align}
	
	Therefore, combining (\ref{signal_term}) and (\ref{interference_term}), we can gain the preliminary expressions of the signal and interference terms with $ \mathbb{E}\left\{\left\|\mathbf{g}_{l,k}\right\|^{2}\right\} $, $\mathbb{E}\left\{\left\|\widehat{\mathbf{g}}_{l,k}\right\|^{2}\right\} $, $\mathbb{E}\left\{\left|\mathbf{g}_{k}^{H} \widehat{\mathbf{g}}_{k}\right|^{2}\right\}$, and $\mathbb{E}\left\{\left|\mathbf{g}_{k}^{H} \widehat{\mathbf{g}}_{i}\right|^{2}\right\}$, where $ \mathbb{E}\left\{\left\|\mathbf{g}_{l,k}\right\|^{2}\right\} $ and $\mathbb{E}\left\{\left\|\widehat{\mathbf{g}}_{l,k}\right\|^{2}\right\} $ have been given in (\ref{ggk}) and (\ref{g^g^k}).
	When the RIS is the ideal hardware and the number of the quantization bits is infinite, i.e., $\widehat{\bf\Phi}$ becomes ${\bf\Phi}$ and $\widehat{\bf g}_k^{H}$ becomes ${\bf g}_k^{H}$, the expressions of the signal term in (\ref{signal_term}) and the interference term in (\ref{interference_term}) equal to the results in \cite[Theorem 1]{fullversion}. 
	Using the same method as \cite[Appendix A]{fullversion}, we can treat $\widehat{\bf\Phi}$ as ${\bf\Phi}$ in derivations and obtain the results involving $\widehat{\bf\Phi}$ for the terms $\mathbb{E}\left\{\left|\mathbf{g}_{k}^{H} \widehat{\mathbf{g}}_{k}\right|^{2}\right\}$ and $\mathbb{E}\left\{\left|\mathbf{g}_{k}^{H} \widehat{\mathbf{g}}_{i}\right|^{2}\right\}$. 
	% For the terms $\mathbb{E}\left\{\left|\mathbf{g}_{k}^{H} \widehat{\mathbf{g}}_{k}\right|^{2}\right\}$ and $\mathbb{E}\left\{\left|\mathbf{g}_{k}^{H} \widehat{\mathbf{g}}_{i}\right|^{2}\right\}$, given the independence of $\widetilde{\bf\Phi}$, we can treat $\widehat{\bf\Phi}$ as ${\bf\Phi}$ in derivations and obtain their results involving $\widehat{\bf\Phi}$ according to the derivations in \cite[Appendix A]{fullversion}.
	Therefore, the expectation terms with $\widehat{\bf\Phi}$ can be given by
	\begin{align}
		\begin{array}{l}\label{signal exp}
			\mathbb{E}\left\{\left|{\bf q}_k^{H} \widehat{\bf q}_k\right|^{2}\right\}  %等价式1
			{= \sum\limits_{l_{1}=1}^{L} \sum\limits_{l_{2}=1}^{L} \sum\limits_{s_{1}=1}^{S} \sum\limits_{s_{2}=1}^{S} \sum\limits_{s_{3}=1}^{S} \sum\limits_{s_{4}=1}^{S} \sqrt{c_{l_{1},s_{1},k}c_{l_{1},s_{2},k}c_{l_{2},s_{3},k}c_{l_{2},s_{4},k} \delta_{l_{1},s_{1}}\delta_{l_{1},s_{2}}
					\delta_{l_{2},s_{3}}\delta_{l_{2},s_{4}}\varepsilon_{s_{1},k}\varepsilon_{s_{2},k}\varepsilon_{s_{3},k}\varepsilon_{s_{4},k}}}\\ {f_{l_{1},s_{1},k}^{H}({\bf\Phi})\mathbb{E}\left\{f_{l_{1},s_{2},k}(\widehat{\bf\Phi})f_{l_{2},s_{3},k}^{H}(\widehat{\bf\Phi})\right\}f_{l_{2},s_{4},k}({\bf\Phi})} {\mathbf{a}_{B}^{H}(l_{1},s_{1})}{\mathbf{a}_{B}(l_{1},s_{2})}{\mathbf{a}_{B}^{H}(l_{2},s_{3})}{\mathbf{a}_{B}(l_{2},s_{4})} \\ +  %式子2
			{\sum\limits_{l_{1}=1}^{L} \sum\limits_{l_{2}=1}^{L} \sum\limits_{s_{1}=1}^{S} \sum\limits_{s_{2}=1}^{S} \sum\limits_{s_{3}=1}^{S}} \Big(2{\sqrt{c_{l_{1},s_{1},k}c_{l_{1},s_{2},k} \delta_{l_{1},s_{1}}\delta_{l_{1},s_{2}} \varepsilon_{s_{1},k}\varepsilon_{s_{2},k}}} c_{l_{2},s_{3},k}B\\ (\delta_{l_{2},s_{3}}+\varepsilon_{s_{3},k}+1) {\mathrm{Re}\left\{f_{l_{1},s_{1},k}^{H}({\bf\Phi}) \mathbb{E}\left\{f_{l_{1},s_{2},k}(\widehat{\bf\Phi})\mathrm{Tr}\left\{\widetilde{\bf\Phi}^{H}_{s_{3}}\right\}\right\} {\mathbf{a}_{B}^{H}(l_{1},s_{1})} {\mathbf{a}_{B}(l_{1},s_{2})} \right\}} \\+  %式子3
			\sqrt{c_{l_{1},s_{1},k}c_{l_{1},s_{2},k}c_{l_{2},s_{3},k}c_{l_{2},s_{1},k} \delta_{l_{1},s_{1}}\delta_{l_{1},s_{2}}
				\delta_{l_{2},s_{3}}\delta_{l_{2},s_{1}} \varepsilon_{s_{2},k}\varepsilon_{s_{3},k}}  \Big(\mathbb{E}\left\{f_{l_{2},s_{3},k}^{H}(\widehat{\bf\Phi})f_{l_{1},s_{2},k}(\widehat{\bf\Phi})\right\} \\ +f_{l_{2},s_{3},k}^{H}({\bf\Phi})f_{l_{1},s_{2},k}({\bf\Phi})\Big) {\mathbf{a}_{B}^{H}(l_{2},s_{3})} {\mathbf{a}_{B}(l_{2},s_{1})}
			{\mathbf{a}_{R}^{H}(l_{2},s_{1})} {\mathbf{a}_{R}(l_{1},s_{1})} {\mathbf{a}_{B}^{H}(l_{1},s_{1})} {\mathbf{a}_{B}(l_{1},s_{2})}\Big) \\+ %式子4
			{\sum\limits_{l_{1}=1}^{L} \sum\limits_{l_{2}=1}^{L} \sum\limits_{s_{1}=1}^{S} \sum\limits_{s_{2}=1}^{S} \Big(2\sqrt{c_{l_{1},s_{1},k}c_{l_{1},s_{2},k}\delta_{l_{1},s_{1}}\delta_{l_{1},s_{2}} \varepsilon_{s_{1},k}\varepsilon_{s_{2},k}}B  \Big(c_{l_{2},s_{2},k} \mathrm{Re}\Big\{\Big(\mathbb{E}\Big\{f_{l_{1},s_{1},k}^{H}(\widehat{\bf\Phi})}\\f_{l_{1},s_{2},k}(\widehat{\bf\Phi})\Big\}+f_{l_{1},s_{1},k}^{H}({\bf\Phi}) f_{l_{1},s_{2},k}({\bf\Phi})\Big) {\mathbf{a}_{B}^{H}(l_{1},s_{1})} {\mathbf{a}_{B}(l_{1},s_{2})}\Big\}
			+ \gamma_{l_{2},k}\mathrm{sinc}\left(\kappa_r\pi\right) f_{l_{1},s_{1},k}^{H}({\bf\Phi})\\ f_{l_{1},s_{2},k}({\bf\Phi})
			{\mathbf{a}_{B}^{H}(l_{1},s_{1})} {\mathbf{a}_{B}(l_{1},s_{2})}\Big)
			+ %式子5
			{\sqrt{c_{l_{1},s_{1},k}c_{l_{1},s_{2},k} c_{l_{2},s_{2},k}c_{l_{2},s_{1},k}\delta_{l_{1},s_{1}} \delta_{l_{1},s_{2}}\delta_{l_{2},s_{1}} \delta_{l_{2},s_{2}}}}\\ {\mathbf{a}_{B}^{H}(l_{1},s_{1})} {\mathbf{a}_{B}(l_{1},s_{2})}  {{\mathbf{a}_{R}^{H}(l_{1},s_{2})}
				{\mathbf{a}_{R}(l_{2},s_{2})} {\mathbf{a}_{B}^{H}(l_{2},s_{2})} {\mathbf{a}_{B}(l_{2},s_{1})}  {\mathbf{a}_{R}^{H}(l_{2},s_{1})} {\mathbf{a}_{R}(l_{1},s_{1})}} \\ + %式子6 
			{c_{l_{1},s_{1},k}c_{l_{2},s_{2},k}}B^{2}\mathbb{E}\left\{\mathrm{Tr}\left\{\widetilde{\bf\Phi}_{s_{1}}\right\}\mathrm{Tr}\left\{\widetilde{\bf\Phi}^{H}_{s_{2}}\right\}\right\}\big(2(\delta_{l_{1},s_{1}}\varepsilon_{s_{2},k}+ \delta_{l_{1},s_{1}}+\varepsilon_{s_{1},k})+ \delta_{l_{1},s_{1}}\delta_{l_{2},s_{2}}+
			\varepsilon_{s_{1},k}\varepsilon_{s_{2},k}+1\big)\Big) 
			\\ + %式子7
			{\sum\limits_{l=1}^{L} \sum\limits_{s_{1}=1}^{S} \sum\limits_{s_{2}=1}^{S} \sum\limits_{s_{3}=1}^{S}}
			\sqrt{c_{l,s_{1},k}c_{l,s_{3},k} \delta_{l,s_{1}}\delta_{l,s_{3}} \varepsilon_{s_{1},k}\varepsilon_{s_{3},k}}  c_{l,s_{2},k}R(\varepsilon_{s_{2},k}+1)\\ \left(\mathbb{E}\left\{f_{l,s_{1},k}^{H}(\widehat{\bf\Phi}) f_{l,s_{3},k}(\widehat{\bf\Phi})\right\}+f_{l,s_{1},k}^{H}({\bf\Phi}) f_{l,s_{3},k}({\bf\Phi})\right) {\mathbf{a}_{B}^{H}(l,s_{1})} {\mathbf{a}_{B}(l,s_{3})}
			\\ + %式子8
			{\sum\limits_{l_{1}=1}^{L} \sum\limits_{l_{2}=1}^{L} \sum\limits_{s=1}^{S}}   c_{l_{1},s,k}B^{2}R\big(c_{l_{2},s,k}(2\delta_{l_{1},s}+2\varepsilon_{s,k}+1)+2\gamma_{l_{2},k}(\delta_{l_{1},s}+\varepsilon_{s,k}+1)\mathrm{sinc}\left(\kappa_r\pi\right)\big)
			\\ +  %式子9
			{\sum\limits_{l=1}^{L} \sum\limits_{s_{1}=1}^{S} \sum\limits_{s_{2}=1}^{S}}\Big( c_{l,s_{1},k} c_{l,s_{2},k}B R^{2}\big( 2(\delta_{l,s_{1}}\varepsilon_{s_{2},k}+ \delta_{l,s_{1}}+\varepsilon_{s_{1},k})+ \varepsilon_{s_{1},k}\varepsilon_{s_{2},k}+1\big)
			\\ + %式子10
			\sqrt{c_{l,s_{1},k}c_{l,s_{2},k}\delta_{l,s_{1}}\delta_{l,s_{2}}\varepsilon_{s_{1},k}\varepsilon _{s_{2},k}} \gamma_{l,k} \left(\mathbb{E}\left\{{f_{l,s_{1},k}^{H}(\widehat{\bf\Phi})}{f_{l,s_{2},k}(\widehat{\bf\Phi})}\right\}+{f_{l,s_{1},k}^{H}({\bf\Phi})}{f_{l,s_{2},k}({\bf\Phi})}\right) \\{\mathbf{a}_{B}^{H}(l,s_{1})} {\mathbf{a}_{B}(l,s_{2})}
			+ %式子11
			2\sqrt{c_{l,s_{1},k}c_{l,s_{2},k} \delta_{l,s_{1}}\delta_{l,s_{2}} \varepsilon_{s_{1},k}\varepsilon_{s_{2},k}} c_{l,s_{2},k} \mathrm{Re}\Big\{
			\Big(\mathbb{E}\left\{f_{l,s_{1},k}^{H}(\widehat{\bf\Phi}) f_{l,s_{2},k}(\widehat{\bf\Phi})\right\}\\+f_{l,s_{1},k}^{H}({\bf\Phi})f_{l,s_{2},k}({\bf\Phi})\Big) {\mathbf{a}_{B}^{H}(l,s_{1})} {\mathbf{a}_{B}(l,s_{2})}\Big\}\Big) 
			\\ + %式子12
			{\sum\limits_{l=1}^{L} \sum\limits_{s=1}^{S}}  c_{l,s,k}BR\big( 2(\delta_{l,s}+\varepsilon_{s,k})(c_{l,s,k}+\gamma_{l,k})+c_{l,s,k}+2\gamma_{l,k}\big)
			+ %式子13
			\left(\sum\limits_{l=1}^{L}\gamma_{l,k}B\right)^{2}+ \sum\limits_{l=1}^{L} \gamma_{l,k}^{2}B,
		\end{array}
	\end{align}
	\begin{align}\label{interference exp}
		\begin{array}{l}
			\mathbb{E}\left\{\left|{\bf q}_k^{H} \widehat{\bf q}_i\right|^{2}\right\}
			\\%等价式1
			{= \sum\limits_{l_{1}=1}^{L} \sum\limits_{l_{2}=1}^{L} \sum\limits_{s_{1}=1}^{S} \sum\limits_{s_{2}=1}^{S} \sum\limits_{s_{3}=1}^{S} \sum\limits_{s_{4}=1}^{S} \sqrt{c_{l_{1},s_{1},k}c_{l_{1},s_{2},i}c_{l_{2},s_{3},i}c_{l_{2},s_{4},k} \delta_{l_{1},s_{1}}\delta_{l_{1},s_{2}}
					\delta_{l_{2},s_{3}}\delta_{l_{2},s_{4}}}}\\\sqrt{\varepsilon_{s_{1},k}\varepsilon_{s_{2},i}\varepsilon_{s_{3},i}\varepsilon_{s_{4},k}}  f_{l_{1},s_{1},k}^{H}({\bf\Phi})\mathbb{E}\left\{f_{l_{1},s_{2},i}(\widehat{\bf\Phi})f_{l_{2},s_{3},i}^{H}(\widehat{\bf\Phi})\right\}f_{l_{2},s_{4},k}({\bf\Phi}) {\mathbf{a}_{B}^{H}(l_{1},s_{1})}{\mathbf{a}_{B}(l_{1},s_{2})}\\{\mathbf{a}_{B}^{H}(l_{2},s_{3})}{\mathbf{a}_{B}(l_{2},s_{4})} \\ +  %式子2
			{\sum\limits_{l_{1}=1}^{L} \sum\limits_{l_{2}=1}^{L} \sum\limits_{s_{1}=1}^{S} \sum\limits_{s_{2}=1}^{S} \sum\limits_{s_{3}=1}^{S}} \Big(2{\sqrt{c_{l_{1},s_{1},k}c_{l_{1},s_{2},i}c_{l_{2},s_{3},i}c_{l_{2},s_{3},k} \delta_{l_{1},s_{1}}\delta_{l_{1},s_{2}} \varepsilon_{s_{1},k}\varepsilon_{s_{2},i} \varepsilon_{s_{3},i}\varepsilon_{s_{3},k}}}B\\ {\mathrm{Re}\left\{f_{l_{1},s_{1},k}^{H}({\bf\Phi}){\mathbf{a}_{B}^{H}(l_{1},s_{1})} {\mathbf{a}_{B}(l_{1},s_{2})} \mathbb{E}\left\{f_{l_{1},s_{2},i}(\widehat{\bf\Phi}) \overline{\mathbf{h}}_{s_{3},i}^{H}\widetilde{\bf\Phi}_{s_{3}}^{H}\right\} \overline{\mathbf{h}}_{s_{3},k} \right\}} \\ %式子3
			+\sqrt{\delta_{l_{1},s_{1}}\delta_{l_{1},s_{2}}
				\delta_{l_{2},s_{2}}\delta_{l_{2},s_{3}}} \Big(\sqrt{c_{l_{1},s_{1},k}c_{l_{1},s_{2},i}c_{l_{2},s_{2},i}c_{l_{2},s_{3},k} \varepsilon_{s_{1},k}\varepsilon_{s_{3},k}} f_{l_{1},s_{1},k}^{H}({\bf\Phi}) f_{l_{2},s_{3},k}({\bf\Phi})\\ %换行
			+\sqrt{c_{l_{1},s_{1},i}c_{l_{1},s_{2},k}c_{l_{2},s_{2},k}c_{l_{2},s_{3},i} \varepsilon_{s_{1},i}\varepsilon_{s_{3},i}} \mathbb{E}\left\{f_{l_{1},s_{1},i}^{H}(\widehat{\bf\Phi}) f_{l_{2},s_{3},i}(\widehat{\bf\Phi})\right\}\Big) {\mathbf{a}_{B}^{H}(l_{1},s_{1})} {\mathbf{a}_{B}(l_{1},s_{2})}\\ %换行
			{\mathbf{a}_{R}^{H}(l_{1},s_{2})} {\mathbf{a}_{R}(l_{2},s_{2})} {\mathbf{a}_{B}^{H}(l_{2},s_{2})} {\mathbf{a}_{B}(l_{2},s_{3})}
			\Big) \\+ %式子4
			\sum\limits_{l_{1}=1}^{L} \sum\limits_{l_{2}=1}^{L} \sum\limits_{s_{1}=1}^{S} \sum\limits_{s_{2}=1}^{S} \Big(2\sqrt{c_{l_{1},s_{1},k}c_{l_{1},s_{2},i}\delta_{l_{1},s_{1}}\delta_{l_{1},s_{2}}}B\mathrm{Re}\Big\{\Big(\sqrt{c_{l_{2},s_{2},i}c_{l_{2},s_{2},k}\varepsilon_{s_{1},k}\varepsilon_{s_{2},k}}   \\f_{l_{1},s_{1},k}^{H}({\bf\Phi}) f_{l_{1},s_{2},k}({\bf\Phi}) + \sqrt{c_{l_{2},s_{1},i}c_{l_{2},s_{1},k} \varepsilon_{s_{1},i}\varepsilon_{s_{2},i}} \mathbb{E}\left\{f_{l_{1},s_{1},i}^{H}(\widehat{\bf\Phi}) f_{l_{1},s_{2},i}(\widehat{\bf\Phi})\right\}\Big) {\mathbf{a}_{B}^{H}(l_{1},s_{1})} {\mathbf{a}_{B}(l_{1},s_{2})}\Big\}  \\+ %式子5
			{\sqrt{c_{l_{1},s_{1},k}c_{l_{1},s_{2},i} c_{l_{2},s_{2},i}c_{l_{2},s_{1},k}\delta_{l_{1},s_{1}} \delta_{l_{1},s_{2}}\delta_{l_{2},s_{1}} \delta_{l_{2},s_{2}}} {\mathbf{a}_{B}^{H}(l_{1},s_{1})} {\mathbf{a}_{B}(l_{1},s_{2})}} \\ {{\mathbf{a}_{R}^{H}(l_{1},s_{2})}
				{\mathbf{a}_{R}(l_{2},s_{2})} {\mathbf{a}_{B}^{H}(l_{2},s_{2})} {\mathbf{a}_{B}(l_{2},s_{1})}  {\mathbf{a}_{R}^{H}(l_{2},s_{1})} {\mathbf{a}_{R}(l_{1},s_{1})}} \\ + %式子6 
			\sqrt{c_{l_{1},s_{1},k}c_{l_{1},s_{2},i} c_{l_{2},s_{2},i}c_{l_{2},s_{1},k} \varepsilon_{s_{1},k}\varepsilon_{s_{1},i}
				\varepsilon_{s_{2},i}\varepsilon_{s_{2},k}}B^{2}\overline{\mathbf{h}}_{s_{1},k}^{H}
			\mathbb{E}\left\{\widetilde{\bf\Phi}_{s_{1}} \overline{\mathbf{h}}_{s_{1},i} \overline{\mathbf{h}}_{s_{2},i}^{H}\widetilde{\bf\Phi}^{H}_{s_{2}}\right\} \overline{\mathbf{h}}_{s_{2},k}\Big) \\ + %式子7
			{\sum\limits_{l=1}^{L} \sum\limits_{s_{1}=1}^{S} \sum\limits_{s_{2}=1}^{S} \sum\limits_{s_{3}=1}^{S}}
			\Big(\sqrt{c_{l,s_{1},k}c_{l,s_{3},k} \delta_{l,s_{1}}\delta_{l,s_{3}} \varepsilon_{s_{1},k}\varepsilon_{s_{3},k}}  c_{l,s_{2},i}R(\varepsilon_{s_{2},i}+1) f_{l,s_{1},k}^{H}({\bf\Phi}) f_{l,s_{3},k}({\bf\Phi})\\ {\mathbf{a}_{B}^{H}(l,s_{1})} {\mathbf{a}_{B}(l,s_{3})} + %式子8
			\sqrt{c_{l,s_{2},i}c_{l,s_{3},i} \delta_{l,s_{2}}\delta_{l,s_{3}} \varepsilon_{s_{2},i}\varepsilon_{s_{3},i}}  c_{l,s_{1},k}R(\varepsilon_{s_{1},k}+1) \mathbb{E}\left\{f_{l,s_{3},i}^{H}(\widehat{\bf\Phi}) f_{l,s_{2},i}(\widehat{\bf\Phi})\right\}\\ {\mathbf{a}_{B}^{H}(l,s_{3})} {\mathbf{a}_{B}(l,s_{2})}\Big)
			\\+
			{\sum\limits_{l_{1}=1}^{L} \sum\limits_{l_{2}=1}^{L} \sum\limits_{s=1}^{S}} \sqrt{c_{l_{1},s,k}c_{l_{1},s,i} c_{l_{2},s,i}c_{l_{2},s,k}} B^{2} R (2\delta_{l_{1},s}+\varepsilon_{s,k}+\varepsilon_{s,i}+1) \\ +  %式子10
			{\sum\limits_{l=1}^{L} \sum\limits_{s_{1}=1}^{S} \sum\limits_{s_{2}=1}^{S}} \Big( 
			\sqrt{\delta_{l,s_{1}}\delta_{l,s_{2}}} \Big(\sqrt{c_{l,s_{1},k}c_{l,s_{2},k}\varepsilon _{s_{1},k}\varepsilon _{s_{2},k}} \gamma_{l,i} {f_{l,s_{1},k}^{H}({\bf\Phi})}{f_{l,s_{2},k}({\bf\Phi})}
			\\+ \sqrt{c_{l,s_{1},i}c_{l,s_{2},i}\varepsilon _{s_{1},i}\varepsilon _{s_{2},i}} \gamma_{l,k} \mathbb{E}\left\{{f_{l,s_{1},i}^{H}(\widehat{\bf\Phi})}{f_{l,s_{2},i}(\widehat{\bf\Phi})}\right\}\Big) {\mathbf{a}_{B}^{H}(l,s_{1})} {\mathbf{a}_{B}(l,s_{2})}
			\\+
			c_{l,s_{1},k} c_{l,s_{2},i} B R^{2}\big( \delta_{l,s_{2}}\left(\varepsilon_{s_{1},k}+1\right)+ \left(\delta_{l,s_{1}}+\varepsilon_{s_{1},k}+1\right)\left(\varepsilon_{s_{2},i}+1\right)\big)\Big)
			\\+ %式子12
			{\sum\limits_{l=1}^L \sum\limits_{s=1}^S} 
			B R \big(c_{l,s,k} (\delta _{l,s}+\varepsilon _{s,k}+1)\gamma_{l,i}+c_{l,s,i} (\delta _{l,s}+\varepsilon _{s,i}+1)\gamma_{l,k}\big)
			\\+ %式子13
			{\sum\limits_{l=1}^L \gamma_{l,k}\gamma_{l,i}B}.
		\end{array}
	\end{align}
	
	Noting that $\mathbb{E}\big\{\widetilde{\bf\Phi}_{s}\big\}=\mathbb{E}\big\{\widetilde{\bf\Phi}_{s}^{H}\big\}=\mathrm{sinc}\left(\kappa_r\pi\right)\mathbf{I}_{R}$ and using the identity $\mathbb{E}\big\{\widetilde{\bf\Phi}_{s}{\bf W}\widetilde{\bf\Phi}_{s}^{H}\big\}=\mathrm{sinc}^{2}\left(\kappa_r\pi\right){\bf W}+\left(1-\mathrm{sinc}^{2}\left(\kappa_r\pi\right)\right){\mathrm{diag}}\left\{{\bf W}\right\}$ for an arbitrary square matrix ${\bf W}$ \cite[Eq. (10)]{9785990}, we can calculate the term $ {\sum\limits_{l_{1}=1}^L\sum\limits_{l_{2}=1}^L \sum\limits_{s_{1}=1 }^S \sum\limits_{s_{2}=1 }^S}\mathbb{E}\left\{{{f_{l_{1},s_{1},k}^{H}(\widehat{\bf\Phi})}{f_{l_{2},s_{2},i}(\widehat{\bf\Phi})}}\right\}$ as follows
	\begin{align}\label{B_begin}
		\begin{array}{l}
			{\sum\limits_{l_{1}=1}^L\sum\limits_{l_{2}=1}^L \sum\limits_{s_{1}=1 }^S \sum\limits_{s_{2}=1 }^S}\mathbb{E}\left\{{{f_{l_{1},s_{1},i}^{H}(\widehat{\bf\Phi})}{f_{l_{2},s_{2},i}(\widehat{\bf\Phi})}}\right\}
			\\=
			{\sum\limits_{l_{1}=1}^L\sum\limits_{l_{2}=1}^L \sum\limits_{s=1}^S}\mathbb{E}\left\{{{f_{l_{1},s,i}^{H}(\widehat{\bf\Phi})}{f_{l_{2},s,i}(\widehat{\bf\Phi})}}\right\}+
			{\sum\limits_{l_{1}=1}^L\sum\limits_{l_{2}=1}^L \sum\limits_{s_{1}=1 }^S \sum\limits_{s_{2}=1 \atop s_{2} \neq s_{1} }^S}\mathbb{E}\left\{{{f_{l_{1},s_{1},i}^{H}(\widehat{\bf\Phi})}{f_{l_{2},s_{2},i}(\widehat{\bf\Phi})}}\right\}
			\\=
			{\sum\limits_{l_{1}=1}^L\sum\limits_{l_{2}=1}^L \sum\limits_{s=1}^S}\overline{\mathbf{h}}^{H}_{s,i}{\bf\Phi}^{H}_{s}\mathbb{E}\left\{{\widetilde{\bf\Phi}^{H}_{s}}\mathbf{a}_{R}(l_{1},s) \mathbf{a}_{R}^{H}(l_{2},s){\widetilde{\bf\Phi}_{s}}\right\}{\bf\Phi}_{s}  \overline{\mathbf{h}}_{s,i}\\+
			{\sum\limits_{l_{1}=1}^L\sum\limits_{l_{2}=1}^L \sum\limits_{s_{1}=1 }^S \sum\limits_{s_{2}=1 \atop s_{2} \neq s_{1} }^S}\overline{\mathbf{h}}^{H}_{s_{1},i}{\bf\Phi}^{H}_{s_{1}}\mathbb{E}\left\{{\widetilde{\bf\Phi}^{H}_{s_{1}}}\right\}\mathbf{a}_{R}(l_{1},s_{1}) \mathbf{a}_{R}^{H}(l_{2},s_{2})\mathbb{E}\left\{{\widetilde{\bf\Phi}_{s_{2}}}\right\}{\bf\Phi}_{s_{2}}  \overline{\mathbf{h}}_{s_{2},i}
			\\=		{\sum\limits_{l_{1}=1}^L\sum\limits_{l_{2}=1}^L \sum\limits_{s=1}^S}\Big(\mathrm{sinc}^{2}\left(\kappa_r\pi\right)\overline{\mathbf{h}}^{H}_{s,i}{\bf\Phi}^{H}_{s}\mathbf{a}_{R}(l_{1},s) \mathbf{a}_{R}^{H}(l_{2},s){\bf\Phi}_{s}  \overline{\mathbf{h}}_{s,i}+\left(1-\mathrm{sinc}^{2}\left(\kappa_r\pi\right)\right)\overline{\mathbf{h}}^{H}_{s,i}{\mathrm{diag}}\left\{\mathbf{a}_{R}(l_{1},s)\right\}\\{\mathrm{diag}}\left\{\mathbf{a}_{R}^{H}(l_{2},s)\right\} \overline{\mathbf{h}}_{s,i}\Big)
			+{\sum\limits_{l_{1}=1}^L\sum\limits_{l_{2}=1}^L \sum\limits_{s_{1}=1 }^S \sum\limits_{s_{2}=1 \atop s_{2} \neq s_{1} }^S}\mathrm{sinc}^{2}\left(\kappa_r\pi\right)\overline{\mathbf{h}}^{H}_{s_{1},i}{\bf\Phi}^{H}_{s_{1}}\mathbf{a}_{R}(l_{1},s_{1}) \mathbf{a}_{R}^{H}(l_{2},s_{2}){\bf\Phi}_{s_{2}}  \overline{\mathbf{h}}_{s_{2},i}
			\\=		{\sum\limits_{l_{1}=1}^L\sum\limits_{l_{2}=1}^L \sum\limits_{s=1}^S}\Big(\mathrm{sinc}^{2}\left(\kappa_r\pi\right){{f_{l_{1},s,i}^{H}({\bf\Phi})}{f_{l_{2},s,i}({\bf\Phi})}}+\left(1-\mathrm{sinc}^{2}\left(\kappa_r\pi\right)\right)\mathbf{a}_{R}^{H}(l_{2},s)\mathbf{a}_{R}(l_{1},s)\Big)\\
			+{\sum\limits_{l_{1}=1}^L\sum\limits_{l_{2}=1}^L \sum\limits_{s_{1}=1 }^S \sum\limits_{s_{2}=1 \atop s_{2} \neq s_{1} }^S}\mathrm{sinc}^{2}\left(\kappa_r\pi\right){{f_{l_{1},s_{1},i}^{H}({\bf\Phi})}{f_{l_{2},s_{2},i}({\bf\Phi})}}
			\\=			{\sum\limits_{l_{1}=1}^L\sum\limits_{l_{2}=1}^L \sum\limits_{s=1}^S}\left(1-\mathrm{sinc}^{2}\left(\kappa_r\pi\right)\right)\mathbf{a}_{R}^{H}(l_{2},s)\mathbf{a}_{R}(l_{1},s)
			\\+{\sum\limits_{l_{1}=1}^L\sum\limits_{l_{2}=1}^L \sum\limits_{s_{1}=1 }^S \sum\limits_{s_{2}=1}^S}\mathrm{sinc}^{2}\left(\kappa_r\pi\right){{f_{l_{1},s_{1},i}^{H}({\bf\Phi})}{f_{l_{2},s_{2},i}({\bf\Phi})}},
		\end{array}
	\end{align} 
	and the term $ {\sum\limits_{s_{1}=1 }^S \sum\limits_{s_{2}=1 }^S} \mathbb{E}\left\{\widetilde{\bf\Phi}_{s_{1}} \overline{\mathbf{h}}_{s_{1},i} \overline{\mathbf{h}}_{s_{2},i}^{H}\widetilde{\bf\Phi}^{H}_{s_{2}}\right\}$ can be derived as 
	\begin{align}
		\begin{array}{l}
			{\sum\limits_{s_{1}=1 }^S \sum\limits_{s_{2}=1 }^S}\mathbb{E}\left\{\widetilde{\bf\Phi}_{s_{1}} \overline{\mathbf{h}}_{s_{1},i} \overline{\mathbf{h}}_{s_{2},i}^{H}\widetilde{\bf\Phi}^{H}_{s_{2}}\right\}
			\\=
			{\sum\limits_{s_{1}=1 }^S \sum\limits_{s_{2}=1 \atop s_{2} \neq s_{1}}^S} \mathbb{E}\left\{\widetilde{\bf\Phi}_{s_{1}}\right\} \overline{\mathbf{h}}_{s_{1},i} \overline{\mathbf{h}}_{s_{2},i}^{H}\mathbb{E}\left\{\widetilde{\bf\Phi}^{H}_{s_{2}}\right\}+{\sum\limits_{s=1 }^S}\mathbb{E}\left\{\widetilde{\bf\Phi}_{s} \overline{\mathbf{h}}_{s,i} \overline{\mathbf{h}}_{s,i}^{H}\widetilde{\bf\Phi}^{H}_{s}\right\}
			\\=
			{\sum\limits_{s_{1}=1 }^S \sum\limits_{s_{2}=1 \atop s_{2} \neq s_{1}}^S}\mathrm{sinc}^{2}\left(\kappa_r\pi\right)  \overline{\mathbf{h}}_{s_{1},i} \overline{\mathbf{h}}_{s_{2},i}^{H}+{\sum\limits_{s=1 }^S}\left(\mathrm{sinc}^{2}\left(\kappa_r\pi\right) \overline{\mathbf{h}}_{s,i} \overline{\mathbf{h}}_{s,i}^{H}+\left(1-\mathrm{sinc}^{2}\left(\kappa_r\pi\right)\right){\mathrm{diag}}\left\{ \overline{\mathbf{h}}_{s,i} \overline{\mathbf{h}}_{s,i}^{H}\right\}\right)
			\\=
			{\sum\limits_{s_{1}=1 }^S \sum\limits_{s_{2}=1}^S}\mathrm{sinc}^{2}\left(\kappa_r\pi\right)  \overline{\mathbf{h}}_{s_{1},i} \overline{\mathbf{h}}_{s_{2},i}^{H}+{\sum\limits_{s=1 }^S}\left(1-\mathrm{sinc}^{2}\left(\kappa_r\pi\right)\right)\mathbf{I}_{R}.
		\end{array}
	\end{align}
	
	Also, using the independence of each element and $\mathbb{E}\left\{e^{\pm j\widetilde{\theta}_{s,r}}\right\}=\mathrm{sinc}\left(\kappa_r\pi\right)$, $\forall s, r$, we can derive the term $ {\sum\limits_{s_{1}=1 }^S \sum\limits_{s_{2}=1 }^S}\mathbb{E}\left\{f_{l,s_{1},k}(\widehat{\bf\Phi})\mathrm{Tr}\left\{\widetilde{\bf\Phi}^{H}_{s_{2}}\right\}\right\}$ as follows
	\begin{align}
		\begin{array}{l}
			{\sum\limits_{s_{1}=1 }^S \sum\limits_{s_{2}=1 }^S}\mathbb{E}\left\{f_{l,s_{1},k}(\widehat{\bf\Phi})\mathrm{Tr}\left\{\widetilde{\bf\Phi}^{H}_{s_{2}}\right\}\right\}
			\\=
			{\sum\limits_{s=1 }^S}\mathbb{E}\left\{f_{l,s,k}(\widehat{\bf\Phi})\mathrm{Tr}\left\{\widetilde{\bf\Phi}^{H}_{s}\right\}\right\}+
			{\sum\limits_{s_{1}=1 }^S \sum\limits_{s_{2}=1 \atop s_{2} \neq s_{1}  }^S}\mathbb{E}\left\{f_{l,s_{1},k}(\widehat{\bf\Phi})\mathrm{Tr}\left\{\widetilde{\bf\Phi}^{H}_{s_{2}}\right\}\right\}
			\\=
			{\sum\limits_{s=1 }^S} \mathbb{E}\left\{\sum\limits_{r_{1}=1}^{R} e^{j\left(\zeta_{r_{1}}^{l,s,k}+\theta_{s,r_{1}}+\widetilde{\theta}_{s,r_{1}}\right)}\sum\limits_{r_{2}=1}^{R} e^{-j\widetilde{\theta}_{s,r_{2}}}\right\}+
			{\sum\limits_{s_{1}=1 }^S\sum\limits_{s_{2}=1 \atop s_{2} \neq s_{1}  }^S}\mathbb{E}\left\{\sum\limits_{r_{1}=1}^{R} e^{j\left(\zeta_{r_{1}}^{l,s_{1},k}+\theta_{s_{1},r_{1}}+\widetilde{\theta}_{s_{1},r_{1}}\right)}
			\sum\limits_{r_{2}=1}^{R} e^{-j\widetilde{\theta}_{s_{2},r_{2}}}\right\}
			\\=
			{\sum\limits_{s=1 }^S} \mathbb{E}\left\{\sum\limits_{r=1}^{R} e^{j\left(\zeta_{r}^{l,s,k}+\theta_{s,r}+\widetilde{\theta}_{s,r}\right)} e^{-j\widetilde{\theta}_{s,r}}+\sum\limits_{r_{1}=1}^{R} e^{j\left(\zeta_{r_{1}}^{l,s,k}+\theta_{s,r_{1}}+\widetilde{\theta}_{s,r_{1}}\right)}\sum\limits_{r_{2}=1 \atop s_{2} \neq s_{1}}^{R} e^{-j\widetilde{\theta}_{s,r_{2}}}\right\}
			\\+{\sum\limits_{s_{1}=1 }^S}\sum\limits_{r_{1}=1}^{R} e^{j\left(\zeta_{r_{1}}^{l,s_{1},k}+\theta_{s_{1},r_{1}}\right)}
			\mathbb{E}\left\{e^{j\widetilde{\theta}_{s_{1},r_{1}}}\right\}
			\sum\limits_{s_{2}=1 \atop s_{2} \neq s_{1}  }^S\sum\limits_{r_{2}=1}^{R} \mathbb{E}\left\{e^{-j\widetilde{\theta}_{s_{2},r_{2}}}\right\}
			\\=
			{\sum\limits_{s=1 }^S}\big(f_{l,s,k}({\bf\Phi})+(R-1)\mathrm{sinc}^{2}\left(\kappa_r\pi\right)f_{l,s,k}({\bf\Phi})\big)+
			{\sum\limits_{s_{1}=1 }^S \sum\limits_{s_{2}=1 \atop s_{2} \neq s_{1}  }^S}R\mathrm{sinc}^{2}\left(\kappa_r\pi\right)f_{l,s_{1},k}({\bf\Phi})
			\\=
			{\sum\limits_{s=1 }^S}\big(1-\mathrm{sinc}^{2}\left(\kappa_r\pi\right)\big)f_{l,s,k}({\bf\Phi})+
			{\sum\limits_{s_{1}=1 }^S \sum\limits_{s_{2}=1}^S}R\mathrm{sinc}^{2}\left(\kappa_r\pi\right)f_{l,s_{1},k}({\bf\Phi}),
		\end{array}
	\end{align} 
	and the term ${\sum\limits_{s_{1}=1 }^S \sum\limits_{s_{2}=1 }^S}\mathbb{E}\left\{\mathrm{Tr}\left\{\widetilde{\bf\Phi}_{s_{1}}\right\}\mathrm{Tr}\left\{\widetilde{\bf\Phi}^{H}_{s_{2}}\right\}\right\}$ can be given by 
	\begin{align}\label{B_end}
		\begin{array}{l}
			{\sum\limits_{s_{1}=1 }^S \sum\limits_{s_{2}=1 }^S}\mathbb{E}\left\{\mathrm{Tr}\left\{\widetilde{\bf\Phi}_{s_{1}}\right\}\mathrm{Tr}\left\{\widetilde{\bf\Phi}^{H}_{s_{2}}\right\}\right\}
			\\=
			{\sum\limits_{s=1 }^S}\mathbb{E}\left\{\mathrm{Tr}\left\{\widetilde{\bf\Phi}_{s}\right\}\mathrm{Tr}\left\{\widetilde{\bf\Phi}^{H}_{s}\right\}\right\}+
			{\sum\limits_{s_{1}=1 }^S \sum\limits_{s_{2}=1 \atop s_{2} \neq s_{1}  }^S}\mathbb{E}\left\{\mathrm{Tr}\left\{\widetilde{\bf\Phi}_{s_{1}}\right\}\mathrm{Tr}\left\{\widetilde{\bf\Phi}^{H}_{s_{2}}\right\}\right\}
			\\=
			{\sum\limits_{s=1 }^S} \mathbb{E}\left\{\sum\limits_{r_{1}=1}^{R} e^{j\widetilde{\theta}_{s,r_{1}}}\sum\limits_{r_{2}=1}^{R} e^{-j\widetilde{\theta}_{s,r_{2}}}\right\}+
			{\sum\limits_{s_{1}=1 }^S\sum\limits_{s_{2}=1 \atop s_{2} \neq s_{1}  }^S}\mathbb{E}\left\{\sum\limits_{r_{1}=1}^{R} e^{j\widetilde{\theta}_{s_{1},r_{1}}}
			\sum\limits_{r_{2}=1}^{R} e^{-j\widetilde{\theta}_{s_{2},r_{2}}}\right\}
			\\=
			{\sum\limits_{s=1 }^S} \mathbb{E}\left\{\sum\limits_{r=1}^{R}1+\sum\limits_{r_{1}=1}^{R} e^{j\widetilde{\theta}_{s,r_{1}}}\sum\limits_{r_{2}=1 \atop r_{2} \neq r_{1}}^{R} e^{-j\widetilde{\theta}_{s,r_{2}}}\right\}
			+{\sum\limits_{s_{1}=1 }^S}\sum\limits_{r_{1}=1}^{R}
			\mathbb{E}\left\{e^{j\widetilde{\theta}_{s_{1},r_{1}}}\right\}
			\sum\limits_{s_{2}=1 \atop s_{2} \neq s_{1}  }^S\sum\limits_{r_{2}=1}^{R} \mathbb{E}\left\{e^{-j\widetilde{\theta}_{s_{2},r_{2}}}\right\}
			\\=
			{\sum\limits_{s=1 }^S}\big(R+R(R-1)\mathrm{sinc}^{2}\left(\kappa_r\pi\right)\big)+
			{\sum\limits_{s_{1}=1 }^S \sum\limits_{s_{2}=1 \atop s_{2} \neq s_{1}  }^S}R^{2}\mathrm{sinc}^{2}\left(\kappa_r\pi\right)
			\\=
			{\sum\limits_{s=1 }^S}R\big(1-\mathrm{sinc}^{2}\left(\kappa_r\pi\right)\big)+
			{\sum\limits_{s_{1}=1 }^S \sum\limits_{s_{2}=1}^S}R^{2}\mathrm{sinc}^{2}\left(\kappa_r\pi\right).
		\end{array}
	\end{align}
	
	Based on the above derivations in (\ref{B_begin}) $\sim$ (\ref{B_end}), 
	%we can obtain the expressions of the three terms with phase noise of desired signal, noise and multi-user interference. Then, 
	we can obtain the closed-form expressions of the desired signal and multi-user interference terms by calculating the expectation terms in (\ref{signal exp}) and (\ref{interference exp}). After some direct simplifications, the terms ${E}_{k}^{(\mathrm{signal})}({\bf\Phi})$ and $I_{k i}({\bf\Phi})$ are given by 
	\begin{align}\label{signal}
		&{E}_{k}^{(\mathrm{signal})}({\bf\Phi})=\mathbb{E}\left\{\left|{\bf q}_k^{H} \widehat{\bf q}_k\right|^{2}\right\} \nonumber\\ & %等价式1
		= \sum\limits_{l_{1}=1}^{L} \sum\limits_{l_{2}=1}^{L} \sum\limits_{s_{1}=1}^{S} \sum\limits_{s_{2}=1}^{S} \sum\limits_{s_{3}=1}^{S} \sum\limits_{s_{4}=1}^{S} \sqrt{c_{l_{1},s_{1},k}c_{l_{1},s_{2},k}c_{l_{2},s_{3},k}c_{l_{2},s_{4},k} \delta_{l_{1},s_{1}}\delta_{l_{1},s_{2}}
			\delta_{l_{2},s_{3}}\delta_{l_{2},s_{4}}\varepsilon_{s_{1},k}}  \nonumber\\ &\sqrt{\varepsilon_{s_{2},k}\varepsilon_{s_{3},k}\varepsilon_{s_{4},k}}\mathrm{sinc}^{2}\left(\kappa_r\pi\right) {f_{l_{1},s_{1},k}^{H}({\bf\Phi})f_{l_{1},s_{2},k}({\bf\Phi})f_{l_{2},s_{3},k}^{H}({\bf\Phi})f_{l_{2},s_{4},k}({\bf\Phi})} {\mathbf{a}_{B}^{H}(l_{1},s_{1})}{\mathbf{a}_{B}(l_{1},s_{2})}{\mathbf{a}_{B}^{H}(l_{2},s_{3})} \nonumber\\ &{\mathbf{a}_{B}(l_{2},s_{4})} +  %式子2
		{\sum\limits_{l_{1}=1}^{L} \sum\limits_{l_{2}=1}^{L} \sum\limits_{s_{1}=1}^{S} \sum\limits_{s_{2}=1}^{S} \sum\limits_{s_{3}=1}^{S}} \Big(2{\sqrt{c_{l_{1},s_{1},k}c_{l_{1},s_{2},k} \delta_{l_{1},s_{1}}\delta_{l_{1},s_{2}} \varepsilon_{s_{1},k}\varepsilon_{s_{2},k}}} c_{l_{2},s_{3},k}BR\nonumber\\ & (\delta_{l_{2},s_{3}}+\varepsilon_{s_{3},k}+1)\mathrm{sinc}^{2}\left(\kappa_r\pi\right) {\mathrm{Re}\left\{f_{l_{1},s_{1},k}^{H}({\bf\Phi})f_{l_{1},s_{2},k}({\bf\Phi}) {\mathbf{a}_{B}^{H}(l_{1},s_{1})} {\mathbf{a}_{B}(l_{1},s_{2})} \right\}} \nonumber\\ &+  %式子3
		\sqrt{c_{l_{1},s_{1},k}c_{l_{1},s_{2},k}c_{l_{2},s_{3},k}c_{l_{2},s_{1},k} \delta_{l_{1},s_{1}}\delta_{l_{1},s_{2}}
			\delta_{l_{2},s_{3}}\delta_{l_{2},s_{1}} \varepsilon_{s_{2},k}\varepsilon_{s_{3},k}}\left(1+\varepsilon_{s_{1},k}+(1-\varepsilon_{s_{1},k})\mathrm{sinc}^{2}\left(\kappa_r\pi\right)\right) \nonumber\\ & f_{l_{2},s_{3},k}^{H}({\bf\Phi})f_{l_{1},s_{2},k}({\bf\Phi}) {\mathbf{a}_{B}^{H}(l_{2},s_{3})} {\mathbf{a}_{B}(l_{2},s_{1})}
		{\mathbf{a}_{R}^{H}(l_{2},s_{1})} {\mathbf{a}_{R}(l_{1},s_{1})} {\mathbf{a}_{B}^{H}(l_{1},s_{1})} {\mathbf{a}_{B}(l_{1},s_{2})}\Big) \nonumber\\ &+ %式子4
		\sum\limits_{l_{1}=1}^{L} \sum\limits_{l_{2}=1}^{L} \sum\limits_{s_{1}=1}^{S} \sum\limits_{s_{2}=1}^{S} \Big(2\sqrt{c_{l_{1},s_{1},k}c_{l_{1},s_{2},k}\delta_{l_{1},s_{1}}\delta_{l_{1},s_{2}} \varepsilon_{s_{1},k}\varepsilon_{s_{2},k}}B  \Big(c_{l_{2},s_{2},k}\big(2+(\delta_{l_{2},s_{2}}+\varepsilon_{s_{2},k})\nonumber\\ &\left(1-\mathrm{sinc}^{2}\left(\kappa_r\pi\right)\right)\big) \mathrm{Re}\Big\{f_{l_{1},s_{1},k}^{H}({\bf\Phi}) f_{l_{1},s_{2},k}({\bf\Phi}){\mathbf{a}_{B}^{H}(l_{1},s_{1})} {\mathbf{a}_{B}(l_{1},s_{2})}\Big\}
		+ \gamma_{l_{2},k}\mathrm{sinc}\left(\kappa_r\pi\right)
		f_{l_{1},s_{1},k}^{H}({\bf\Phi})f_{l_{1},s_{2},k}({\bf\Phi})
		\nonumber\\ &{\mathbf{a}_{B}^{H}(l_{1},s_{1})} {\mathbf{a}_{B}(l_{1},s_{2})}\Big)+ %式子5
		\sqrt{c_{l_{1},s_{1},k}c_{l_{1},s_{2},k} c_{l_{2},s_{2},k}c_{l_{2},s_{1},k}\delta_{l_{1},s_{1}} \delta_{l_{1},s_{2}}\delta_{l_{2},s_{1}} \delta_{l_{2},s_{2}}}\left(1+\varepsilon_{s_{2},k}\left(1-\mathrm{sinc}^{2}\left(\kappa_r\pi\right)\right)\right) \nonumber\\ &{\mathbf{a}_{B}^{H}(l_{1},s_{1})} {\mathbf{a}_{B}(l_{1},s_{2})}  {{\mathbf{a}_{R}^{H}(l_{1},s_{2})}
			{\mathbf{a}_{R}(l_{2},s_{2})} {\mathbf{a}_{B}^{H}(l_{2},s_{2})} {\mathbf{a}_{B}(l_{2},s_{1})}  {\mathbf{a}_{R}^{H}(l_{2},s_{1})} {\mathbf{a}_{R}(l_{1},s_{1})}} \nonumber\\ & + %式子6 
		{{c_{l_{1},s_{1},k}c_{l_{2},s_{2},k}}B^{2}R^{2}\mathrm{sinc}^{2}\left(\kappa_r\pi\right)\big(2(\delta_{l_{1},s_{1}}\varepsilon_{s_{2},k}+ \delta_{l_{1},s_{1}}+\varepsilon_{s_{1},k})+ \delta_{l_{1},s_{1}}\delta_{l_{2},s_{2}}+
			\varepsilon_{s_{1},k}\varepsilon_{s_{2},k}+1\big)}\Big)
		\nonumber\\ &+ %式子7
		{\sum\limits_{l=1}^{L} \sum\limits_{s_{1}=1}^{S} \sum\limits_{s_{2}=1}^{S} \sum\limits_{s_{3}=1}^{S}}
		\sqrt{c_{l,s_{1},k}c_{l,s_{3},k} \delta_{l,s_{1}}\delta_{l,s_{3}} \varepsilon_{s_{1},k}\varepsilon_{s_{3},k}}  c_{l,s_{2},k}R(\varepsilon_{s_{2},k}+1) \left(1+\mathrm{sinc}^{2}\left(\kappa_r\pi\right)\right)\nonumber\\ &f_{l,s_{1},k}^{H}({\bf\Phi})f_{l,s_{3},k}({\bf\Phi}) {\mathbf{a}_{B}^{H}(l,s_{1})} {\mathbf{a}_{B}(l,s_{3})}
		+ %式子8
		{\sum\limits_{l_{1}=1}^{L} \sum\limits_{l_{2}=1}^{L} \sum\limits_{s=1}^{S}}   c_{l_{1},s,k}B^{2}R\Big(c_{l_{2},s,k}\big((4\delta_{l_{1},s}\varepsilon_{s,k}+ \delta_{l_{1},s}\delta_{l_{2},s}\nonumber\\ &+\varepsilon_{s,k}\varepsilon_{s,k})\left(1-\mathrm{sinc}^{2}\left(\kappa_r\pi\right)\right)+(2\delta_{l_{1},s}+2\varepsilon_{s,k}+1)\left(2-\mathrm{sinc}^{2}\left(\kappa_r\pi\right)\right)\big)+2\gamma_{l_{2},k}(\delta_{l_{1},s}+\varepsilon_{s,k}+1)\mathrm{sinc}\left(\kappa_r\pi\right)\Big) %式子13
		\nonumber\\ & +  %式子9
		{\sum\limits_{l=1}^{L} \sum\limits_{s_{1}=1}^{S} \sum\limits_{s_{2}=1}^{S}}\Big( c_{l,s_{1},k} c_{l,s_{2},k}B R^{2} \big(2\delta_{l,s_{1}}\varepsilon_{s_{1},k}\left(1-\mathrm{sinc}^{2}\left(\kappa_r\pi\right)\right)+ 2\delta_{l,s_{1}}+\varepsilon_{s_{1},k}+1\big)(\varepsilon_{s_{2},k}+1)
		\nonumber\\ &  %式子10
		+\sqrt{c_{l,s_{1},k}c_{l,s_{2},k}\delta_{l,s_{1}}\delta_{l,s_{2}}\varepsilon_{s_{1},k}\varepsilon _{s_{2},k}} \left(c_{l,s_{1},k}+c_{l,s_{2},k}+\gamma_{l,k}\right) \left(1+\mathrm{sinc}^{2}\left(\kappa_r\pi\right)\right){f_{l,s_{1},k}^{H}({\bf\Phi})}{f_{l,s_{2},k}({\bf\Phi})} {\mathbf{a}_{B}^{H}(l,s_{1})} 
		\nonumber\\ &{\mathbf{a}_{B}(l,s_{2})}
		\Big) +  %式子9
		{\sum\limits_{l=1}^{L} \sum\limits_{s=1}^{S}}  c_{l,s,k}BR\Big(\delta_{l,s}\varepsilon_{s,k}(2c_{l,s,k}+\gamma_{l,k})\left(1-\mathrm{sinc}^{2}\left(\kappa_r\pi\right)\right)+ 2(\delta_{l,s}+\varepsilon_{s,k})(c_{l,s,k}+\gamma_{l,k})	\nonumber\\ &+c_{l,s,k}+2\gamma_{l,k}\Big)	+\left(\sum\limits_{l=1}^{L}\gamma_{l,k}B\right)^{2}+ \sum\limits_{l=1}^{L} \gamma_{l,k}^{2}B
		,
	\end{align}
	\begin{align}\label{interference}
		&I_{k i}({\bf\Phi})
		%等价式1
		=\mathbb{E}\left\{\left|{\bf q}_k^{H} \widehat{\bf q}_i\right|^{2}\right\}
		\nonumber\\ &= \sum\limits_{l_{1}=1}^{L} \sum\limits_{l_{2}=1}^{L} \sum\limits_{s_{1}=1}^{S} \sum\limits_{s_{2}=1}^{S} \sum\limits_{s_{3}=1}^{S} \sum\limits_{s_{4}=1}^{S} \sqrt{c_{l_{1},s_{1},k}c_{l_{1},s_{2},i}c_{l_{2},s_{3},i}c_{l_{2},s_{4},k}\delta_{l_{1},s_{1}}\delta_{l_{1},s_{2}}
			\delta_{l_{2},s_{3}}\delta_{l_{2},s_{4}}\varepsilon_{s_{1},k}}\nonumber\\ &\sqrt{\varepsilon_{s_{2},i}\varepsilon_{s_{3},i}\varepsilon_{s_{4},k}}\mathrm{sinc}^{2}\left(\kappa_r\pi\right)  f_{l_{1},s_{1},k}^{H}({\bf\Phi})f_{l_{1},s_{2},i}({\bf\Phi})f_{l_{2},s_{3},i}^{H}({\bf\Phi})f_{l_{2},s_{4},k}({\bf\Phi}) {\mathbf{a}_{B}^{H}(l_{1},s_{1})}{\mathbf{a}_{B}(l_{1},s_{2})}{\mathbf{a}_{B}^{H}(l_{2},s_{3})}{\mathbf{a}_{B}(l_{2},s_{4})} \nonumber\\ & +  %式子2
		{\sum\limits_{l_{1}=1}^{L} \sum\limits_{l_{2}=1}^{L} \sum\limits_{s_{1}=1}^{S} \sum\limits_{s_{2}=1}^{S} \sum\limits_{s_{3}=1}^{S}} \Big(2{\sqrt{c_{l_{1},s_{1},k}c_{l_{1},s_{2},i}c_{l_{2},s_{3},i}c_{l_{2},s_{3},k} \delta_{l_{1},s_{1}}\delta_{l_{1},s_{2}} \varepsilon_{s_{1},k}\varepsilon_{s_{2},i} \varepsilon_{s_{3},i}\varepsilon_{s_{3},k}}}B\nonumber\\ &\mathrm{sinc}^{2}\left(\kappa_r\pi\right) {\mathrm{Re}\left\{f_{l_{1},s_{1},k}^{H}({\bf\Phi})f_{l_{1},s_{2},i}({\bf\Phi}) {\mathbf{a}_{B}^{H}(l_{1},s_{1})} {\mathbf{a}_{B}(l_{1},s_{2})} \overline{\mathbf{h}}_{s_{3},i}^{H}\overline{\mathbf{h}}_{s_{3},k} \right\}}  %式子3
		+\sqrt{\delta_{l_{1},s_{1}}\delta_{l_{1},s_{2}}\delta_{l_{2},s_{2}}\delta_{l_{2},s_{3}}}\Big(\sqrt{c_{l_{1},s_{1},k}}\nonumber\\ & \sqrt{c_{l_{1},s_{2},i}c_{l_{2},s_{2},i}c_{l_{2},s_{3},k} \varepsilon_{s_{1},k}\varepsilon_{s_{3},k}}\left(1+\varepsilon_{s_{2},i}\left(1-\mathrm{sinc}^{2}\left(\kappa_r\pi\right)\right)\right)f_{l_{1},s_{1},k}^{H}({\bf\Phi}) f_{l_{2},s_{3},k}({\bf\Phi}) +\sqrt{c_{l_{1},s_{1},i}c_{l_{1},s_{2},k}c_{l_{2},s_{2},k}}
		\nonumber\\ &\sqrt{c_{l_{2},s_{3},i} \varepsilon_{s_{1},i}\varepsilon_{s_{3},i}}\mathrm{sinc}^{2}\left(\kappa_r\pi\right) f_{l_{1},s_{1},i}^{H}({\bf\Phi}) f_{l_{2},s_{3},i}({\bf\Phi})\Big) {\mathbf{a}_{B}^{H}(l_{1},s_{1})} {\mathbf{a}_{B}(l_{1},s_{2})} %换行
		{\mathbf{a}_{R}^{H}(l_{1},s_{2})} {\mathbf{a}_{R}(l_{2},s_{2})} {\mathbf{a}_{B}^{H}(l_{2},s_{2})}  \nonumber\\ &{\mathbf{a}_{B}(l_{2},s_{3})}
		\Big)+ %式子4
		\sum\limits_{l_{1}=1}^{L} \sum\limits_{l_{2}=1}^{L} \sum\limits_{s_{1}=1}^{S} \sum\limits_{s_{2}=1}^{S} \Big(2\sqrt{c_{l_{1},s_{1},k}c_{l_{1},s_{2},i}\delta_{l_{1},s_{1}}\delta_{l_{1},s_{2}}}B\mathrm{Re}\Big\{\Big(\sqrt{c_{l_{2},s_{2},i}c_{l_{2},s_{2},k}\varepsilon_{s_{1},k}\varepsilon_{s_{2},k}}   \nonumber\\ &\big(\left(1-\mathrm{sinc}^{2}\left(\kappa_r\pi\right)\right)\varepsilon_{s_{2},i}+1\big)f_{l_{1},s_{1},k}^{H}({\bf\Phi}) f_{l_{1},s_{2},k}({\bf\Phi}) + \sqrt{c_{l_{2},s_{1},i}c_{l_{2},s_{1},k} \varepsilon_{s_{1},i}\varepsilon_{s_{2},i}}\mathrm{sinc}^{2}\left(\kappa_r\pi\right)f_{l_{1},s_{1},i}^{H}({\bf\Phi}) f_{l_{1},s_{2},i}({\bf\Phi})\Big)  \nonumber\\ & {\mathbf{a}_{B}^{H}(l_{1},s_{1})} {\mathbf{a}_{B}(l_{1},s_{2})}\Big\}+ %式子5
		\sqrt{c_{l_{1},s_{1},k}c_{l_{1},s_{2},i} c_{l_{2},s_{2},i}c_{l_{2},s_{1},k}}\Big(\sqrt{\delta_{l_{1},s_{1}} \delta_{l_{1},s_{2}}\delta_{l_{2},s_{1}} \delta_{l_{2},s_{2}}}\big(\left(1-\mathrm{sinc}^{2}\left(\kappa_r\pi\right)\right)\varepsilon_{s_{2},i}+1\big) \nonumber\\ & {\mathbf{a}_{B}^{H}(l_{1},s_{1})} {\mathbf{a}_{B}(l_{1},s_{2})} {{\mathbf{a}_{R}^{H}(l_{1},s_{2})}
			{\mathbf{a}_{R}(l_{2},s_{2})} {\mathbf{a}_{B}^{H}(l_{2},s_{2})} {\mathbf{a}_{B}(l_{2},s_{1})}  {\mathbf{a}_{R}^{H}(l_{2},s_{1})} {\mathbf{a}_{R}(l_{1},s_{1})}}  + %式子6 
		\sqrt{ \varepsilon_{s_{1},k}\varepsilon_{s_{1},i}
			\varepsilon_{s_{2},i}\varepsilon_{s_{2},k}}B^{2} \nonumber\\ &\mathrm{sinc}^{2}\left(\kappa_r\pi\right)\overline{\mathbf{h}}_{s_{1},k}^{H}\overline{\mathbf{h}}_{s_{1},i} \overline{\mathbf{h}}_{s_{2},i}^{H}\overline{\mathbf{h}}_{s_{2},k}\Big)\Big) + %式子7
		{\sum\limits_{l=1}^{L} \sum\limits_{s_{1}=1}^{S} \sum\limits_{s_{2}=1}^{S} \sum\limits_{s_{3}=1}^{S}}
		\Big(\sqrt{c_{l,s_{1},k}c_{l,s_{3},k} \delta_{l,s_{1}}\delta_{l,s_{3}} \varepsilon_{s_{1},k}\varepsilon_{s_{3},k}}  c_{l,s_{2},i}R \nonumber\\ &(\varepsilon_{s_{2},i}+1) f_{l,s_{1},k}^{H}({\bf\Phi}) f_{l,s_{3},k}({\bf\Phi}) {\mathbf{a}_{B}^{H}(l,s_{1})} {\mathbf{a}_{B}(l,s_{3})}+ %式子8
		\sqrt{c_{l,s_{2},i}c_{l,s_{3},i}\delta_{l,s_{2}}\delta_{l,s_{3}}\varepsilon_{s_{2},i}\varepsilon_{s_{3},i}}  c_{l,s_{1},k}R(\varepsilon_{s_{1},k}+1)\mathrm{sinc}^{2}\left(\kappa_r\pi\right) 
		\nonumber\\ &f_{l,s_{3},i}^{H}({\bf\Phi}) f_{l,s_{2},i}({\bf\Phi}) {\mathbf{a}_{B}^{H}(l,s_{3})} {\mathbf{a}_{B}(l,s_{2})}\Big)
		+{\sum\limits_{l_{1}=1}^{L} \sum\limits_{l_{2}=1}^{L} \sum\limits_{s=1}^{S}} \sqrt{c_{l_{1},s,k}c_{l_{1},s,i} c_{l_{2},s,i}c_{l_{2},s,k}} B^{2} R \Big(\left(2\delta_{l_{1},s}+\varepsilon_{s,k}\right) \nonumber\\ &\left(1+\varepsilon_{s,i}\left(1-\mathrm{sinc}^{2}\left(\kappa_r\pi\right)\right)\right)+\varepsilon_{s,i}+1\Big) +  %式子10
		{\sum\limits_{l=1}^{L} \sum\limits_{s_{1}=1}^{S} \sum\limits_{s_{2}=1}^{S}} \Big( 
		\sqrt{\delta_{l,s_{1}}\delta_{l,s_{2}}} \Big(\sqrt{c_{l,s_{1},k}c_{l,s_{2},k}\varepsilon _{s_{1},k}\varepsilon _{s_{2},k}} \gamma_{l,i} \nonumber\\ &{f_{l,s_{1},k}^{H}({\bf\Phi})}{f_{l,s_{2},k}({\bf\Phi})}
		+ \sqrt{c_{l,s_{1},i}c_{l,s_{2},i}\varepsilon _{s_{1},i}\varepsilon _{s_{2},i}} \gamma_{l,k}\mathrm{sinc}^{2}\left(\kappa_r\pi\right) {f_{l,s_{1},i}^{H}({\bf\Phi})}{f_{l,s_{2},i}({\bf\Phi})}\Big) {\mathbf{a}_{B}^{H}(l,s_{1})} {\mathbf{a}_{B}(l,s_{2})}
		\nonumber\\ &+
		c_{l,s_{1},k} c_{l,s_{2},i} B R^{2}\big( \delta_{l,s_{2}}(\varepsilon_{s_{1},k}+1)\left(1+\varepsilon_{s_{2},i}\left(1-\mathrm{sinc}^{2}\left(\kappa_r\pi\right)\right)\right)+ (\delta_{l,s_{1}}+\varepsilon_{s_{1},k}+1)(\varepsilon_{s_{2},i}+1)\big)\Big)
		\nonumber\\ &+
		{\sum\limits_{l=1}^L \sum\limits_{s=1}^S} 
		B R \Big(c_{l,s,k} (\delta _{l,s}+\varepsilon _{s,k}+1)\gamma_{l,i}+c_{l,s,i} \big(\delta_{l,s}\left(1+\varepsilon_{s,i}\left(1-\mathrm{sinc}^{2}\left(\kappa_r\pi\right)\right)\right)+\varepsilon _{s,i}+1\big)\gamma_{l,k}\Big)
		\nonumber\\ &+{\sum\limits_{l=1}^L \gamma_{l,k}\gamma_{l,i}B},
	\end{align}
	
	\subsection[${\eta}$]{Derivations of $\mathbb{E}\left\{\left|{\mathbf{q}}_{k}^{H}\widehat{\mathbf{q}}_{i}{\mathbb{\eta}}_{\it i,t}\right|^{2}\right\}$ and $\mathbb{E}\left\{\left|{\mathbf{q}}_{k}^{H}{\bm{\eta}}_{\it r}\right|^{2}\right\}$}\label{subsection3}
	%To derive the closed-form expression of the achievable rate, 
	In this subsection, we derive the HWI term $E_{k}^{(\mathrm{hwi})}({\bf\Phi})=\sum\limits_{i=1}^{K}\mathbb{E}\left\{\left|{\mathbf{q}}_{k}^{H}\widehat{\mathbf{q}}_{i}{\mathbb{\eta}}_{\it i,t}\right|^{2}\right\}+\mathbb{E}\left\{\left|{\mathbf{q}}_{k}^{H}{\bm{\eta}}_{\it r}\right|^{2}\right\}$. Firstly, we need to present some necessary preliminary results.  Since $\widetilde{\mathbf{Z}}_{l,s}$ and $ \widetilde{\mathbf{h}}_{s,k} $ are independent of each other and have zero means, $ \forall l\in \mathcal{L}$, $s\in \mathcal{S}$, $k\in \mathcal{K}$, we can derive the term $\mathbb{E}\left\{\widehat{\mathbf{g}}_{l,s,k}\widehat{\mathbf{g}}_{l,s,k}^{H}\right\}$ by extracting the non-zero terms in the expansion as follows
	\begin{align}\label{ggi^H}
		\begin{array}{l}
			\mathbb{E}\left\{\widehat{\mathbf{g}}_{l,s,k}\widehat{\mathbf{g}}_{l,s,k}^{H}\right\}
			\\=\mathbb{E}\left\{\sum\limits_{\omega_{1}=1}^{4}\widehat{\mathbf{g}}_{l,s,k}^{\omega_{1}}\sum\limits_{\omega_{2}=1}^{4}\left(\widehat{\mathbf{g}}_{l,s,k}^{\omega_{2}}\right)^{H}\right\}
			{\mathop  = \limits^{\left( a \right)} }\mathbb{E}\left\{\sum\limits_{\omega=1}^{4}\widehat{\mathbf{g}}_{l,s,k}^{\omega}\left(\widehat{\mathbf{g}}_{l,s,k}^{\omega}\right)^{H}\right\}
			\\=c_{l,s,k}\mathbb{E}\Big\{\delta_{l,s} \varepsilon_{s,k}\overline{\mathbf{Z}}_{l,s} \widehat{\bf \Phi}_{s}\overline{\mathbf{h}}_{s,k}\overline{\mathbf{h}}^{H}_{s,k}\widehat{\bf \Phi}^{H}_{s}\overline{\mathbf{Z}}^{H}_{l,s} + \delta_{l,s}\overline{\mathbf{Z}}_{l,s} \widehat{\bf \Phi}_{s}\widetilde{\mathbf{h}}_{s,k}\widetilde{\mathbf{h}}^{H}_{s,k}\widehat{\bf \Phi}^{H}_{s}\overline{\mathbf{Z}}^{H}_{l,s}  \\\quad\quad+\varepsilon_{s,k} \widetilde{\mathbf{Z}}_{l,s} \widehat{\bf \Phi}_{s}\overline{\mathbf{h}}_{s,k}\overline{\mathbf{h}}^{H}_{s,k}\widehat{\bf \Phi}^{H}_{s}\widetilde{\mathbf{Z}}^{H}_{l,s}+\widetilde{\mathbf{Z}}_{l,s} \widehat{\bf \Phi}_{s}\widetilde{\mathbf{h}}_{s,k}\widetilde{\mathbf{h}}^{H}_{s,k}\widehat{\bf \Phi}^{H}_{s}\widetilde{\mathbf{Z}}^{H}_{l,s}\Big\}
			\\{\mathop  = \limits^{\left( b \right)} } c_{l,s,k}\bigg(\delta_{l,s} \varepsilon_{s,k}\mathbb{E}\left\{\left|f_{l,s,k}(\widehat{\bf \Phi})\right|^{2}\right\} \mathbf{a}_{B}(l,s)\mathbf{a}^{H}_{B}(l,s) + \delta_{l,s}\overline{\mathbf{Z}}_{l,s} \overline{\mathbf{Z}}^{H}_{l,s}  \\\quad\quad+\varepsilon_{s,k}  \mathbb{E}\left\{\mathrm{Tr}\left\{\widehat{\bf\Phi}_{s}\overline{\mathbf{h}}_{s,k}\overline{\mathbf{h}}^{H}_{s,k}\widehat{\bf \Phi}^{H}_{s}\right\}\mathbf{I}_{B}\right\}+\mathbb{E}\left\{\widetilde{\mathbf{Z}}_{l,s} \widetilde{\mathbf{Z}}^{H}_{l,s}\right\}\bigg)
			\\=c_{l,s,k}\Big(\delta_{l,s}R \left(\varepsilon_{s,k}\left(1-\mathrm{sinc}^{2}\left(\kappa_r\pi\right)\right)+1\right) \mathbf{a}_{B}(l,s)\mathbf{a}^{H}_{B}(l,s)  +\left(\varepsilon_{s,k}+1 \right) R\mathbf{I}_{B}\Big),
		\end{array}
	\end{align}
	where $(a)$ exploits the independence and the zero-mean properties of $\widetilde{\mathbf{Z}}_{l,s}$ and $\widetilde{\mathbf{h}}_{s,k}$, and $(b)$ follows by exploiting the identities
	\begin{align}
		\begin{array}{l}
			\widehat{\bf\Phi}_{s}\widehat{\bf\Phi}_{s}^H=\widehat{\bf\Phi}_{s}^H\widehat{\bf\Phi}_{s}={\bf I}_{R}, \mathbb{E}\left\{\widetilde{\mathbf{h}}_{s,k} \widetilde{\mathbf{h}}_{s,k}^{H}\right\}={\bf I}_{R},
			\mathbb{E}\left\{\widetilde{\mathbf{h}}_{s,k}^{H}\widetilde{\mathbf{h}}_{s,k}\right\}=\overline{\mathbf{h}}_{s,k}^{H} \overline{\mathbf{h}}_{s,k}=R, \\
			\mathbb{E}\left\{\widetilde{\mathbf{Z}}_{l,s}^{H} \widetilde{\mathbf{Z}}_{l,s}\right\}=B \mathbf{I}_{R}, \mathbb{E}\left\{ \widetilde{\mathbf{Z}}_{l,s}\widetilde{\mathbf{Z}}_{l,s}^{H}\right\}=R \mathbf{I}_{B},
			\mathbb{E}\left\{\widetilde{\mathbf{Z}}_{l,s} \mathbf{W}\widetilde{\mathbf{Z}}_{l,s}^{H}\right\}=\operatorname{Tr}\{\mathbf{W}\} \mathbf{I}_{B}.
		\end{array}
	\end{align}
	
	Besides, for arbitrary $s_{1}$, $s_{2}$,  and $s_{1}\neq s_{2}$, we have 
	\begin{align}\label{ggi^H_end}
		\begin{array}{l}
			\mathbb{E}\left\{\widehat{\mathbf{g}}_{l,s_{1},k}\widehat{\mathbf{g}}_{l,s_{2},k}^{H}\right\}
			\\=\mathbb{E}\left\{\sum\limits_{\omega_{1}=1}^{4}\widehat{\mathbf{g}}_{l,s_{1},k}^{\omega_{1}}\sum\limits_{\omega_{2}=1}^{4}\left(\widehat{\mathbf{g}}_{l,s_{2},k}^{\omega_{2}}\right)^{H}\right\}
			=\mathbb{E}\left\{\sum\limits_{\omega=1}^{4}\widehat{\mathbf{g}}_{l,s_{1},k}^{\omega}\left(\widehat{\mathbf{g}}_{l,s_{2},k}^{\omega}\right)^{H}\right\}
			\\=\sqrt{c_{l,s_{1},k}c_{l,s_{2},k}}\mathbb{E}\Big\{\sqrt{\delta_{l,s_{1}} \delta_{l,s_{2}} \varepsilon_{s_{1},k}\varepsilon_{s_{2},k}} \;\overline{\mathbf{Z}}_{l,s_{1}} \widehat{\bf \Phi}_{s_{1}}\overline{\mathbf{h}}_{s_{1},k}\overline{\mathbf{h}}^{H}_{s_{2},k}\widehat{\bf \Phi}^{H}_{s_{2}}\overline{\mathbf{Z}}^{H}_{l,s_{2}} + \sqrt{\delta_{l,s_{1}}\delta_{l,s_{2}}}\overline{\mathbf{Z}}_{l,s_{1}} \widehat{\bf \Phi}_{s_{1}}\\\widetilde{\mathbf{h}}_{s_{1},k}\widetilde{\mathbf{h}}^{H}_{s_{2},k}\widehat{\bf \Phi}^{H}_{s_{2}}\overline{\mathbf{Z}}^{H}_{l,s_{2}} +\sqrt{\varepsilon_{s_{1},k}\varepsilon_{s_{2},k}} \widetilde{\mathbf{Z}}_{l,s_{1}} \widehat{\bf \Phi}_{s_{1}}\overline{\mathbf{h}}_{s_{1},k}\overline{\mathbf{h}}^{H}_{s_{2},k}\widehat{\bf \Phi}^{H}_{s_{2}}\widetilde{\mathbf{Z}}^{H}_{l,s_{2}}+\widetilde{\mathbf{Z}}_{l,s_{1}} \widehat{\bf \Phi}_{s_{1}}\widetilde{\mathbf{h}}_{s_{1},k}\widetilde{\mathbf{h}}^{H}_{s_{2},k}\widehat{\bf \Phi}^{H}_{s_{2}}\widetilde{\mathbf{Z}}^{H}_{l,s_{2}}\Big\}
			\\{\mathop  = \limits^{\left( c \right)} } \sqrt{c_{l,s_{1},k}c_{l,s_{2},k}\delta_{l,s_{1}} \delta_{l,s_{2}} \varepsilon_{s_{1},k}\varepsilon_{s_{2},k}} \mathbb{E}\left\{f_{l,s_{1},k}(\widehat{\bf \Phi})f_{l,s_{2},k}^{H}(\widehat{\bf \Phi})\right\} \mathbf{a}_{B}(l,s_{1})\mathbf{a}^{H}_{B}(l,s_{2})
			\\=
			\sqrt{c_{l,s_{1},k}c_{l,s_{2},k}\delta_{l,s_{1}} \delta_{l,s_{2}} \varepsilon_{s_{1},k}\varepsilon_{s_{2},k}} f_{l,s_{1},k}({\bf \Phi})f_{l,s_{2},k}^{H}({\bf \Phi}) \mathbf{a}_{B}(l,s_{1})\mathbf{a}^{H}_{B}(l,s_{2})\mathrm{sinc}^{2}\left(\kappa_r\pi\right),
		\end{array}
	\end{align}
	where $(c)$ exploits the independence and the zero-mean properties of $\widetilde{\mathbf{Z}}_{l,s_{1}}$, $\widetilde{\mathbf{Z}}_{l,s_{2}}$, $\widetilde{\mathbf{h}}_{s_{1},k}$, and $\widetilde{\mathbf{h}}_{s_{2},k}$, $ \forall s_{1} \neq s_{2}$. 
	
	Based on (\ref{ggi^H}) $\sim$ (\ref{ggi^H_end}), the term $\mathbb{E}\left\{\widehat{\mathbf{q}}_{l,k}\widehat{\mathbf{q}}_{l,k}^{H}\right\}$ can be calculated as follows
	\begin{align}\label{qqi^H}
		\begin{array}{l}
			%\mathbb{E}\left\{\left\|\widehat{\mathbf{q}}_{k}\right\|^{2}\right\} =
			%		\mathbb{E}\left\{\left\|\widehat{\mathbf{g}}_{k}\right\|^{2}\right\}+\mathbb{E}\left\{\left\|{\mathbf{d}}_{k}\right\|^{2}\right\}=\sum\limits_{l=1}^{L}\left(\mathbb{E}\left\{\left\|\widehat{\mathbf{g}}_{l,k}\right\|^{2}\right\}+\mathbb{E}\left\{\left\|{\mathbf{d}}_{l,k}\right\|^{2}\right\}\right)
			\mathbb{E}\left\{\widehat{\mathbf{q}}_{l,k}\widehat{\mathbf{q}}_{l,k}^{H}\right\}=\mathbb{E}\left\{\left(\widehat{\mathbf{g}}_{l,k}+\mathbf{d}_{l,k}\right)\left(\widehat{\mathbf{g}}_{l,k}+\mathbf{d}_{l,k}\right)^{H}\right\}=\mathbb{E}\left\{\widehat{\mathbf{g}}_{l,k}\widehat{\mathbf{g}}_{l,k}^{H}\right\}+\mathbb{E}\left\{{\mathbf{d}}_{l,k}{\mathbf{d}}_{l,k}^{H}\right\}
			\\=\sum\limits_{s=1}^{S}\mathbb{E}\left\{\widehat{\mathbf{g}}_{l,s,k}\widehat{\mathbf{g}}_{l,s,k}^{H}\right\}+\sum\limits_{s_{1}=1}^{S}       \sum\limits_{s_{2}=1\atop s_{2}\neq s_{1}}^{S} \mathbb{E}\left\{\widehat{\mathbf{g}}_{l,s_{1},k}\widehat{\mathbf{g}}_{l,s_{2},k}^{H}\right\}+\gamma_{l,k}\mathbf{I}_{B}
			\\=\gamma_{l,k}\mathbf{I}_{B}+\sum\limits_{s=1}^{S}c_{l,s,k}\Big(\delta_{l,s}R \left(\varepsilon_{s,k}\left(1-\mathrm{sinc}^{2}\left(\kappa_r\pi\right)\right)+1\right) \mathbf{a}_{B}(l,s)\mathbf{a}^{H}_{B}(l,s)  +\left(\varepsilon_{s,k}+1 \right) R\mathbf{I}_{B}\Big)
			\\+\sum\limits_{s_{1}=1}^{S}       \sum\limits_{s_{2}=1}^{S} \sqrt{c_{l,s_{1},k}c_{l,s_{2},k}\delta_{l,s_{1}} \delta_{l,s_{2}} \varepsilon_{s_{1},k}\varepsilon_{s_{2},k}} f_{l,s_{1},k}({\bf \Phi}) f_{l,s_{2},k}^{H}({\bf \Phi}) \mathbf{a}_{B}(l,s_{1})\mathbf{a}^{H}_{B}(l,s_{2})\mathrm{sinc}^{2}\left(\kappa_r\pi\right).
		\end{array}
	\end{align}
	
	Then, based on (\ref{qqi^H}), we can derive the term $ \mathbb{E}\left\{{\mathbf{q}}_{l,k}{\mathbf{q}}_{l,k}^{H}\right\} $ as 
	\begin{align}\label{ggk^H}
		\begin{array}{l}
			\mathbb{E}\left\{{\mathbf{q}}_{l,k}{\mathbf{q}}_{l,k}^{H}\right\}
			= { \sum\limits_{s_{1}=1 }^S \sum\limits_{s_{2}=1 }^S} {\sqrt{c_{l,s_{1},k}c_{l,s_{2},k}\delta _{l,s_{1}}\delta _{l,s_{2}}\varepsilon _{s_{1},k}\varepsilon _{s_{2},k}} {f_{l,s_{1},k}({\bf\Phi})}{f_{l,s_{2},k}^{H}({\bf\Phi})}} \\{\mathbf{a}_{B}(l,s_{1})} {\mathbf{a}_{B}^{H}(l,s_{2})} + 
			\sum\limits_{s=1}^{S}c_{l,s,k}R\big(\delta_{l,s} \mathbf{a}_{B}(l,s)\mathbf{a}^{H}_{B}(l,s)  +\left(\varepsilon_{s,k}+1 \right) \mathbf{I}_{B}\big)+\gamma_{l,k}\mathbf{I}_{B}.
		\end{array}
	\end{align}
	
	Finally, with the aid of (\ref{ggk}) and (\ref{ggi^H}) $\sim$ (\ref{ggk^H}), we can derive the HWI term as
	\begin{align}
		\begin{array}{l}
			E_{k}^{(\mathrm{hwi})}({\bf\Phi})
			\\=\sum\nolimits_{i=1}^{K}\mathbb{E}\left\{\left|{\mathbf{q}}_{k}^{H}\widehat{\mathbf{q}}_{i}{\mathbb{\eta}}_{\it i,t}\right|^{2}\right\}+\mathbb{E}\left\{\left|{\mathbf{q}}_{k}^{H}{\bm{\eta}}_{\it r}\right|^{2}\right\}
			\\=\kappa_u^{2}  \left(p_{k}\mathbb{E}\left\{\left| {\mathbf{q}}_{k}^{H}\widehat{\mathbf{q}}_{k}\right|^{2}\right\}+\sum\nolimits_{i=1,i\neq k}^{K}p_{i}\mathbb{E}\left\{\left| {\mathbf{q}}_{k}^{H}\widehat{\mathbf{q}}_{i}\right|^{2}\right\}\right)+\mathbb{E}\left\{{\mathbf{q}}_{k}^{H}{\bm{\eta}}_{\it r}{\bm{\eta}}^{H}_{\it r}{\mathbf{q}}_{k}\right\}
			\\=\kappa_u^{2}  \left(p_{k}{E}_{k}^{(\mathrm{signal})}({\bf\Phi})+\sum\nolimits_{i=1,i\neq k}^{K}p_{i}I_{k i}({\bf\Phi})\right)+\mathbb{E}\left\{{\mathbf{q}}_{k}^{H}\left(\kappa_b^{2}\sum_{i=1}^{K}\widetilde{diag}\{\mathbb{E}\{\widehat{\mathbf{y}}_{i}\widehat{\mathbf{y}}_{i}^H\}\}\right){\mathbf{q}}_{k}\right\},
		\end{array}
	\end{align}
	where
	\begin{align}
		\begin{array}{l}
			\mathbb{E}\left\{{\mathbf{q}}_{k}^{H}\left(\kappa_b^{2}\sum_{i=1}^{K}\widetilde{diag}\{\mathbb{E}\{\widehat{\mathbf{y}}_{i}\widehat{\mathbf{y}}_{i}^H\}\}\right){\mathbf{q}}_{k}\right\}
			\\=\mathbb{E}\left\{{\mathbf{q}}_{k}^{H}\left(\kappa_b^{2}\sum_{i=1}^{K}\widetilde{diag}\left\{\mathbb{E}\left\{\widehat{\mathbf{q}}_{i}\widehat{\mathbf{q}}_{i}^{H}\left|\sqrt{p_i}x_i + {\mathbb{\eta}}_{i,t}\right|^{2}\right\}\right\}\right){\mathbf{q}}_{k}\right\}
			\\=\mathbb{E}\left\{{\mathbf{q}}_{k}^{H}\left(\kappa_b^{2}\sum_{i=1}^{K}\left(1+\kappa_u^{2}\right)p_{i}\widetilde{diag}\left\{\mathbb{E}\left\{\widehat{\mathbf{q}}_{i}\widehat{\mathbf{q}}_{i}^{H}\right\}\right\}\right){\mathbf{q}}_{k}\right\}
			\\=\sum_{l=1}^{L}\mathbb{E}\left\{{\mathbf{q}}_{l,k}^{H}\left(\kappa_b^{2}\sum_{i=1}^{K}\left(1+\kappa_u^{2}\right)p_{i}\widetilde{diag}\left\{\mathbb{E}\left\{\widehat{\mathbf{q}}_{l,i}\widehat{\mathbf{q}}_{l,i}^{H}\right\}\right\}\right){\mathbf{q}}_{l,k}\right\}
			\\=\kappa_b^{2}\left(1+\kappa_u^{2}\right)\sum_{l=1}^{L}\sum_{i=1}^{K}p_{i}\mathbb{E}\left\{{\mathbf{q}}_{l,k}^{H}\left(\widetilde{diag}\left\{\mathbb{E}\left\{\widehat{\mathbf{q}}_{l,i}\widehat{\mathbf{q}}_{l,i}^{H}\right\}\right\}\right){\mathbf{q}}_{l,k}\right\},
		\end{array}
	\end{align}
	and
	\begin{align}
		\begin{array}{l}
			\mathbb{E}\left\{{\mathbf{q}}_{l,k}^{H}\left(\widetilde{diag}\left\{\mathbb{E}\left\{\widehat{\mathbf{q}}_{l,i}\widehat{\mathbf{q}}_{l,i}^{H}\right\}\right\}\right){\mathbf{q}}_{l,k}\right\}
			\\=\mathbb{E}\left\{{\mathbf{q}}_{l,k}^{H}\left(\gamma_{l,i}\mathbf{I}_{B}+\sum\limits_{s=1}^{S}c_{l,s,i}R\Big(\delta_{l,s} \left(\varepsilon_{s,i}\left(1-\mathrm{sinc}^{2}\left(\kappa_r\pi\right)\right)+1\right)  +\varepsilon_{s,i}+1  \Big)\mathbf{I}_{B}\right){\mathbf{q}}_{l,k}\right\}
			\\+\mathbb{E}\bigg\{{\mathbf{q}}_{l,k}^{H}\widetilde{diag}\bigg\{\sum\limits_{s_{1}=1}^{S}       \sum\limits_{s_{2}=1}^{S} \sqrt{c_{l,s_{1},i}c_{l,s_{2},i}\delta_{l,s_{1}} \delta_{l,s_{2}} \varepsilon_{s_{1},i}\varepsilon_{s_{2},i}} f_{l,s_{1},i}({\bf \Phi}) f_{l,s_{2},i}^{H}({\bf \Phi}) \mathbf{a}_{B}(l,s_{1})\mathbf{a}^{H}_{B}(l,s_{2})\\\mathrm{sinc}^{2}\left(\kappa_r\pi\right)\bigg\}{\mathbf{q}}_{l,k}\bigg\}
			\\=\mathbb{E}\left\{{\mathbf{q}}_{l,k}^{H}{\mathbf{q}}_{l,k}\right\}\left(\gamma_{l,i}+\sum\limits_{s=1}^{S}c_{l,s,i}R\Big(\delta_{l,s} \left(\varepsilon_{s,i}\left(1-\mathrm{sinc}^{2}\left(\kappa_r\pi\right)\right)+1\right)  +\varepsilon_{s,i}+1  \Big)\right)
			\\+\left(\sum\limits_{s_{1}=1}^{S}        \sqrt{c_{l,s_{1},i} \delta_{l,s_{1}} \varepsilon_{s_{1},i}} f_{l,s_{1},i}^{H}({\bf \Phi}) \mathbf{a}^{H}_{B}(l,s_{1})\right)\widetilde{diag}\left\{\mathbb{E}\left\{{\mathbf{q}}_{l,k}{\mathbf{q}}_{l,k}^{H}\right\}\right\}  \bigg(\sum\limits_{s_{2}=1}^{S} \sqrt{c_{l,s_{2},i}\delta_{l,s_{2}}  \varepsilon_{s_{2},i}} \\f_{l,s_{2},i}({\bf \Phi}) \mathbf{a}_{B}(l,s_{2})\bigg)\mathrm{sinc}^{2}\left(\kappa_r\pi\right)
			\\=\bigg({\sum\limits_{s_{1}=1 }^S \sum\limits_{s_{2}=1 }^S} {\sqrt{c_{l,s_{1},k}c_{l,s_{2},k}\delta _{l,s_{1}}\delta _{l,s_{2}}\varepsilon _{s_{1},k}\varepsilon _{s_{2},k}} {f_{l,s_{1},k}^{H}({\bf\Phi})}{f_{l,s_{2},k}({\bf\Phi})}} {\mathbf{a}_{B}^{H}(l,s_{1})} {\mathbf{a}_{B}(l,s_{2})} \\+ { \sum\limits_{s=1}^S c_{l,s,k} B R (\delta _{l,s}+\varepsilon _{s,k}+1)} + {\gamma_{l,k}B}\bigg)\left(\gamma_{l,i}+\sum\limits_{s=1}^{S}c_{l,s,i}R\Big(\delta_{l,s} \left(\varepsilon_{s,i}\left(1-\mathrm{sinc}^{2}\left(\kappa_r\pi\right)\right)+1\right)  +\varepsilon_{s,i}+1  \Big)\right)
			\\+\left(\sum\limits_{s_{1}=1}^{S}        \sqrt{c_{l,s_{1},i} \delta_{l,s_{1}} \varepsilon_{s_{1},i}} f_{l,s_{1},i}^{H}({\bf \Phi}) \mathbf{a}^{H}_{B}(l,s_{1})\right)\bigg({ \sum\limits_{s=1}^S c_{l,s,k}  R (\delta _{l,s}+\varepsilon _{s,k}+1)} + {\gamma_{l,k}}\bigg)\mathbf{I}_{B}    \\\bigg(\sum\limits_{s_{2}=1}^{S} \sqrt{c_{l,s_{2},i}\delta_{l,s_{2}}  \varepsilon_{s_{2},i}} f_{l,s_{2},i}({\bf \Phi}) \mathbf{a}_{B}(l,s_{2})\bigg)\mathrm{sinc}^{2}\left(\kappa_r\pi\right)
			+\left(\sum\limits_{s_{1}=1}^{S}        \sqrt{c_{l,s_{1},i} \delta_{l,s_{1}} \varepsilon_{s_{1},i}} f_{l,s_{1},i}^{H}({\bf \Phi}) \mathbf{a}^{H}_{B}(l,s_{1})\right)\\\widetilde{diag}\bigg\{\sum\limits_{s_{3}=1}^{S}       \sum\limits_{s_{4}=1}^{S} \sqrt{c_{l,s_{3},k}c_{l,s_{4},k}\delta_{l,s_{3}} \delta_{l,s_{4}} \varepsilon_{s_{3},k}\varepsilon_{s_{4},k}} f_{l,s_{3},k}({\bf \Phi}) f_{l,s_{4},k}^{H}({\bf \Phi}) \mathbf{a}_{B}(l,s_{3})\mathbf{a}^{H}_{B}(l,s_{4})\bigg\}  \\  \bigg(\sum\limits_{s_{2}=1}^{S} \sqrt{c_{l,s_{2},i}\delta_{l,s_{2}}  \varepsilon_{s_{2},i}} f_{l,s_{2},i}({\bf \Phi}) \mathbf{a}_{B}(l,s_{2})\bigg)\mathrm{sinc}^{2}\left(\kappa_r\pi\right)
			\\=\bigg({\sum\limits_{s_{1}=1 }^S \sum\limits_{s_{2}=1 }^S} {\sqrt{c_{l,s_{1},k}c_{l,s_{2},k}\delta _{l,s_{1}}\delta _{l,s_{2}}\varepsilon _{s_{1},k}\varepsilon _{s_{2},k}} {f_{l,s_{1},k}^{H}({\bf\Phi})}{f_{l,s_{2},k}({\bf\Phi})}} {\mathbf{a}_{B}^{H}(l,s_{1})} {\mathbf{a}_{B}(l,s_{2})} \\+ { \sum\limits_{s=1}^S c_{l,s,k} B R (\delta _{l,s}+\varepsilon _{s,k}+1)} + {\gamma_{l,k}B}\bigg)\bigg(\gamma_{l,i}+\sum\limits_{s=1}^{S}c_{l,s,i}R\Big(\delta_{l,s} \left(\varepsilon_{s,i}\left(1-\mathrm{sinc}^{2}\left(\kappa_r\pi\right)\right)+1\right)  +\varepsilon_{s,i}+1  \Big)\bigg)\\
			+\left(\sum\limits_{s_{1}=1}^{S} \sum\limits_{s_{2}=1}^{S}\sqrt{c_{l,s_{1},i}c_{l,s_{2},i} \delta_{l,s_{1}}\delta_{l,s_{2}} \varepsilon_{s_{1},i}\varepsilon_{s_{2},i}} f_{l,s_{1},i}^{H}({\bf \Phi})f_{l,s_{2},i}({\bf \Phi}) \mathbf{a}^{H}_{B}(l,s_{1}) \mathbf{a}_{B}(l,s_{2})  \right) \\\bigg({ \sum\limits_{s=1}^S c_{l,s,k}  R (\delta _{l,s}+\varepsilon _{s,k}+1)} + {\gamma_{l,k}}\bigg) \mathrm{sinc}^{2}\left(\kappa_r\pi\right)
			+\left(\sum\limits_{s_{1}=1}^{S}        \sqrt{c_{l,s_{1},i} \delta_{l,s_{1}} \varepsilon_{s_{1},i}} f_{l,s_{1},i}^{H}({\bf \Phi}) \mathbf{a}^{H}_{B}(l,s_{1})\right)\\\widetilde{diag}\bigg\{\sum\limits_{s_{3}=1}^{S}       \sum\limits_{s_{4}=1}^{S} \sqrt{c_{l,s_{3},k}c_{l,s_{4},k}\delta_{l,s_{3}} \delta_{l,s_{4}} \varepsilon_{s_{3},k}\varepsilon_{s_{4},k}} f_{l,s_{3},k}({\bf \Phi}) f_{l,s_{4},k}^{H}({\bf \Phi}) \mathbf{a}_{B}(l,s_{3})\mathbf{a}^{H}_{B}(l,s_{4})\bigg\}  \\  \bigg(\sum\limits_{s_{2}=1}^{S} \sqrt{c_{l,s_{2},i}\delta_{l,s_{2}}  \varepsilon_{s_{2},i}} f_{l,s_{2},i}({\bf \Phi}) \mathbf{a}_{B}(l,s_{2})\bigg)\mathrm{sinc}^{2}\left(\kappa_r\pi\right).
		\end{array}
	\end{align}
	Therefore, the HWI term can be given by
	\begin{align}\label{HWI_begin}
		&E_{k}^{(\mathrm{hwi})}({\bf\Phi})
		=\kappa_u^{2}  \left(p_{k}{E}_{k}^{(\mathrm{signal})}({\bf\Phi})+\sum\limits_{i=1,i\neq k}^{K}p_{i}I_{k i}({\bf\Phi})\right)+\kappa_b^{2}\left(1+\kappa_u^{2}\right)\sum\limits_{l=1}^{L}\sum\limits_{i=1}^{K}p_{i}
		\nonumber\\ &
		\bigg(\bigg( {\sum\limits_{s_{1}=1 }^S \sum\limits_{s_{2}=1 }^S} \sqrt{c_{l,s_{1},k}c_{l,s_{2},k}\delta _{l,s_{1}}\delta _{l,s_{2}}\varepsilon _{s_{1},k}\varepsilon _{s_{2},k}} {f_{l,s_{1},k}^{H}({\bf\Phi})}{f_{l,s_{2},k}({\bf\Phi})} {\mathbf{a}_{B}^{H}(l,s_{1})} {\mathbf{a}_{B}(l,s_{2})} 
		\nonumber\\ &+{\gamma_{l,k}B}+{ \sum\limits_{s=1}^S c_{l,s,k} B R (\delta _{l,s}+\varepsilon _{s,k}+1)} \bigg)\bigg(\gamma_{l,i}+\sum\limits_{s=1}^{S}c_{l,s,i}R\Big(\delta_{l,s} \left(\varepsilon_{s,i}\left(1-\mathrm{sinc}^{2}\left(\kappa_r\pi\right)\right)+1\right)  +\varepsilon_{s,i}+1  \Big)\bigg)\nonumber\\ &
		+\bigg({ \sum\limits_{s=1}^S c_{l,s,k}  R (\delta _{l,s}+\varepsilon _{s,k}+1)} + {\gamma_{l,k}}\bigg) \mathrm{sinc}^{2}\left(\kappa_r\pi\right)\sum\limits_{s_{1}=1}^{S} \sum\limits_{s_{2}=1}^{S}\sqrt{c_{l,s_{1},i}c_{l,s_{2},i} \delta_{l,s_{1}}\delta_{l,s_{2}} \varepsilon_{s_{1},i}\varepsilon_{s_{2},i}} f_{l,s_{1},i}^{H}({\bf \Phi}) \nonumber\\ &f_{l,s_{2},i}({\bf \Phi})\mathbf{a}^{H}_{B}(l,s_{1}) \mathbf{a}_{B}(l,s_{2})   
		+\sum\limits_{s_{1}=1}^{S}\sum\limits_{s_{2}=1}^{S} \sum\limits_{s_{3}=1}^{S}  \sum\limits_{s_{4}=1}^{S}  \sqrt{c_{l,s_{1},i}c_{l,s_{2},i}c_{l,s_{3},k}c_{l,s_{4},k} \delta_{l,s_{1}}\delta_{l,s_{2}}\delta_{l,s_{3}} \delta_{l,s_{4}}}\nonumber\\ &\sqrt{\varepsilon_{s_{1},i}\varepsilon_{s_{2},i} \varepsilon_{s_{3},k}\varepsilon_{s_{4},k}}\mathrm{sinc}^{2}\left(\kappa_r\pi\right)f_{l,s_{1},i}^{H}({\bf \Phi})f_{l,s_{2},i}({\bf \Phi})f_{l,s_{3},k}({\bf \Phi}) f_{l,s_{4},k}^{H}({\bf \Phi})\nonumber\\ & \mathbf{a}^{H}_{B}(l,s_{1})\widetilde{diag}\left\{  \mathbf{a}_{B}(l,s_{3})\mathbf{a}^{H}_{B}(l,s_{4})\right\}    \mathbf{a}_{B}(l,s_{2})\bigg),
	\end{align}
	
	\section{}\label{appB}
	In the section, we transform the rate expression $R_{k}$ in (\ref{rate}) to get a tractable object function. To this end, we transform the summation of matrix products containing matrix $\mathbf{\Phi}_{s}$ into the matrix product only containing matrix $\mathbf{\Phi}$. To facilitate the transformation of the rate expression, we present definitions of some formulas that will appear in the results of the expression transformation.
	
	We define  
	${{\bf \overline Z}^{1}_{l,s}}\triangleq\sqrt {\frac{{{\delta _{l,s}\beta_{l,s}}}}{{{\delta _{l,s}} + 1}}} {{\bf \overline Z}_{l,s}}\in {{\mathbb{C}}^{B \times R}}$, ${{\bf \overline Z}^{2,k}_{l,s}}\triangleq\sqrt {{c_{l,s,k}\delta_{l,s}}}\;{{\bf \overline Z}_{l,s}}\in {{\mathbb{C}}^{B \times R}}$, ${{\bf \overline Z}^{3,k}_{l,s}}\triangleq\sqrt {{c_{l,s,k}\delta_{l,s}\varepsilon_{s,k}}}\\{{\bf \overline Z}_{l,s}}\in {{\mathbb{C}}^{B \times R}}$, ${{\bf \overline Z}^{4,k}_{l,s}}\triangleq c_{l,s,k}{{\bf \overline Z}^{1}_{l,s}}\in {{\mathbb{C}}^{B \times R}}$, and ${{\bf  \overline h}^{1}_{s,k}}\triangleq\sqrt {\frac{{{\varepsilon _{s,k}\alpha _{s,k}}}}{{{\varepsilon _{s,k}} + 1}}} {{\bf  \overline h}_{s,k}}\in {{\mathbb{C}}^{R \times 1}}$. Also, we define $\overline{\bf {H}}^{1} = [{\overline{\bf h}^{1}_1},{\overline{\bf h}^{1}_2},...,{\overline{\bf h}^{1}_K} ]\in {{\mathbb{C}}^{SR \times K}}$, ${\left(\overline{\bf h}_k^{1}\right)^{T}} = \left[ {{\left(\overline{\bf h}_{1,k}^{1}\right)^{T}},{\left(\overline{\bf h}_{2,k}^{1}\right)^{T}},...,{\left(\overline{\bf h}_{S,k}^{1}\right)^{T}}} \right]\in {{\mathbb{C}}^{SR \times 1}}$, $\overline{\mathbf{Z}}^{1}=\left[ \overline{\mathbf{Z}}^{1}_1,\overline{\mathbf{Z}}^{1}_2,\ldots,\overline{\mathbf{Z}}^{1}_S\right]\in {{\mathbb{C}}^{LB \times SR}}$, $ \left(\overline{\mathbf{Z}}_s^{1}\right)^{T} = \left[{\left(\overline{\bf Z}_{1,s}^{1}\right)^{T}},{\left(\overline{\bf Z}_{2,s}^{1}\right)^{T}},...,{\left(\overline{\bf Z}_{L,s}^{1}\right)^{T}} \right]\in {{\mathbb{C}}^{LB \times R}}$, and  $\overline{\mathbf{Z}}^{2,k}$, $\overline{\mathbf{Z}}^{3,k}$, $\overline{\mathbf{Z}}^{4,k}$ have the same form as $\overline{\mathbf{Z}}^{1}$.
	
	To further simplify the expression, we define some matrices as follows ${\mathbf U}_{s}^{1,k}\triangleq{\sum\limits_{l=1}^{L}}c_{l,s,k}{\mathbf I}_{R}$, ${\mathbf U}_{s}^{2,k}\triangleq{\sum\limits_{l=1}^{L}}c_{l,s,k}\delta_{l,s}{\mathbf I}_{R} $, ${\mathbf U}_{s}^{3,k}\triangleq{\sum\limits_{l=1}^{L}}c_{l,s,k}\varepsilon_{s,k}{\mathbf I}_{R}$, ${\mathbf U}_{s}^{4,k,i}\triangleq{\sum\limits_{l=1}^{L}}\sqrt{c_{l,s,k}c_{l,s,i}\varepsilon_{s,i}\varepsilon_{s,k}}{\mathbf I}_{R} \in {{\mathbb{C}}^{R \times R}}$. Additionally, we define $ {\mathbf U }^{1,k} = {\mathrm{diag}}\Big\{\mathrm{diag}^{T}\left\{{{\mathbf U}^{1,k}_1}\right\},\mathrm{diag}^{T}\left\{{\mathbf U}^{1,k}_2\right\},...,\mathrm{diag}^{T}\left\{{\mathbf U}^{1,k}_S\right\} \Big\} \in {{\mathbb{C}}^{SR \times SR}}$, and ${\mathbf U }^{2,k}$, ${\mathbf U }^{3,k}$, ${\mathbf U }^{4,k,i}$ have the same form as ${\mathbf U }^{1,k}$.
	
	For ease of description, we first provide the detailed process for the noise term $E_{k}^{(\mathrm{noise})}({\bf\Phi})$ in (\ref{noise}). Then, the noise term $E_{k}^{(\mathrm{noise})}({\bf\Phi})$ in (\ref{noise}) can be transformed into $E_{k}^{(\mathrm{noise,new})}({\bf\Phi})$, as follows
	\begin{align}\label{noise_new}
		\begin{array}{l}
			E_{k}^{(\mathrm{noise,new})}({\bf\Phi})
			={\sum\limits_{l=1}^{L}\sum\limits_{s_{1}=1 }^S \sum\limits_{s_{2}=1 }^S} {\sqrt{c_{l,s_{1},k}c_{l,s_{2},k}\delta _{l,s_{1}}\delta _{l,s_{2}}\varepsilon _{s_{1},k}\varepsilon _{s_{2},k}} {f_{l,s_{1},k}^{H}({\bf\Phi})}{f_{l,s_{2},k}({\bf\Phi})}} \\{\mathbf{a}_{B}^{H}(l,s_{1})} {\mathbf{a}_{B}(l,s_{2})} + { \sum\limits_{l=1}^{L}\sum\limits_{s=1}^S c_{l,s,k} B R (\delta _{l,s}+\varepsilon _{s,k}+1)}+\sum\limits_{l=1}^{L}\gamma_{l,k} B\\
			%		={\sum\limits_{l=1}^{L}\sum\limits_{s_{1}=1 }^S \sum\limits_{s_{2}=1 }^S} {\sqrt{c_{l,s_{1},k}c_{l,s_{2},k}\delta _{l,s_{1}}\delta _{l,s_{2}}\varepsilon _{s_{1},k}\varepsilon _{s_{2},k}} { \overline{\mathbf{h}}^{H}_{s_{1},k}{\bf\Phi}^{H}_{s_{1}}\mathbf{a}_{R}(l,s_{1}) }{\mathbf{a}_{R}^{H}(l,s_{2}) {\bf\Phi}_{s_{2}} }} \\\overline{\mathbf{h}}_{s_{2},k}{\mathbf{a}_{B}^{H}(l,s_{1})} {\mathbf{a}_{B}(l,s_{2})} + { \sum\limits_{l=1}^{L}\sum\limits_{s=1}^S c_{l,s,k} B R (\delta _{l,s}+\varepsilon _{s,k}+1)}+\sum\limits_{l=1}^{L}\gamma_{l,k} B\\
			={\sum\limits_{l=1}^{L}\sum\limits_{s_{1}=1 }^S \sum\limits_{s_{2}=1 }^S} {\sqrt{c_{l,s_{1},k}c_{l,s_{2},k}\delta _{l,s_{1}}\delta _{l,s_{2}}\varepsilon _{s_{1},k}\varepsilon _{s_{2},k}} \;{ \overline{\mathbf{h}}^{H}_{s_{1},k}{\bf\Phi}^{H}_{s_{1}}\mathbf{a}_{R}(l,s_{1}) }{\mathbf{a}_{B}^{H}(l,s_{1})}} \\ {\mathbf{a}_{B}(l,s_{2})}{\mathbf{a}_{R}^{H}(l,s_{2}) {\bf\Phi}_{s_{2}} \overline{\mathbf{h}}_{s_{2},k}} + { \sum\limits_{l=1}^{L}\sum\limits_{s=1}^S c_{l,s,k} B R (\delta _{l,s}+\varepsilon _{s,k}+1)}+\sum\limits_{l=1}^{L}\gamma_{l,k} B\\
			={\sum\limits_{l=1}^{L}\sum\limits_{s_{1}=1 }^S \sum\limits_{s_{2}=1 }^S}\left(\overline{\mathbf{h}}^{1}_{s_{1},k}\right)^{H}{\mathbf\Phi}_{s_{1}}^{H}\left({\bf \overline Z}_{l,s_{1}}^{1}\right)^{H} {{\bf \overline Z}_{l,s_{2}}^{1}}{\mathbf\Phi}_{s_{2}} \overline{\mathbf{h}}^{1}_{s_{2},k}  \\+ { \sum\limits_{l=1}^{L}\sum\limits_{s=1}^S c_{l,s,k} B R (\delta _{l,s}+\varepsilon _{s,k}+1)}+\sum\limits_{l=1}^{L}\gamma_{l,k} B\\
			=\left(\overline{\mathbf{h}}^{1}_{k}\right)^{H}{\mathbf\Phi}^{H}\left({\bf \overline Z}^{1}\right)^{H} {{\bf \overline Z}^{1}}{\mathbf\Phi} \overline{\mathbf{h}}^{1}_{k} + 
			{ \sum\limits_{l=1}^{L}\sum\limits_{s=1}^S c_{l,s,k} B R (\delta _{l,s}+\varepsilon _{s,k}+1)}+\sum\limits_{l=1}^{L}\gamma_{l,k} B.
		\end{array}
	\end{align}
	
	Based on the above derivations, we can find that only the terms containing the matrix $\mathbf{\Phi}_{s}$ need to be transformed in our expressions. Therefore, using the similar method, we focus on the matrix products containing matrix $\mathbf{\Phi}_{s}$ and transform  them into the matrix product only containing matrix $\mathbf{\Phi}$ for the signal term ${E}_{k}^{(\mathrm{signal})}({\bf\Phi})$, the interference term $I_{k i}({\bf\Phi})$, and the HWI term $E_{k}^{(\mathrm{hwi})}({\bf\Phi})$. 
	For example, some of the formulas in the signal term ${E}_{k}^{(\mathrm{signal})}({\bf\Phi})$ can be transformed as follows
	\begin{align}
		\begin{array}{l}
			\sum\limits_{l_{1}=1}^{L} \sum\limits_{l_{2}=1}^{L} \sum\limits_{s_{1}=1}^{S} \sum\limits_{s_{2}=1}^{S} \sum\limits_{s_{3}=1}^{S} \sum\limits_{s_{4}=1}^{S} \sqrt{c_{l_{1},s_{1},k}c_{l_{1},s_{2},k}c_{l_{2},s_{3},k}c_{l_{2},s_{4},k} \delta_{l_{1},s_{1}}\delta_{l_{1},s_{2}}
				\delta_{l_{2},s_{3}}\delta_{l_{2},s_{4}}\varepsilon_{s_{1},k}\varepsilon_{s_{2},k}\varepsilon_{s_{3},k}\varepsilon_{s_{4},k}}  \\\mathrm{sinc}^{2}\left(\kappa_r\pi\right) {f_{l_{1},s_{1},k}^{H}({\bf\Phi})f_{l_{1},s_{2},k}({\bf\Phi})f_{l_{2},s_{3},k}^{H}({\bf\Phi})f_{l_{2},s_{4},k}({\bf\Phi})} {\mathbf{a}_{B}^{H}(l_{1},s_{1})}{\mathbf{a}_{B}(l_{1},s_{2})}{\mathbf{a}_{B}^{H}(l_{2},s_{3})}{\mathbf{a}_{B}(l_{2},s_{4})}
			\\=\sum\limits_{l_{1}=1}^{L} \sum\limits_{l_{2}=1}^{L} \sum\limits_{s_{1}=1}^{S} \sum\limits_{s_{2}=1}^{S} \sum\limits_{s_{3}=1}^{S} \sum\limits_{s_{4}=1}^{S}\sqrt{c_{l_{1},s_{1},k}c_{l_{1},s_{2},k}c_{l_{2},s_{3},k}c_{l_{2},s_{4},k} \delta_{l_{1},s_{1}}\delta_{l_{1},s_{2}}\delta_{l_{2},s_{3}}\delta_{l_{2},s_{4}}\varepsilon_{s_{1},k}\varepsilon_{s_{2},k}\varepsilon_{s_{3},k}\varepsilon_{s_{4},k}} \\\mathrm{sinc}^{2}\left(\kappa_r\pi\right)\overline{\mathbf{h}}_{s_{1},k}^{H}{\mathbf\Phi}^{H}_{s_{1}}{\bf \overline Z}_{l_{1},s_{1}}^{H} {{\bf \overline Z}_{l_{1},s_{2}}}{\mathbf\Phi}_{s_{2}} \overline{\mathbf{h}}_{s_{2},k}
			\overline{\mathbf{h}}_{s_{3},k}^{H}{\mathbf\Phi}^{H}_{s_{3}}{\bf \overline Z}_{l_{2},s_{3}}^{H} {{\bf \overline Z}_{l_{2},s_{4}}}{\mathbf\Phi}_{s_{4}} \overline{\mathbf{h}}_{s_{4},k}
			\\=\sum\limits_{l_{1}=1}^{L} \sum\limits_{l_{2}=1}^{L} \sum\limits_{s_{1}=1}^{S} \sum\limits_{s_{2}=1}^{S} \sum\limits_{s_{3}=1}^{S} \sum\limits_{s_{4}=1}^{S}\mathrm{sinc}^{2}\left(\kappa_r\pi\right) \\\left(\overline{\mathbf{h}}^{1}_{s_{1},k}\right)^{H}{\mathbf\Phi}^{H}_{s_{1}}\left({\bf \overline Z}^{1}_{l_{1},s_{1}}\right)^{H} {{\bf \overline Z}^{1}_{l_{1},s_{2}}}{\mathbf\Phi}_{s_{2}} \overline{\mathbf{h}}^{1}_{s_{2},k}
			\left(\overline{\mathbf{h}}^{1}_{s_{3},k}\right)^{H}{\mathbf\Phi}^{H}_{s_{3}}\left({\bf \overline Z}^{1}_{l_{2},s_{3}}\right)^{H} {{\bf \overline Z}^{1}_{l_{2},s_{4}}}{\mathbf\Phi}_{s_{4}} \overline{\mathbf{h}}^{1}_{s_{4},k}
			\\=\mathrm{sinc}^{2}\left(\kappa_r\pi\right) \left(\overline{\mathbf{h}}^{1}_{k}\right)^{H}{\mathbf\Phi}^{H}\left({\bf \overline Z}^{1}\right)^{H} {{\bf \overline Z}^{1}}{\mathbf\Phi} \overline{\mathbf{h}}^{1}_{k}
			\left(\overline{\mathbf{h}}^{1}_{k}\right)^{H}{\mathbf\Phi}^{H}\left({\bf \overline Z}^{1}\right)^{H} {{\bf \overline Z}^{1}}{\mathbf\Phi}\overline{\mathbf{h}}^{1}_{k}.
		\end{array}
	\end{align}
	
	Due to the complicated form of the four terms in the rate expression, the detailed processes of transforming formulas are simple but very long and cumbersome. Therefore, to save space, we only present the final results in the following. Then, the transformed and tractable rate expression $r_{k}(\mathbf{\Phi})$ can be given by 
	\begin{align}\label{r_k}
		r_{k}(\mathbf{\Phi}) \approx \log _{2}\left(1+\frac{p_k {E}_{k}^{(\mathrm{signal,new})}({\bf\Phi})}{ \sum\limits_{i=1, i \neq k}^{K} p_i I_{k i}^{(\mathrm{new})}({\bf\Phi})+E_{k}^{(\mathrm{hwi,new})}({\bf\Phi})+\sigma^{2} E_{k}^{(\mathrm{\mathrm{noise,new}})}({\bf\Phi})}\right),
	\end{align}
	where the noise term $E_{k}^{(\mathrm{\mathrm{noise,new}})}({\bf\Phi})$ has been given in (\ref{noise_new}), and the signal term ${E}_{k}^{(\mathrm{signal,new})}({\bf\Phi})$ is
	\begin{align}\label{signal_new}
		{E}_{k}^{(\mathrm{signal,new})}({\bf\Phi})={E}_{k}^{(\mathrm{signal,1})}+{E}_{k}^{(\mathrm{signal,2})}({\bf\Phi}),
	\end{align}
	\begin{align}\label{signal1}
		\begin{array}{l}
			{E}_{k}^{(\mathrm{signal,1})}\\ %等价式1
			= 
			\sum\limits_{l_{1}=1}^{L} \sum\limits_{l_{2}=1}^{L} \sum\limits_{s_{1}=1}^{S} \sum\limits_{s_{2}=1}^{S} \Big(
			\sqrt{c_{l_{1},s_{1},k}c_{l_{1},s_{2},k} c_{l_{2},s_{2},k}c_{l_{2},s_{1},k}\delta_{l_{1},s_{1}} \delta_{l_{1},s_{2}}\delta_{l_{2},s_{1}} \delta_{l_{2},s_{2}}}\\\left(1+\varepsilon_{s_{2},k}\left(1-\mathrm{sinc}^{2}\left(\kappa_r\pi\right)\right)\right) {\mathbf{a}_{B}^{H}(l_{1},s_{1})} {{\bf \overline Z}_{l_{1},s_{2}}}{{\bf \overline Z}^{H}_{l_{2},s_{2}}} {{\bf \overline Z}_{l_{2},s_{1}}} {\mathbf{a}_{R}(l_{1},s_{1})} \\ + %式子6 
			{{c_{l_{1},s_{1},k}c_{l_{2},s_{2},k}}B^{2}R^{2}\mathrm{sinc}^{2}\left(\kappa_r\pi\right)\big(2(\delta_{l_{1},s_{1}}\varepsilon_{s_{2},k}+ \delta_{l_{1},s_{1}}+\varepsilon_{s_{1},k})+ \delta_{l_{1},s_{1}}\delta_{l_{2},s_{2}}+
				\varepsilon_{s_{1},k}\varepsilon_{s_{2},k}+1\big)}\Big)
			\\+ %式子4
			{\sum\limits_{l_{1}=1}^{L} \sum\limits_{l_{2}=1}^{L} \sum\limits_{s=1}^{S}}   c_{l_{1},s,k}B^{2}R\Big(c_{l_{2},s,k}\big((4\delta_{l_{1},s}\varepsilon_{s,k}+ \delta_{l_{1},s}\delta_{l_{2},s}+\varepsilon_{s,k}\varepsilon_{s,k})\left(1-\mathrm{sinc}^{2}\left(\kappa_r\pi\right)\right)\\+(2\delta_{l_{1},s}+2\varepsilon_{s,k}+1)\left(2-\mathrm{sinc}^{2}\left(\kappa_r\pi\right)\right)\big)+2\gamma_{l_{2},k}(\delta_{l_{1},s}+\varepsilon_{s,k}+1)\mathrm{sinc}\left(\kappa_r\pi\right)\Big)
			\\ +  %式子9
			{\sum\limits_{l=1}^{L} \sum\limits_{s_{1}=1}^{S} \sum\limits_{s_{2}=1}^{S}} c_{l,s_{1},k} c_{l,s_{2},k}B R^{2} \big(2\delta_{l,s_{1}}\varepsilon_{s_{1},k}\left(1-\mathrm{sinc}^{2}\left(\kappa_r\pi\right)\right)+ 2\delta_{l,s_{1}}+\varepsilon_{s_{1},k}+1\big)(\varepsilon_{s_{2},k}+1)
			\\ +  %式子9
			{\sum\limits_{l=1}^{L} \sum\limits_{s=1}^{S}}  c_{l,s,k}BR\Big(\delta_{l,s}\varepsilon_{s,k}(2c_{l,s,k}+\gamma_{l,k})\left(1-\mathrm{sinc}^{2}\left(\kappa_r\pi\right)\right)+ 2(\delta_{l,s}+\varepsilon_{s,k})(c_{l,s,k}+\gamma_{l,k})\\+c_{l,s,k}+2\gamma_{l,k}\Big)
			+ %式子13
			\left(\sum\limits_{l=1}^{L}\gamma_{l,k}B\right)^{2}+ \sum\limits_{l=1}^{L} \gamma_{l,k}^{2}B,
		\end{array}
	\end{align}
	\begin{align}\label{signal2}
		\begin{array}{l}
			{E}_{k}^{(\mathrm{signal,2})}({\bf\Phi})\\ %等价式1
			= \mathrm{sinc}^{2}\left(\kappa_r\pi\right) \left(\overline{\mathbf{h}}^{1}_{k}\right)^{H}{\mathbf\Phi}^{H}\left({\bf \overline Z}^{1}\right)^{H} {{\bf \overline Z}^{1}}{\mathbf\Phi} \overline{\mathbf{h}}^{1}_{k}\left(\overline{\mathbf{h}}^{1}_{k}\right)^{H}{\mathbf\Phi}^{H}\left({\bf \overline Z}^{1}\right)^{H} {{\bf \overline Z}^{1}}{\mathbf\Phi}\overline{\mathbf{h}}^{1}_{k}
			+2B\bigg(\sum\limits_{l=1}^{L}\gamma_{l,k}\mathrm{sinc}\left(\kappa_r\pi\right)\\+{\sum\limits_{l=1}^{L}  \sum\limits_{s=1}^{S}}  c_{l,s,k}R (\delta_{l,s}+\varepsilon_{s,k}+1)\mathrm{sinc}^{2}\left(\kappa_r\pi\right)\bigg) \left(\overline{\mathbf{h}}^{1}_{k}\right)^{H}{\mathbf\Phi}^{H}\left({\bf \overline Z}^{1}\right)^{H} {{\bf \overline Z}^{1}}{\mathbf\Phi}\overline{\mathbf{h}}^{1}_{k}  %式子3
			+ \left(\overline{\mathbf{h}}^{1}_{k}\right)^{H}{\mathbf\Phi}^{H}\left({\bf \overline Z}^{1}\right)^{H}\\\left(\left(1+\mathrm{sinc}^{2}\left(\kappa_r\pi\right)\right){\bf \overline Z}^{2,k}\left({\bf \overline Z}^{2,k}\right)^{H}+\left(1-\mathrm{sinc}^{2}\left(\kappa_r\pi\right)\right){\bf \overline Z}^{3,k}\left({\bf \overline Z}^{3,k}\right)^{H}\right) {{\bf \overline Z}^{1}}{\mathbf\Phi}\overline{\mathbf{h}}^{1}_{k}
			+B\Big(\left(\overline{\mathbf{h}}^{1}_{k}\right)^{H}\\{\mathbf\Phi}^{H}\left({\bf \overline Z}^{1}\right)^{H} {{\bf \overline Z}^{1}}{\mathbf\Phi}\left(2{\mathbf U }^{1,k}+\left({\mathbf U }^{2,k}+{\mathbf U }^{3,k}\right)\left(1-\mathrm{sinc}^{2}\left(\kappa_r\pi\right)\right)\right)\overline{\mathbf{h}}^{1}_{k}
			+\left(\overline{\mathbf{h}}^{1}_{k}\right)^{H}\big(2{\mathbf U }^{1,k}\\+\left({\mathbf U }^{2,k}+{\mathbf U }^{3,k}\right)\left(1-\mathrm{sinc}^{2}\left(\kappa_r\pi\right)\right)\big){\mathbf\Phi}^{H}\left({\bf \overline Z}^{1}\right)^{H} {{\bf \overline Z}^{1}}{\mathbf\Phi}\overline{\mathbf{h}}^{1}_{k}\Big)
			+\left(1+\mathrm{sinc}^{2}\left(\kappa_r\pi\right)\right)\\\left(\overline{\mathbf{h}}^{1}_{k}\right)^{H}{\mathbf\Phi}^{H}\bigg(\left({\bf \overline Z}^{4,k}\right)^{H} {{\bf \overline Z}^{1}}+\left({\bf \overline Z}^{1}\right)^{H} {{\bf \overline Z}^{4,k}}+\left({\bf \overline Z}^{1}\right)^{H}{\mathbf V}^{2,k} {{\bf \overline Z}^{1}}+R\left({\bf \overline Z}^{1}\right)^{H}{\mathbf V }^{1,k} {{\bf \overline Z}^{1}}\bigg){\mathbf\Phi} \overline{\mathbf{h}}^{1}_{k}.
		\end{array}
	\end{align}
	Also, the interference term $I_{k i}^{(\mathrm{new})}({\bf\Phi})$ is
	\begin{align}\label{interference_new}
		I_{k i}^{(\mathrm{new})}({\bf\Phi})=I_{k i}^{(1)}+I_{k i}^{(2)}({\bf\Phi}),
	\end{align}
	where
	\begin{align}\label{interference1}
		\begin{array}{l}
			I_{k i}^{(1)}
			\\=
			\sum\limits_{l_{1}=1}^{L} \sum\limits_{l_{2}=1}^{L} \sum\limits_{s_{1}=1}^{S} \sum\limits_{s_{2}=1}^{S} \Big(
			\sqrt{c_{l_{1},s_{1},k}c_{l_{1},s_{2},i} c_{l_{2},s_{2},i}c_{l_{2},s_{1},k}\delta_{l_{1},s_{1}} \delta_{l_{1},s_{2}}\delta_{l_{2},s_{1}} \delta_{l_{2},s_{2}}}\\\big(\left(1-\mathrm{sinc}^{2}\left(\kappa_r\pi\right)\right)\varepsilon_{s_{2},i}+1\big) {\mathbf{a}_{B}^{H}(l_{1},s_{1})} {{\bf \overline Z}_{l_{1},s_{2}}}{{\bf \overline Z}^{H}_{l_{2},s_{2}}} {{\bf \overline Z}_{l_{2},s_{1}}} {\mathbf{a}_{R}(l_{1},s_{1})} \\ + %式子6 
			\sqrt{c_{l_{1},s_{1},k}c_{l_{1},s_{2},i} c_{l_{2},s_{2},i}c_{l_{2},s_{1},k} \varepsilon_{s_{1},k}\varepsilon_{s_{1},i}
				\varepsilon_{s_{2},i}\varepsilon_{s_{2},k}}B^{2}\mathrm{sinc}^{2}\left(\kappa_r\pi\right)\overline{\mathbf{h}}_{s_{1},k}^{H}\overline{\mathbf{h}}_{s_{1},i} \overline{\mathbf{h}}_{s_{2},i}^{H}\overline{\mathbf{h}}_{s_{2},k}\Big) \\ + %式子3
			{\sum\limits_{l_{1}=1}^{L} \sum\limits_{l_{2}=1}^{L} \sum\limits_{s=1}^{S}} \sqrt{c_{l_{1},s,k}c_{l_{1},s,i} c_{l_{2},s,i}c_{l_{2},s,k}} B^{2} R \Big(\left(2\delta_{l_{1},s}+\varepsilon_{s,k}\right)\left(1+\varepsilon_{s,i}\left(1-\mathrm{sinc}^{2}\left(\kappa_r\pi\right)\right)\right)\\+\varepsilon_{s,i}+1\Big)  +  %式子10
			{\sum\limits_{l=1}^{L} \sum\limits_{s_{1}=1}^{S} \sum\limits_{s_{2}=1}^{S}}  
			c_{l,s_{1},k} c_{l,s_{2},i} B R^{2}\Big( \delta_{l,s_{2}}(\varepsilon_{s_{1},k}+1)\left(1+\varepsilon_{s_{2},i}\left(1-\mathrm{sinc}^{2}\left(\kappa_r\pi\right)\right)\right)\\+ (\delta_{l,s_{1}}+\varepsilon_{s_{1},k}+1)(\varepsilon_{s_{2},i}+1)\Big)
			+
			{\sum\limits_{l=1}^L \sum\limits_{s=1}^S} 
			B R \Big(c_{l,s,k} (\delta _{l,s}+\varepsilon _{s,k}+1)\gamma_{l,i}\\+c_{l,s,i} \left(\delta_{l,s}\left(1+\varepsilon_{s,i}\left(1-\mathrm{sinc}^{2}\left(\kappa_r\pi\right)\right)\right)+\varepsilon _{s,i}+1\right)\gamma_{l,k}\Big)
			+{\sum\limits_{l=1}^L \gamma_{l,k}\gamma_{l,i}B},
		\end{array}
	\end{align}
	\begin{align}\label{interference2}
		\begin{array}{l}
			I_{k i}^{(2)}({\bf\Phi})
			\\=\mathrm{sinc}^{2}\left(\kappa_r\pi\right) \left(\overline{\mathbf{h}}^{1}_{k}\right)^{H}{\mathbf\Phi}^{H}\left({\bf \overline Z}^{1}\right)^{H} {{\bf \overline Z}^{1}}{\mathbf\Phi} \overline{\mathbf{h}}^{1}_{i}
			\left(\overline{\mathbf{h}}^{1}_{i}\right)^{H}{\mathbf\Phi}^{H}\left({\bf \overline Z}^{1}\right)^{H} {{\bf \overline Z}^{1}}{\mathbf\Phi}\overline{\mathbf{h}}^{1}_{k}
			\\+B\Big( \left(\overline{\mathbf{h}}^{1}_{k}\right)^{H}{\mathbf\Phi}^{H}\left({\bf \overline Z}^{1}\right)^{H} {{\bf \overline Z}^{1}}{\mathbf\Phi} \Big(\left(\left(1-\mathrm{sinc}^{2}\left(\kappa_r\pi\right)\right){\mathbf U}^{3,i}+{\mathbf U}^{1,i}\right)\overline{\mathbf{h}}^{1}_{k}\\+\mathrm{sinc}^{2}\left(\kappa_r\pi\right)\overline{\mathbf{h}}^{1}_{i} \overline{\mathbf{h}}_{i}^{H}{\mathbf U}^{4,k,i}\overline{\mathbf{h}}_{k}\Big)+\mathrm{sinc}^{2}\left(\kappa_r\pi\right)\left(\overline{\mathbf{h}}^{1}_{i}\right)^{H}{\mathbf U}^{1,k}{\mathbf\Phi}^{H}\left({\bf \overline Z}^{1}\right)^{H} {{\bf \overline Z}^{1}}{\mathbf\Phi} \overline{\mathbf{h}}^{1}_{i}\Big)
			\\+B\Big( \left(\overline{\mathbf{h}}^{1}_{k}\right)^{H}\left(\left(1-\mathrm{sinc}^{2}\left(\kappa_r\pi\right)\right){\mathbf U}^{3,i}+{\mathbf U}^{1,i}\right){\mathbf\Phi}^{H}\left({\bf \overline Z}^{1}\right)^{H} {{\bf \overline Z}^{1}}{\mathbf\Phi} \overline{\mathbf{h}}^{1}_{k}+\mathrm{sinc}^{2}\left(\kappa_r\pi\right)\\ \left(\overline{\mathbf{h}}^{1}_{i}\right)^{H}{\mathbf\Phi}^{H}\left({\bf \overline Z}^{1}\right)^{H} {{\bf \overline Z}^{1}}{\mathbf\Phi}\left(\overline{\mathbf{h}}^{1}_{k} \overline{\mathbf{h}}_{k}^{H}{\mathbf U}^{4,k,i}\overline{\mathbf{h}}_{i}+{\mathbf U}^{1,k} \overline{\mathbf{h}}^{1}_{i}\right)\Big)
			\\+ \left(\overline{\mathbf{h}}^{1}_{k}\right)^{H}{\mathbf\Phi}^{H}\left({\bf \overline Z}^{1}\right)^{H} \bigg(R{\mathbf V}^{1,i}+{\mathbf V}^{2,i}+{{\bf \overline Z}^{2,i}}\left({{\bf \overline Z}^{2,i}}\right)^{H}+\left(1-\mathrm{sinc}^{2}\left(\kappa_r\pi\right)\right){{\bf \overline Z}^{3,i}}\left({{\bf \overline Z}^{3,i}}\right)^{H}\bigg){{\bf \overline Z}^{1}}{\mathbf\Phi}\overline{\mathbf{h}}^{1}_{k}
			\\+\mathrm{sinc}^{2}\left(\kappa_r\pi\right) \left(\overline{\mathbf{h}}^{1}_{i}\right)^{H}{\mathbf\Phi}^{H}\left({\bf \overline Z}^{1}\right)^{H} \left(R{\mathbf V}^{1,k}+{\mathbf V}^{2,k}+{{\bf \overline Z}^{2,k}}\left({{\bf \overline Z}^{2,k}}\right)^{H}\right){{\bf \overline Z}^{1}}{\mathbf\Phi} \overline{\mathbf{h}}^{1}_{i}.
		\end{array}
	\end{align}
	The HWI term $E_{k}^{(\mathrm{hwi,new})}({\bf\Phi})$ can be given by
	\begin{align}\label{HWI_new}
		\begin{array}{l}
			E_{k}^{(\mathrm{hwi,new})}({\bf\Phi})
			=\kappa_u^{2}  \left(p_{k}{E}_{k}^{(\mathrm{signal,new})}({\bf\Phi})+\sum\nolimits_{i=1,i\neq k}^{K}p_{i}I_{k i}^{(\mathrm{new})}({\bf\Phi})\right)+E_{k}^{(\mathrm{hwi,1})}({\bf\Phi}),
		\end{array}
	\end{align}
	where
	\begin{align}\label{HWI_new_1}
		\begin{array}{l}
			E_{k}^{(\mathrm{hwi,1})}({\bf\Phi})=\kappa_b^{2}\left(1+\kappa_u^{2}\right)\sum_{i=1}^{K}p_{i}\bigg(
			\sum\limits_{l=1}^L\bigg({\gamma_{l,k}B}+{ \sum\limits_{s=1}^S c_{l,s,k} B R (\delta _{l,s}+\varepsilon _{s,k}+1)}\bigg)
			\\\bigg(\gamma_{l,i}+\sum\limits_{s=1}^{S}c_{l,s,i}R\Big(\delta_{l,s} \left(\varepsilon_{s,i}\left(1-\mathrm{sinc}^{2}\left(\kappa_r\pi\right)\right)+1\right)  +\varepsilon_{s,i}+1  \Big)\bigg)\\
			+\left(\overline{\mathbf{h}}^{1}_{k}\right)^{H}{\mathbf\Phi}^{H}\left({\bf \overline Z}^{1}\right)^{H} \Big({\mathbf V }^{2,i}+R\Big({\mathbf V }^{1,i}+{\mathbf V }^{3,i}+{\mathbf V }^{4,i}\left(1-\mathrm{sinc}^{2}\left(\kappa_r\pi\right)\right)\Big)\Big){{\bf \overline Z}^{1}}{\mathbf\Phi} \overline{\mathbf{h}}^{1}_{k}
			\\
			+\mathrm{sinc}^{2}\left(\kappa_r\pi\right)\left(\overline{\mathbf{h}}^{1}_{i}\right)^{H}{\mathbf\Phi}^{H}\left({\bf \overline Z}^{1}\right)^{H} \bigg({\mathbf V }^{2,k}+R\left({\mathbf V }^{1,k}+{\mathbf V }^{3,k}\right)\\+\widetilde{diag}\left\{{{\bf \overline Z}^{1}}{\mathbf\Phi} \overline{\mathbf{h}}^{1}_{k}\left(\overline{\mathbf{h}}^{1}_{k}\right)^{H}{\mathbf\Phi}^{H}\left({\bf \overline Z}^{1}\right)^{H}\right\}\bigg){{\bf \overline Z}^{1}}{\mathbf\Phi} \overline{\mathbf{h}}^{1}_{i} \bigg).
		\end{array}
	\end{align}
	
	\section{}\label{appC}
	In this section, we need to calculate the gradients of $r_{k}(\boldsymbol{\theta})$ and $\underline{r}_{ k}(\mathbf{\boldsymbol{\theta}})$ in (\ref{r_k_min}), where $r_{k}(\boldsymbol{\theta})$ can be obtained by substituting $\mathbf{\Phi}=\mathrm{diag}  \left(\boldsymbol{v}\right)=\mathrm{diag}  \left( e^ {j \boldsymbol{\theta}}  \right)$ into $r_{k}(\mathbf{\Phi})$ in (\ref{r_k}).  According to Appendix \ref{appB}, we note that $E_{k}^{(\mathrm{\mathrm{noise,new}})}(\boldsymbol{\theta})$, ${E}_{k}^{(\mathrm{signal,new})}(\boldsymbol{\theta})$, $I_{k i}^{(\mathrm{new})}(\boldsymbol{\theta})$, and $E_{k}^{(\mathrm{hwi,new})}(\boldsymbol{\theta})$ have been constructed into the tractable expressions in (\ref{noise_new}), (\ref{signal_new}) $\sim$ (\ref{HWI_new_1}). Based on the similar idea in \cite[Section VI]{9973349}, we first calculate the gradients of the four terms by using an important property in \cite[Lemma 4]{9973349}. For ease of viewing, we provide this property as follows
	\begin{align}\label{gradient_lemma1}
		\begin{aligned}
			\frac{\partial \operatorname{Tr}\left\{\mathbf{A} \boldsymbol{\Phi  } \mathbf{B}  \boldsymbol{\Phi}^{H}\right\}}{\partial  \boldsymbol{\theta}}
			&=j \boldsymbol{\Phi}^{T}\left(\mathbf{A}^{T} \odot \mathbf{B}\right) \boldsymbol{v}^{*}
			-j \boldsymbol{\Phi}^{H}\left(\mathbf{A} \odot \mathbf{B}^{T}\right) \boldsymbol{v} \triangleq \boldsymbol{f}_{d}(\mathbf{A}, \mathbf{B}),
		\end{aligned}
	\end{align}
	where $\mathbf{A}$ and $\mathbf{B}$ are the given deterministic matrices.
	If $\mathbf{A}=\mathbf{A}^{H}, \mathbf{B}=\mathbf{B}^{H}$, we further have
	\begin{align}\label{gradient_lemma2}
		\frac{\partial \operatorname{Tr}\left\{\mathbf{A} \boldsymbol{\Phi } \mathbf{B} \boldsymbol{\Phi}^{H}\right\}}{\partial  \boldsymbol{\theta}}=2 \operatorname{Im}\left\{\boldsymbol{\Phi}^{H}\left(\mathbf{A} \odot \mathbf{B}^{T}\right) \boldsymbol{v}\right\}.
	\end{align}
	
	Based on this property, we first derive the gradient of $E_{k}^{(\mathrm{\mathrm{noise,new}})}(\boldsymbol{\theta})$ as follows 
	\begin{align}\label{gradient_noise}
		&\frac{\partial E_{k}^{(\mathrm{\mathrm{noise,new}})}(\boldsymbol{\theta})}{\partial{\bm \theta}}\nonumber\\
		&=\frac{\partial \left(\overline{\mathbf{h}}^{1}_{k}\right)^{H}{\mathbf\Phi}^{H}\left({\bf \overline Z}^{1}\right)^{H} {{\bf \overline Z}^{1}}{\mathbf\Phi} \overline{\mathbf{h}}^{1}_{k}}{\partial{\bm \theta}} + { \sum\limits_{l=1}^{L}\sum\limits_{s=1}^S c_{l,s,k} B R (\delta _{l,s}+\varepsilon _{s,k}+1)}+\sum\limits_{l=1}^{L}\gamma_{l,k} B\nonumber\\
		&=\frac{\partial \operatorname{Tr}\left\{\left({\bf \overline Z}^{1}\right)^{H} {{\bf \overline Z}^{1}} \boldsymbol{\Phi } \overline{\mathbf{h}}^{1}_{k}\left(\overline{\mathbf{h}}^{1}_{k}\right)^{H} \boldsymbol{\Phi}^{H}\right\}}{\partial{\bm \theta}} + { \sum\limits_{l=1}^{L}\sum\limits_{s=1}^S c_{l,s,k} B R (\delta _{l,s}+\varepsilon _{s,k}+1)}+\sum\limits_{l=1}^{L}\gamma_{l,k} B\nonumber\\
		&=2\operatorname{Im}\left\{\boldsymbol{\Phi}^{H}\left(\left({\bf \overline Z}^{1}\right)^{H} {{\bf \overline Z}^{1}} \odot \left(\overline{\mathbf{h}}^{1}_{k}\left(\overline{\mathbf{h}}^{1}_{k}\right)^{H}\right)^{T}\right) \boldsymbol{v}\right\}\nonumber\\
		&+ { \sum\limits_{l=1}^{L}\sum\limits_{s=1}^S c_{l,s,k} B R (\delta _{l,s}+\varepsilon _{s,k}+1)}+\sum\limits_{l=1}^{L}\gamma_{l,k} B.
	\end{align}
	
	According to the above derivation, we only need to focus on the gradients of the terms containing $\mathbf{\Phi}=\mathrm{diag}  \left( e^ {j \boldsymbol{\theta}}  \right)$. Therefore, we can simplify the gradients of ${E}_{k}^{(\mathrm{signal,new})}(\boldsymbol{\theta})$, $I_{k i}^{(\mathrm{new})}(\boldsymbol{\theta})$, and $E_{k}^{(\mathrm{hwi,new})}(\boldsymbol{\theta})$ as follows
	\begin{align} 
		\frac{\partial {E}_{k}^{(\mathrm{signal,new})}(\boldsymbol{\theta})}{\partial \boldsymbol{\theta}}= {E}_{k}^{(\mathrm{signal,1})}+\frac{\partial {E}_{k}^{(\mathrm{signal,2})}(\boldsymbol{\theta})}{\partial \boldsymbol{\theta}}, 
	\end{align}
	\begin{align} 
		\frac{\partial I_{k i}^{(\mathrm{new})}(\boldsymbol{\theta})}{\partial \boldsymbol{\theta}}=I_{k i}^{(\mathrm{1})}+\frac{\partial I_{k i}^{(\mathrm{2})}(\boldsymbol{\theta})}{\partial \boldsymbol{\theta}}, 
	\end{align}
	\begin{align} 
		\frac{\partial E_{k}^{(\mathrm{hwi,new})}(\boldsymbol{\theta})}{\partial \boldsymbol{\theta}}=\kappa_u^{2}  \left(p_{k}\frac{\partial {E}_{k}^{(\mathrm{signal,new})}(\boldsymbol{\theta})}{\partial \boldsymbol{\theta}}+\sum\nolimits_{i=1,i\neq k}^{K}p_{i}\frac{\partial I_{k i}^{(\mathrm{new})}(\boldsymbol{\theta})}{\partial \boldsymbol{\theta}}\right)+\frac{\partial E_{k}^{(\mathrm{hwi,1})}(\boldsymbol{\theta})}{\partial \boldsymbol{\theta}}, 
	\end{align} 
	where ${E}_{k}^{(\mathrm{signal,1})}$ and $I_{k i}^{(\mathrm{1})}$ have been given in (\ref{signal1}) and (\ref{interference1}). Then, we calculate $\frac{\partial {E}_{k}^{(\mathrm{signal,2})}(\boldsymbol{\theta})}{\partial \boldsymbol{\theta}}$, $\frac{\partial I_{k i}^{(\mathrm{2})}(\boldsymbol{\theta})}{\partial \boldsymbol{\theta}}$, and $\frac{\partial E_{k}^{(\mathrm{hwi,1})}(\boldsymbol{\theta})}{\partial \boldsymbol{\theta}}$ based on the chain rule and the similar method in (\ref{gradient_noise}). For example, the gradient of some formulas in the term ${E}_{k}^{(\mathrm{signal,2})}(\boldsymbol{\theta})$ can be calculated as follows
	\begin{align}
		&\frac{\partial\mathrm{sinc}^{2}\left(\kappa_r\pi\right) \left(\overline{\mathbf{h}}^{1}_{k}\right)^{H}{\mathbf\Phi}^{H}\left({\bf \overline Z}^{1}\right)^{H} {{\bf \overline Z}^{1}}{\mathbf\Phi} \overline{\mathbf{h}}^{1}_{k}\left(\overline{\mathbf{h}}^{1}_{k}\right)^{H}{\mathbf\Phi}^{H}\left({\bf \overline Z}^{1}\right)^{H} {{\bf \overline Z}^{1}}{\mathbf\Phi}\overline{\mathbf{h}}^{1}_{k}}{\partial \boldsymbol{\theta}}\nonumber\\
		&{\mathop  = \limits^{\left( d \right)} }2\mathrm{sinc}^{2}\left(\kappa_r\pi\right)\frac{\partial\operatorname{Tr}\left\{\left(\left({\bf \overline Z}^{1}\right)^{H} {{\bf \overline Z}^{1}}{\mathbf\Phi}\overline{\mathbf{h}}^{1}_{k}\left(\overline{\mathbf{h}}^{1}_{k}\right)^{H}{\mathbf\Phi}^{H}\left({\bf \overline Z}^{1}\right)^{H} {{\bf \overline Z}^{1}}\right) \boldsymbol{\Phi }\left( \overline{\mathbf{h}}^{1}_{k}\left(\overline{\mathbf{h}}^{1}_{k}\right)^{H}\right) \boldsymbol{\Phi}^{H}\right\}}{\partial \boldsymbol{\theta}}\nonumber\\
		&=4\mathrm{sinc}^{2}\left(\kappa_r\pi\right) \operatorname{Im}\left\{\boldsymbol{\Phi}^{H}\left(\left({\bf \overline Z}^{1}\right)^{H} {{\bf \overline Z}^{1}}{\mathbf\Phi}\overline{\mathbf{h}}^{1}_{k}\left(\overline{\mathbf{h}}^{1}_{k}\right)^{H}{\mathbf\Phi}^{H}\left({\bf \overline Z}^{1}\right)^{H} {{\bf \overline Z}^{1}} \odot \left(\overline{\mathbf{h}}^{1}_{k}\left(\overline{\mathbf{h}}^{1}_{k}\right)^{H}\right)^{T}\right) \boldsymbol{v}\right\},
	\end{align}
	where $(d)$ exploits the chain rule, and regards $\left({\bf \overline Z}^{1}\right)^{H} {{\bf \overline Z}^{1}}{\mathbf\Phi}\overline{\mathbf{h}}^{1}_{k}\left(\overline{\mathbf{h}}^{1}_{k}\right)^{H}{\mathbf\Phi}^{H}\left({\bf \overline Z}^{1}\right)^{H} {{\bf \overline Z}^{1}}$ and $\overline{\mathbf{h}}^{1}_{k}\left(\overline{\mathbf{h}}^{1}_{k}\right)^{H}$ as $\mathbf{A}$ and $\mathbf{B}$, respectively. Based on the above derivations and the similar method in \cite[Appendix K]{9973349}, we can get the gradients of $ {E}_{k}^{(\mathrm{signal,2})}(\boldsymbol{\theta})$, $I_{k i}^{(\mathrm{2})}(\boldsymbol{\theta})$, and $ E_{k}^{(\mathrm{hwi,1})}(\boldsymbol{\theta})$. Due to the similar and cumbersome processes of calculating the three terms, we only present the final results of $\frac{\partial {E}_{k}^{(\mathrm{signal,2})}(\boldsymbol{\theta})}{\partial \boldsymbol{\theta}}$, $\frac{\partial I_{k i}^{(\mathrm{2})}(\boldsymbol{\theta})}{\partial \boldsymbol{\theta}}$, and $\frac{\partial E_{k}^{(\mathrm{hwi,1})}(\boldsymbol{\theta})}{\partial \boldsymbol{\theta}}$ in the following.
	
	Based on the chain rule and the similar method in \cite[Section VI]{9973349}, we can get the gradient of $\underline{r}_{ k}(\mathbf{\boldsymbol{\theta}})$ with respect to $\boldsymbol{\theta}$ as follows
	\begin{align}\label{gradient_r_k_min}
		&\frac{\partial \underline{r}_{k}(\boldsymbol{\theta})}{\partial \boldsymbol{\theta}}=\frac{ \sum_{k=1}^{K}\left\{\frac{\exp \left\{-\mu r_{k}(\boldsymbol{\theta})\right\}}{1+\mathrm{SINR}^{\mathrm{new}}_{ k}(\boldsymbol{\theta})} \frac{\partial\mathrm{SINR}^{\mathrm{new}}_{ k}(\boldsymbol{\theta})}{\partial \boldsymbol{\theta}}\right\}}{(\ln 2)\left(\sum_{k=1}^{K} \exp \left\{-\mu r_{k}(\boldsymbol{\theta})\right\}\right)},
	\end{align}
	with
	\begin{align}\label{SINR_new}
		\begin{aligned}
			&\mathrm{SINR}^{\mathrm{new}}_{ k}(\boldsymbol{\theta})=\frac{p_k {E}_{k}^{(\mathrm{signal,new})}(\boldsymbol{\theta})}{ \sum\nolimits_{i=1, i \neq k}^{K} p_i I_{k i}^{(\mathrm{new})}(\boldsymbol{\theta})+E_{k}^{(\mathrm{hwi,new})}(\boldsymbol{\theta})+\sigma^{2} E_{k}^{(\mathrm{\mathrm{noise,new}})}(\boldsymbol{\theta})},
		\end{aligned}
	\end{align}
	and
	\begin{align}\label{gradient_SINR_new}
		\begin{aligned}
			&\frac{\partial \mathrm{SINR}^{\mathrm{new}}_{ k}(\boldsymbol{\theta})}{\partial \boldsymbol{\theta}}=\frac{p_{k} \frac{\partial {E}_{k}^{(\mathrm{signal,new})}(\boldsymbol{\theta})}{\partial \boldsymbol{\theta}}}{\sum\nolimits_{i=1, i \neq k}^{K} p_i I_{k i}^{(\mathrm{new})}(\boldsymbol{\theta})+E_{k}^{(\mathrm{hwi,new})}(\boldsymbol{\theta})+\sigma^{2} E_{k}^{(\mathrm{\mathrm{noise,new}})}(\boldsymbol{\theta})} \\
			&-p_{k} {E}_{k}^{(\mathrm{signal,new})}(\boldsymbol{\theta}) \frac{ \sum_{i=1, i \neq k}^{K} p_{i}\frac{\partial I_{k i}^{(\mathrm{new})}(\boldsymbol{\theta})}{\partial \boldsymbol{\theta}}+\frac{\partial E_{k}^{(\mathrm{hwi,new})}(\boldsymbol{\theta})}{\partial \boldsymbol{\theta}}+\sigma^{2} \frac{\partial E_{k}^{(\mathrm{\mathrm{noise,new}})}(\boldsymbol{\theta})}{\partial \boldsymbol{\theta}}}{\left(\sum\nolimits_{i=1, i \neq k}^{K} p_i I_{k i}^{(\mathrm{new})}(\boldsymbol{\theta})+E_{k}^{(\mathrm{hwi,new})}(\boldsymbol{\theta})+\sigma^{2} E_{k}^{(\mathrm{\mathrm{noise,new}})}(\boldsymbol{\theta})\right)^{2}}.
		\end{aligned}
	\end{align}
	
	%Therefore, the gradient of $\underline{r}_{k}(\boldsymbol{\theta})$ can be obtained by calculating $\frac{\partial {E}_{k}^{(\mathrm{signal,new})}(\boldsymbol{\theta})}{\partial \boldsymbol{\theta}}$, $\frac{\partial I_{k i}^{(\mathrm{new})}(\boldsymbol{\theta})}{\partial \boldsymbol{\theta}}$, $\frac{\partial E_{k}^{(\mathrm{hwi,new})}(\boldsymbol{\theta})}{\partial \boldsymbol{\theta}}$, and $\frac{\partial E_{k}^{(\mathrm{\mathrm{noise,new}})}(\boldsymbol{\theta})}{\partial \boldsymbol{\theta}}$ in (\ref{gradient_SINR_new}).
	
	The gradient of ${E}_{k}^{(\mathrm{signal,new})}(\boldsymbol{\theta})$ is given by
	\begin{align} 
		\frac{\partial {E}_{k}^{(\mathrm{signal,new})}(\boldsymbol{\theta})}{\partial \boldsymbol{\theta}}= {E}_{k}^{(\mathrm{signal,1})}+\frac{\partial {E}_{k}^{(\mathrm{signal,2})}(\boldsymbol{\theta})}{\partial \boldsymbol{\theta}}, 
	\end{align}
	where ${E}_{k}^{(\mathrm{signal,1})}$ has been given in (\ref{signal1}) and the gradient of ${E}_{k}^{(\mathrm{signal,2})}(\boldsymbol{\theta})$ is
	\begin{align}
		&\frac{\partial {E}_{k}^{(\mathrm{signal,2})}(\boldsymbol{\theta})}{\partial \boldsymbol{\theta}}\nonumber\\
		&= 4\mathrm{sinc}^{2}\left(\kappa_r\pi\right) \operatorname{Im}\left\{\boldsymbol{\Phi}^{H}\left(\left({\bf \overline Z}^{1}\right)^{H} {{\bf \overline Z}^{1}}{\mathbf\Phi}\overline{\mathbf{h}}^{1}_{k}\left(\overline{\mathbf{h}}^{1}_{k}\right)^{H}{\mathbf\Phi}^{H}\left({\bf \overline Z}^{1}\right)^{H} {{\bf \overline Z}^{1}} \odot \left(\overline{\mathbf{h}}^{1}_{k}\left(\overline{\mathbf{h}}^{1}_{k}\right)^{H}\right)^{T}\right) \boldsymbol{v}\right\}\nonumber
		\\&+4B\bigg(\sum\limits_{l=1}^{L}\gamma_{l,k}\mathrm{sinc}\left(\kappa_r\pi\right)+{\sum\limits_{l=1}^{L}  \sum\limits_{s=1}^{S}}  c_{l,s,k}R (\delta_{l,s}+\varepsilon_{s,k}+1)\mathrm{sinc}^{2}\left(\kappa_r\pi\right)\bigg)\nonumber \\&\operatorname{Im}\left\{\boldsymbol{\Phi}^{H}\left(\left({\bf \overline Z}^{1}\right)^{H} {{\bf \overline Z}^{1}} \odot \left(\overline{\mathbf{h}}^{1}_{k}\left(\overline{\mathbf{h}}^{1}_{k}\right)^{H}\right)^{T}\right) \boldsymbol{v}\right\}
		+2\operatorname{Im}\Bigg\{\boldsymbol{\Phi}^{H}\Bigg(\bigg(\left({\bf \overline Z}^{1}\right)^{H}\bigg(\left(1+\mathrm{sinc}^{2}\left(\kappa_r\pi\right)\right)\nonumber\\
		&{\bf \overline Z}^{2,k}\left({\bf \overline Z}^{2,k}\right)^{H}+\left(1-\mathrm{sinc}^{2}\left(\kappa_r\pi\right)\right){\bf \overline Z}^{3,k}\left({\bf \overline Z}^{3,k}\right)^{H}\bigg){{\bf \overline Z}^{1}}\bigg) \odot \left(\overline{\mathbf{h}}^{1}_{k}\left(\overline{\mathbf{h}}^{1}_{k}\right)^{H}\right)^{T}\Bigg) \boldsymbol{v}\Bigg\}\nonumber
		\\&+B\boldsymbol{f}_{d}\left(\left({\bf \overline Z}^{1}\right)^{H} {{\bf \overline Z}^{1}},\left(2{\mathbf U }^{1,k}+\left({\mathbf U }^{2,k}+{\mathbf U }^{3,k}\right)\left(1-\mathrm{sinc}^{2}\left(\kappa_r\pi\right)\right)\right)\overline{\mathbf{h}}^{1}_{k}\left(\overline{\mathbf{h}}^{1}_{k}\right)^{H}\right)\nonumber
		\\&+B\boldsymbol{f}_{d}\left(\left({\bf \overline Z}^{1}\right)^{H} {{\bf \overline Z}^{1}},\overline{\mathbf{h}}^{1}_{k}\left(\overline{\mathbf{h}}^{1}_{k}\right)^{H}\left(2{\mathbf U }^{1,k}+\left({\mathbf U }^{2,k}+{\mathbf U }^{3,k}\right)\left(1-\mathrm{sinc}^{2}\left(\kappa_r\pi\right)\right)\right)\right)\nonumber
		\\&+2\left(1+\mathrm{sinc}^{2}\left(\kappa_r\pi\right)\right)\operatorname{Im}\Bigg\{\boldsymbol{\Phi}^{H}\Bigg(\bigg(\left({\bf \overline Z}^{4,k}\right)^{H} {{\bf \overline Z}^{1}}+\left({\bf \overline Z}^{1}\right)^{H} {{\bf \overline Z}^{4,k}}+\left({\bf \overline Z}^{1}\right)^{H}{\mathbf V}^{2,k} {{\bf \overline Z}^{1}}\nonumber
		\\&+R\left({\bf \overline Z}^{1}\right)^{H}{\mathbf V }^{1,k} {{\bf \overline Z}^{1}}\bigg)\odot \left(\overline{\mathbf{h}}^{1}_{k}\left(\overline{\mathbf{h}}^{1}_{k}\right)^{H}\right)^{T}\Bigg) \boldsymbol{v}\Bigg\}.
	\end{align}
	
	The gradient of $I_{k i}^{(\mathrm{new})}(\boldsymbol{\theta})$ is given by
	\begin{align} 
		\frac{\partial I_{k i}^{(\mathrm{new})}(\boldsymbol{\theta})}{\partial \boldsymbol{\theta}}=I_{k i}^{(\mathrm{1})}+\frac{\partial I_{k i}^{(\mathrm{2})}(\boldsymbol{\theta})}{\partial \boldsymbol{\theta}}, 
	\end{align}
	where $I_{k i}^{(\mathrm{1})}$ has been given in (\ref{interference1}) and the gradient of $I_{k i}^{(\mathrm{2})}(\boldsymbol{\theta})$ is
	\begin{align}
		&\frac{\partial I_{k i}^{(\mathrm{2})}(\boldsymbol{\theta})}{\partial \boldsymbol{\theta}}\nonumber\\
		&=2\mathrm{sinc}^{2}\left(\kappa_r\pi\right) \Bigg( \operatorname{Im}\left\{\boldsymbol{\Phi}^{H}\left(\left({\bf \overline Z}^{1}\right)^{H} {{\bf \overline Z}^{1}}{\mathbf\Phi}\overline{\mathbf{h}}^{1}_{k}\left(\overline{\mathbf{h}}^{1}_{k}\right)^{H}{\mathbf\Phi}^{H}\left({\bf \overline Z}^{1}\right)^{H} {{\bf \overline Z}^{1}} \odot \left(\overline{\mathbf{h}}^{1}_{i}\left(\overline{\mathbf{h}}^{1}_{i}\right)^{H}\right)^{T}\right) \boldsymbol{v}\right\}\nonumber
		\\&+\operatorname{Im}\left\{\boldsymbol{\Phi}^{H}\left(\left({\bf \overline Z}^{1}\right)^{H} {{\bf \overline Z}^{1}}{\mathbf\Phi}\overline{\mathbf{h}}^{1}_{i}\left(\overline{\mathbf{h}}^{1}_{i}\right)^{H}{\mathbf\Phi}^{H}\left({\bf \overline Z}^{1}\right)^{H} {{\bf \overline Z}^{1}} \odot \left(\overline{\mathbf{h}}^{1}_{k}\left(\overline{\mathbf{h}}^{1}_{k}\right)^{H}\right)^{T}\right) \boldsymbol{v}\right\}\Bigg)\nonumber
		\\&+B\bigg(\boldsymbol{f}_{d}\left(\left({\bf \overline Z}^{1}\right)^{H} {{\bf \overline Z}^{1}},\Big(\left(\left(1-\mathrm{sinc}^{2}\left(\kappa_r\pi\right)\right){\mathbf U}^{3,i}+{\mathbf U}^{1,i}\right)\overline{\mathbf{h}}^{1}_{k}+\mathrm{sinc}^{2}\left(\kappa_r\pi\right)\overline{\mathbf{h}}^{1}_{i} \overline{\mathbf{h}}_{i}^{H}{\mathbf U}^{4,k,i}\overline{\mathbf{h}}_{k}\Big)\left(\overline{\mathbf{h}}^{1}_{k}\right)^{H}\right)\nonumber
		\\&+\mathrm{sinc}^{2}\left(\kappa_r\pi\right)\left(\boldsymbol{f}_{d}\left(\left({\bf \overline Z}^{1}\right)^{H} {{\bf \overline Z}^{1}},\overline{\mathbf{h}}^{1}_{i}\left(\overline{\mathbf{h}}^{1}_{i}\right)^{H}{\mathbf U}^{1,k}\right)+\boldsymbol{f}_{d}\left(\left({\bf \overline Z}^{1}\right)^{H} {{\bf \overline Z}^{1}},\left(\overline{\mathbf{h}}^{1}_{k} \overline{\mathbf{h}}_{k}^{H}{\mathbf U}^{4,k,i}\overline{\mathbf{h}}_{i}+{\mathbf U}^{1,k} \overline{\mathbf{h}}^{1}_{i}\right)\left(\overline{\mathbf{h}}^{1}_{i}\right)^{H}\right)\right)\nonumber
		\\&+\boldsymbol{f}_{d}\left(\left({\bf \overline Z}^{1}\right)^{H} {{\bf \overline Z}^{1}},\overline{\mathbf{h}}^{1}_{k}\left(\overline{\mathbf{h}}^{1}_{k}\right)^{H}\left(\left(1-\mathrm{sinc}^{2}\left(\kappa_r\pi\right)\right){\mathbf U}^{3,i}+{\mathbf U}^{1,i}\right)\right)\bigg)+2 \operatorname{Im}\Bigg\{\boldsymbol{\Phi}^{H}\Bigg(\left({\bf \overline Z}^{1}\right)^{H}\nonumber\\
		&\bigg(R{\mathbf V}^{1,i}+{\mathbf V}^{2,i}+{{\bf \overline Z}^{2,i}}\left({{\bf \overline Z}^{2,i}}\right)^{H}+\left(1-\mathrm{sinc}^{2}\left(\kappa_r\pi\right)\right){{\bf \overline Z}^{3,i}}\left({{\bf \overline Z}^{3,i}}\right)^{H}\bigg){{\bf \overline Z}^{1}} \odot \left(\overline{\mathbf{h}}^{1}_{k}\left(\overline{\mathbf{h}}^{1}_{k}\right)^{H}\right)^{T}\Bigg) \boldsymbol{v}\Bigg\}\nonumber
		\\&+2\mathrm{sinc}^{2}\left(\kappa_r\pi\right) \operatorname{Im}\left\{\boldsymbol{\Phi}^{H}\left(\left({\bf \overline Z}^{1}\right)^{H} \left(R{\mathbf V}^{1,k}+{\mathbf V}^{2,k}+{{\bf \overline Z}^{2,k}}\left({{\bf \overline Z}^{2,k}}\right)^{H}\right){{\bf \overline Z}^{1}} \odot \left(\overline{\mathbf{h}}^{1}_{i}\left(\overline{\mathbf{h}}^{1}_{i}\right)^{H}\right)^{T}\right) \boldsymbol{v}\right\}.
	\end{align}
	
	The gradient of $E_{k}^{(\mathrm{hwi,new})}(\boldsymbol{\theta})$ is given by
	\begin{align} 
		\frac{\partial E_{k}^{(\mathrm{hwi,new})}(\boldsymbol{\theta})}{\partial \boldsymbol{\theta}}=\kappa_u^{2}  \left(p_{k}\frac{\partial {E}_{k}^{(\mathrm{signal,new})}(\boldsymbol{\theta})}{\partial \boldsymbol{\theta}}+\sum\nolimits_{i=1,i\neq k}^{K}p_{i}\frac{\partial I_{k i}^{(\mathrm{new})}(\boldsymbol{\theta})}{\partial \boldsymbol{\theta}}\right)+\frac{\partial E_{k}^{(\mathrm{hwi,1})}(\boldsymbol{\theta})}{\partial \boldsymbol{\theta}}, 
	\end{align}
	where
	\begin{align}
		&\frac{\partial E_{k}^{(\mathrm{hwi,1})}(\boldsymbol{\theta})}{\partial \boldsymbol{\theta}}
		=\kappa_b^{2}\left(1+\kappa_u^{2}\right)\sum_{i=1}^{K}p_{i}\Bigg(\nonumber
		\\&2 \operatorname{Im}\left\{\boldsymbol{\Phi}^{H}\left(\left({\bf \overline Z}^{1}\right)^{H} \Big({\mathbf V }^{2,i}+R\Big({\mathbf V }^{1,i}+{\mathbf V }^{3,i}+{\mathbf V }^{4,i}\left(1-\mathrm{sinc}^{2}\left(\kappa_r\pi\right)\right)\Big)\Big){{\bf \overline Z}^{1}} \odot \left(\overline{\mathbf{h}}^{1}_{k}\left(\overline{\mathbf{h}}^{1}_{k}\right)^{H}\right)^{T}\right) \boldsymbol{v}\right\}\nonumber
		\\
		&+2\mathrm{sinc}^{2}\left(\kappa_r\pi\right)\Bigg( \operatorname{Im}\left\{\boldsymbol{\Phi}^{H}\left(\left({\bf \overline Z}^{1}\right)^{H} \Big({\mathbf V }^{2,k}+R\left({\mathbf V }^{1,k}+{\mathbf V }^{3,k}\right)\Big){{\bf \overline Z}^{1}} \odot \left(\overline{\mathbf{h}}^{1}_{i}\left(\overline{\mathbf{h}}^{1}_{i}\right)^{H}\right)^{T}\right) \boldsymbol{v}\right\}\nonumber
		\\&+ \operatorname{Im}\left\{\boldsymbol{\Phi}^{H}\left(\left({\bf \overline Z}^{1}\right)^{H}\widetilde{diag}\left\{{{\bf \overline Z}^{1}}{\mathbf\Phi} \overline{\mathbf{h}}^{1}_{k}\left(\overline{\mathbf{h}}^{1}_{k}\right)^{H}{\mathbf\Phi}^{H}\left({\bf \overline Z}^{1}\right)^{H}\right\}{{\bf \overline Z}^{1}} \odot \left(\overline{\mathbf{h}}^{1}_{i}\left(\overline{\mathbf{h}}^{1}_{i}\right)^{H}\right)^{T}\right) \boldsymbol{v}\right\}\nonumber
		\\&+ \operatorname{Im}\left\{\boldsymbol{\Phi}^{H}\left(\left({\bf \overline Z}^{1}\right)^{H}\widetilde{diag}\left\{{{\bf \overline Z}^{1}}{\mathbf\Phi} \overline{\mathbf{h}}^{1}_{i}\left(\overline{\mathbf{h}}^{1}_{i}\right)^{H}{\mathbf\Phi}^{H}\left({\bf \overline Z}^{1}\right)^{H}\right\}{{\bf \overline Z}^{1}} \odot \left(\overline{\mathbf{h}}^{1}_{k}\left(\overline{\mathbf{h}}^{1}_{k}\right)^{H}\right)^{T}\right) \boldsymbol{v}\right\}\Bigg)\Bigg).
	\end{align}
	
	Similarly, we use the same method to get the gradient of $r_{k}(\boldsymbol{\theta})$ as follows
	\begin{align}\label{gradient_r_k}
		&\frac{\partial r_{k}(\boldsymbol{\theta})}{\partial \boldsymbol{\theta}}=\frac{\frac{\partial\mathrm{SINR}^{\mathrm{new}}_{ k}(\boldsymbol{\theta})}{\partial \boldsymbol{\theta}}}{(\ln 2)\left(1+\mathrm{SINR}^{\mathrm{new}}_{ k}(\boldsymbol{\theta})\right)},
	\end{align}
	where $\mathrm{SINR}^{\mathrm{new}}_{ k}(\boldsymbol{\theta})$ and $\frac{\partial\mathrm{SINR}^{\mathrm{new}}_{ k}(\boldsymbol{\theta})}{\partial \boldsymbol{\theta}}$ have been given in (\ref{SINR_new}) and (\ref{gradient_SINR_new}).

\end{appendices}

%%%%%%%%%%%%%%%%%%%%%%%%%%%%%%%%%%%%% Reference
\bibliographystyle{IEEEtran}
\vspace{-6pt}
\bibliography{myref.bib}
%\end{thebibliography}

\end{document}